\documentclass{pacificfusion}
\usepackage[sort&compress,numbers]{natbib}
\usepackage{xcolor}
\usepackage{appendix}
\usepackage{subcaption}
\usepackage{mwe}
\usepackage{graphicx}
\usepackage{xr}
\usepackage{marvosym}
\usepackage{comment}

\graphicspath{{assets/images/}}

\title{Affordable, manageable, practical, and scalable (AMPS) high-yield and high-gain inertial fusion}

\author[1]{Andrew Alexander}
\author[1]{Laura Robin Benedetti}
\author[1]{Indrani Bhattacharyya}
\author[1]{Jared Bowen}
\author[1]{June Cabatu}
\author[1]{Virgil Cacdac}
\author[1]{Chhavi Chhavi}
\author[1]{Chiatai Chen}
\author[1]{Karen Chen}
\author[1]{Dan Clark}
\author[1]{Jerry Clark}
\author[1]{Tyler Cope}
\author[1]{Will Dannemann}
\author[1]{Scott Davidson}
\author[1]{David DeHaan}
\author[1]{John Dugan}
\author[1]{Mindy Eihusen}
\author[1]{C. Leland Ellison}
\author[1]{Carlos Esquivel}
\author[1]{David Ethridge}
\author[1]{Blake Ferguson}
\author[1]{Bryan Ferguson}
\author[1]{Jon Fry}
\author[1]{Fernando Garcia-Rubio}
\author[1]{Tarun Goyal}
\author[1]{Gary Grim}
\author[1]{Justin Grodman}
\author[1]{Ben Haid}
\author[1]{Fred Howland}
\author[1]{Van Huynh}
\author[1]{Vishal John}
\author[1]{Patrick Knapp}
\author[1]{Isaac Kravitz}
\author[1]{Eric S. Lander}
\author[1]{Samuel Langendorf}
\author[1]{Keith LeChien}
\author[1]{Anthony Link}
\author[1]{Nathan Meezan}
\author[1]{Douglas S. Miller}
\author[1]{Nantas Nardelli}
\author[1]{Queenelle Ogirri}
\author[1]{Jon He Peng}
\author[1]{Alexander Pinto}
\author[1]{Rudolph Powser}
\author[1]{Fritz Roy Puno}
\author[1]{Kenny Quang}
\author[1]{Brett Rahn}
\author[1]{Will Regan}
\author[1]{Kelsey Reichenbach}
\author[1,5]{Adam Reyes}
\author[1]{Courtney Richardson}
\author[1]{David Rose}
\author[1]{Joseph Samaniego}
\author[1]{Paul F. Schmit}
\author[1]{Victor Silva}
\author[1]{Nick Simon}
\author[1]{Shiva Sitaraman}
\author[1]{Hardeep Sullan}
\author[1]{James Trebesch}
\author[1]{Minh Truong}
\author[1]{Carrie Von Muench}
\author[1]{Cory Waltz}
\author[1]{Doug Williams}
\author[1]{Echo Wood}
\author[1]{Sid Wu}
\author[1]{Alex B. Zylstra}

\affil[1]{Pacific Fusion Corporation, Fremont, CA, 94538, USA}

\begin{abstract}

High-yield inertial fusion offers a transformative path to affordable clean firm power and advanced defense capabilities. Recent milestones at large facilities, particularly the National Ignition Facility (NIF), have demonstrated the feasibility of ignition but highlight the need for approaches that can deliver large amounts of energy to fusion targets at much higher efficiency and lower cost. We propose that pulser-driven inertial fusion energy (IFE), which uses high-current pulsed-power technology to compress targets to thermonuclear conditions, can achieve this goal. In this paper, we detail the physics basis for pulser IFE, focusing on magnetized liner inertial fusion (MagLIF), where cylindrical metal liners compress DT fuel under strong magnetic fields and pre-heat. We discuss how the low implosion velocities, direct-drive efficiency, and scalable pulser architecture can achieve ignition-level conditions at low capital cost. Our multi-dimensional simulations, benchmarked against experiments at the Z facility, show that scaling from 20 MA to 50–60 MA of current enables net facility gain. We then introduce our Demonstration System (DS), a pulsed-power driver designed to deliver more than 60 MA and store approximately 80 MJ of energy. The DS is designed to achieve a $1000\times$ increase in effective performance compared to the NIF, delivering approximately $100\times$ greater facility-level energy gain --- and importantly, achieving net facility gain, or $Q_f>1$ --- at just 1/10 the capital cost. We also examine the engineering requirements for repetitive operation, target fabrication, and chamber maintenance, highlighting a practical roadmap to commercial power plants. 
\\ \\

\end{abstract}

\begin{document}
\maketitle

\section{Introduction}
Growing global demand for energy, increasing urgency for clean power sources, \cite{energy_2023} and recent scientific advances in fusion \cite{ARC_2015, NIFLawson_2022, NIFTargetGain_2024} have catalyzed a national conversation regarding the most promising path to commercial fusion energy. It is widely understood that a practical fusion system would be revolutionary for the energy supply, propelling economic growth and improving the feasibility of meeting societal decarbonization goals.

A fusion reaction combines two light nuclei into a heavier nucleus, releasing energy. Stars burn hydrogen into helium under gravitational confinement. Terrestrially, the most readily accessible fusion reaction is to fuse two hydrogen isotopes, deuterium (D) and tritium (T), into helium. DT fusion requires reaching temperatures of several keV (1 keV $\approx 10^7$ K). To extract energy from fusion, the total fusion energy released by the fuel must exceed the energy invested in creating and sustaining the plasma ($Q_p$ for the plasma energy gain factor). This leads to a criterion on the product of the plasma pressure ($P$) and energy confinement time ($\tau$) by Lawson \cite{Lawson_1957}, with $P\tau$ requirements of 10's of $\textrm{atm-s}$.

\begin{figure*}
\centering
\includegraphics[width=5in]{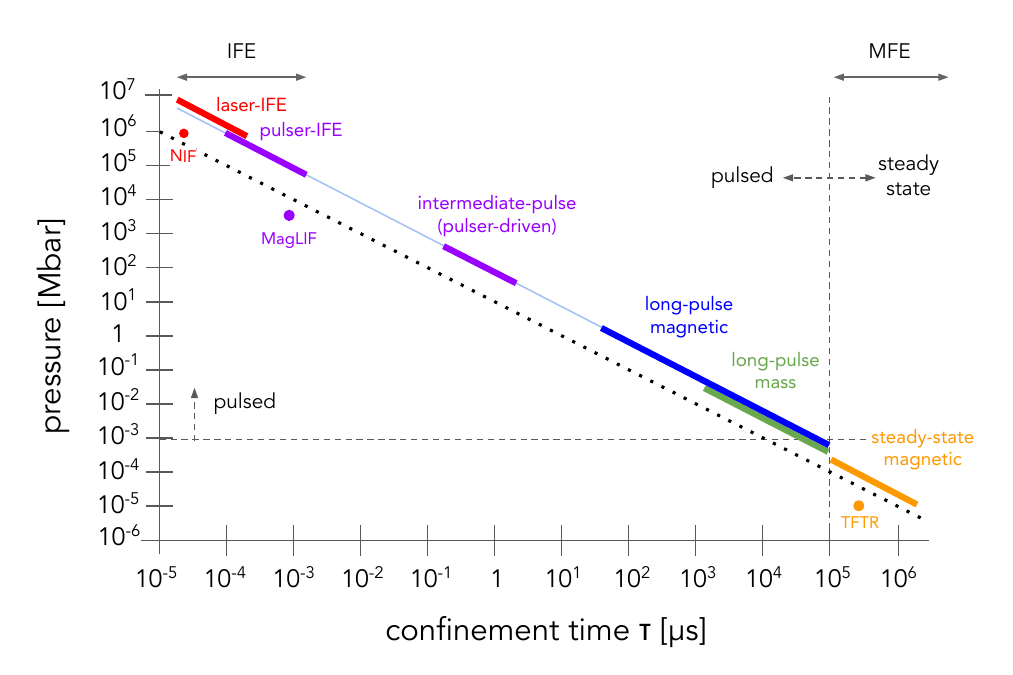}
\caption{\label{fig:SfA}  Parameter space of pressure ($P$) and confinement time ($\tau$) for fusion approaches, adapted from Science for America\cite{SfA}. Pulser-driven fusion can operate in the regimes shown in purple, including inertial fusion (the focus of this paper) and intermediate confinement times.}
\end{figure*}

Fusion has been an active area of basic research for several decades with investment overwhelmingly into two approaches: (1) magnetic fusion energy (MFE), in which a plasma's charged particles are confined by magnetic fields over long periods of time, and (2) inertial fusion energy (IFE), in which a target (small container holding the fuel) is rapidly imploded to high pressure and hydrodynamically confined during the burn by its own inertia. The confinement times for these two approaches differ by 8-12 orders of magnitude: seconds and above for MFE vs. below 10 ns for IFE, with the pressures varying inversely. The full parameter space for pressure ($P$) and confinement time ($\tau$) is shown in Fig. \ref{fig:SfA}. It also includes approaches with intermediate times and pressures between those for MFE and IFE \cite{SfA,JASON,lindemuth2009fundamental}. For MFE, public funding has predominantly supported the tokamak approach with historical investments into major facilities including JET \cite{jet1992fusion} and TFTR \cite{strachan1994fusion} in the 1990s, and ongoing significant investment into the construction of ITER \cite{holtkamp2007overview}. For IFE, public funding has predominantly supported laser-driven inertial confinement fusion (ICF) — that is, using lasers to implode a target. Major facilities with laser-based drivers have been constructed, including the National Ignition Facility (NIF) \cite{miller2004national}, OMEGA \cite{boehly1997initial}, Laser Megajoule \cite{fleurot2005laser}, and others.

The science basis for inertial fusion has advanced significantly in the last few years with the demonstration of ignition and propagating burn on the NIF, which was the first controlled experiment to exceed Lawson's criterion for ignition \cite{NIFLawson_2022} and subsequently demonstrate $Q_{target} > 1$ \cite{NIFTargetGain_2024}. While a groundbreaking scientific demonstration, the NIF is still far from overall net facility energy gain ($Q_f>1$), defined as fusion energy output exceeding the energy stored by the driver; the record facility-level gain achieved on the NIF is $Q_f \sim 0.016$, as discussed later. This is due in part to the laser system's low efficiency ($\sim 0.7 \%$) in converting stored energy to laser light incident on the target, and in part from the inefficiency of indirect-drive in converting incident laser light to heating of the fuel ($\sim 1 \%$), a total efficiency of $<0.01\%$. A more efficient and scalable approach to inertial fusion is needed for commercial power production.

Any realistic fusion energy system will need to be affordable, manageable, practical, and scalable (AMPS). By these terms we mean:
\begin{itemize}
    \item{\textbf{Affordable.} The overnight capital cost (OCC) and operating costs of a fusion energy system must be economically competitive with other power sources.}
    \item{\textbf{Manageable.} The reliability, availability, maintainability, and inspectability (RAMI) of key components must be sufficiently manageable to achieve high system up-time.}
    \item{\textbf{Practical.} The system has a credible scientific and technological basis while being compatible with energy market needs.}
    \item{\textbf{Scalable.} The system is based upon mass-manufacturable components, with no major supply chain bottlenecks, so that power plants can be deployed rapidly.}
\end{itemize}

This paper focuses on an alternative to laser IFE: pulser IFE, which has also seen major progress in the past few years. In pulser IFE, a fast pulsed power driver (risetime $\sim 100$ ns) or \textit{pulser}, magnetically implodes a fusion target to reach fusion conditions (with $\tau$ $\lesssim 1$ ns). The range of parameters for pulser IFE is overlapping and adjacent to those for which ignition was demonstrated on the NIF with lasers (Fig. \ref{fig:SfA}). (As shown in Fig. \ref{fig:SfA}, there are other (non-inertial) pulser-driven approaches to fusion that involve much longer confinement times; these approaches will not be discussed here.)

As discussed in a recent community white paper \cite{pmfe_whitepaper_2024} and elaborated here, pulser IFE is uniquely suited to address the AMPS requirements above. The driver technology for pulser IFE is more efficient than for laser IFE: as discussed below, Pacific Fusion's pulser IFE Demonstration System (DS) is approximately  $200\times$ as efficient at coupling stored energy to fuel internal energy compared to laser indirect-drive on the NIF. This efficiency advantage is due in part to recent developments in pulsed power architectures that enable systems that are highly modular and scalable, and that can be designed for maintainability and practicality. 

\begin{figure}
\centering
\includegraphics[width=3.5in]{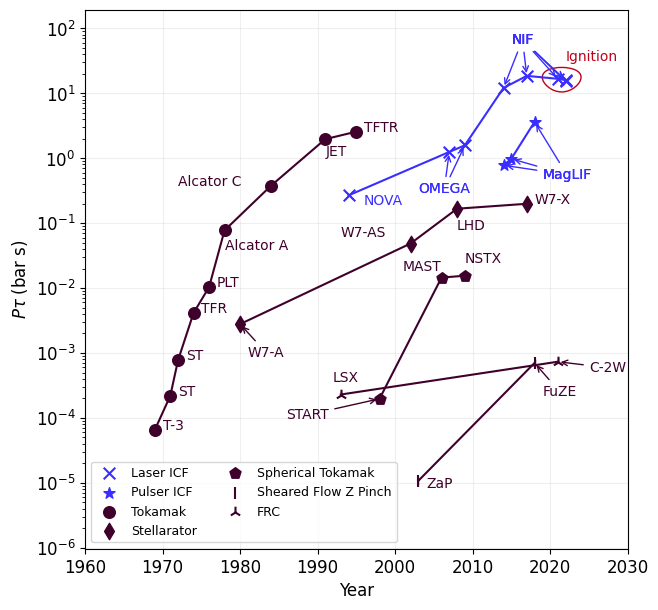}
\caption{\label{fig:Wurzel} Fusion triple product ($nT\tau$, equivalently $P\tau$) for various fusion approaches versus year. Figure derived from Ref.\ \citenum{wurzel2021} using the associated generator at \href{https://github.com/swurzel/lawson-criterion-paper}{https://github.com/swurzel/lawson-criterion-paper}). Several concepts, such as spherical tokamaks, sheared flow z-pinch, and FRCs have newer records that have not yet been incorporated into this analysis. For a description of this dataset and associated definitions, see Ref.~\cite{wurzel2021}.} 
\end{figure}

The technical basis of pulser IFE stems from experiments demonstrating values of $P \tau$ on par with those obtained by laser ICF and tokamaks. Recent work by Wurzel and Hsu \cite{wurzel2021} documented the Lawson product $P \tau$ achieved by various fusion approaches over time. For pulser ICF, the leading experimental precedent was established on the Z facility at Sandia National Laboratories \cite{sinars2020review}, at which the magnetized liner inertial fusion (MagLIF) target concept\cite{slutz2010pulsed,Slutz2012HighGain} is the most explored in published literature. In Fig. \ref{fig:Wurzel}, we adapt Ref. \citenum{wurzel2021} to include more recent MagLIF experiments from Ref. \citenum{Knapp_2022}; subsequent experiments have exceeded these values \cite{lewis2024developments}.

Laser-driven ICF on the NIF has demonstrated the highest $P\tau$ experimentally achieved to date with any fusion concept. MagLIF has demonstrated the second-highest $P\tau$, second only to laser-driven ICF and exceeding even the highest tokamak $P\tau$ to date. Notably, pulser ICF achieved these results with a small fraction of the investment dedicated to tokamaks and laser ICF: according to the 2018 JASON report on \textit{Prospects for Low Cost Fusion Development}, the total funding for pulser ICF has been approximately 1 \% of the total funding for laser ICF and MFE \cite{JASON}.

\subsection{Pacific Fusion's Roadmap}

Pacific Fusion is a privately-funded company founded in 2023 to develop and realize the commercial potential of pulser-driven inertial fusion energy. At a high level, our immediate objectives are to:
\begin{enumerate}
    \item{Demonstrate facility gain ($Q_f>1$)}, i.e., generate more fusion energy output than was stored in the capacitors
    \item{Resolve significant hurdles to building fusion power plants}
\end{enumerate}
Together, in collaboration with our key partners at U.S. National Laboratories and academic institutions, our goal is to achieve these objectives by the end of the decade. The first goal will represent a major milestone in fusion engineering: while inertial fusion has exceeded target gain greater than 1 (NIF, $Q_{target} \sim 2.4$) and magnetic confinement fusion plasma gain has approached this threshold (JET, $Q_p \sim 0.67$), the total per-shot energy input of these facilities is significantly greater than the energy coupled to the plasma, so that the NIF's $Q_f \sim 0.016$ and JET's $Q_f \sim 0.03$ \cite{JET_power}. 

We will accomplish this objective by constructing an inertial fusion driver that is orders of magnitude more efficient than the NIF at delivering energy to a target and to the fusion fuel itself (see Section \ref{sec:efficiency}) and is also consistent with the AMPS goals for delivering economically competitive fusion power (Section \ref{sec:reactor}). Our DS facility (Section \ref{sec:pulserdesign}) will demonstrate $Q_f>1$ and will enable innovation in target physics and design over the current state of the art. DS will deliver $>60$ MA of current to a target in $\sim 100$ ns, consistent with the established theoretical and computational basis for achieving high gain (Section \ref{sec:physics}). An enabling technology for the DS to efficiently deliver large amounts of current at low cost is the impedance-matched Marx generator architecture \cite{stygar2017impedance}, which uses commodity components and eliminates energy-inefficient pulse shaping stages from the prior generation of pulsed power architectures. In parallel to our fusion gain objective, we will advance the technological readiness of key technologies needed for a commercial power generation system. After these goals are achieved, a commercial power generation system will be produced. We will discuss engineering considerations for such commercial systems in this paper, and will elaborate on technoeconomics in subsequent papers. 

\section{Physics basis for high-gain pulser-driven inertial fusion}\label{sec:physics}
\subsection{Physics of pulser-driven inertial fusion}

Pulser-driven inertial fusion provides an opportunity to achieve facility gain $Q_f > 1$ by leveraging the lessons learned from the NIF to access ignition and gain while using a driver that is more efficient than a laser. DT fusion schemes must generate a plasma that is hot enough to have high fusion reactivity (typically $T \gtrsim 10$ keV) and is confined long enough that self-heating by the 3.5 MeV alpha particles released in the fusion reactions surpasses all energy loss mechanisms. In the inertial fusion regime, the hot fuel ``ignites,'' initiating a propagating wave of fusion reactions in which the fuel's pressure significantly increases while the fusion plasma expands into a surrounding reservoir of cooler, denser fuel \cite{Christopherson2019}. 

Pulser-driven inertial fusion relies on the direct magnetic acceleration and compression of targets, typically electrically conducting cylindrical shells or ``liners'' filled with DT fusion fuel.  
Liners are accelerated to lower implosion velocities ($\sim$100 km/s) than laser-driven inertial fusion schemes ($\gtrsim$350 km/s\cite{NIFLawson_2022}) to accommodate the typically longer pulsed-power time scales ($\gtrsim$ 100 ns) compared to that of high-power lasers ($\sim$10 ns) while managing hydrodynamic instabilities. Externally supplied energy (``preheat'') and thermally-insulating magnetic fields are injected into the fuel cavity prior to implosion or as the implosion proceeds, conditioning the hot spot to reach high temperatures and ignite at lower stagnation (peak) pressures ($\sim$10 Gbar) than laser-driven implosions ($\sim$100 Gbar). The ability for pulser ICF to reach ignition conditions and net target gain at substantially lower pressures and correspondingly higher target masses represents a key advantage relative to laser ICF. 

One example of this approach is the MagLIF (Magnetized Liner Inertial Fusion) concept developed at Sandia National Laboratories\cite{slutz2010pulsed}.  The MagLIF concept provides an excellent baseline for scaling pulser-driven ICF to ignition and high gain for five reasons: 
\begin{itemize}
    \item \textbf{Sound theoretical basis.} There is a sound theoretical basis for inertial fusion of magnetized fuel~\cite{Basko2000}.
    \item \textbf{Efficient energy delivery.} Energy can be delivered to MagLIF targets with high efficiency.
    \item \textbf{High-fidelity simulations.} MagLIF-type experiments can be studied with high-fidelity simulations.
    \item \textbf{Paths to ignition and high gain.} 
   There are documented scaling paths to ignition and high gain ~\cite{Slutz2012HighGain,Slutz2016Scaling,SchmitRuiz2020,Ruiz2023Iscaling}.
    \item \textbf{Experimental demonstration.} High fusion performance has been experimentally demonstrated, and there is a demonstrated method to determine proximity to ignition~\cite{Knapp_2022}.
\end{itemize}

These points are discussed in turn in the following five sections.

\subsection{Theoretical basis for pulser-driven inertial fusion} \label{sec:theory}
As with all inertial fusion concepts, the core objective is to assemble a hot DT mass and confine it long enough for fusion reactions to self-heat and ignite the DT fuel. Assuming a cylindrical DT fuel arrangement, the evolution of the DT internal energy, $\hat{E}_{\rm DT}$, is governed by an equation like
\begin{equation}
    \frac{d\hat{E}_{\rm DT}}{dt} = \hat{W}_{\alpha} - \hat{W}_{PdV} - \hat{W}_r - \hat{W}_{ce} - \hat{W}_{ci} - \hat{W}_{\rm end}, \label{eq:DT_powbal}
\end{equation}
with net heating of the DT fuel occurring when $d\hat{E}_{\rm DT}/dt>0$. The energies and powers in the above equation are expressed per unit length of the cylindrical fuel assembly, denoted by the hat notation. Alpha-particle heating of the fuel scales as 
\begin{equation}
    \hat{W}_\alpha  \propto \eta_\alpha (\rho R)^2 s(T), \label{eq:p_a}
\end{equation}
where $\eta_\alpha\le 1$ is alpha-particle coupling efficiency to the fuel plasma, described in Ref.~\cite{Ruiz2023Theory}. The characteristic DT fuel mass density and radius at peak compression are given by $\rho$ and $R$, respectively. The term $s(T)$ captures the temperature dependence of the Maxwell-averaged DT reactivity $\langle\sigma v\rangle$, which can be approximated $s(T)\propto T^3$ in the 4-12~keV range.\cite{SchmitRuiz2020}.  The remaining terms on the right hand side of Eq.~\ref{eq:DT_powbal} represent the different mechanisms channeling energy between the hot DT fuel and its surrounding environment, which include the isentropic work associated with the fuel displacing the liner ($\hat{W}_{PdV}$), radiation losses ($\hat{W}_r$), electron and ion heat conduction losses ($\hat{W}_{ce}$ and $\hat{W}_{ci}$), and bulk flows of hot fuel out the open end(s) of the liner ($\hat{W}_{\rm end}$).  For quantitative descriptions of these terms with respect to the underlying plasma conditions, see Appendix~\ref{sec:ignitionmetric}.

Self-heating of DT fuel by fusion reactions occurs when $\hat{W}_\alpha$ overcomes all the energy transport mechanisms represented in Eq~\ref{eq:DT_powbal}. In MagLIF, insulating magnetic fields are introduced in the fuel to inhibit thermal conduction from the hot DT plasma to the cold liner walls providing the inertial confinement.\cite{slutz2010pulsed}  This magnetic insulation is essential to sustaining the internal energy of the hot plasma during the $\sim$100~ns liner implosions typical of pulser ICF. 

\begin{figure}
    \centering
    \includegraphics[width=0.8\textwidth]{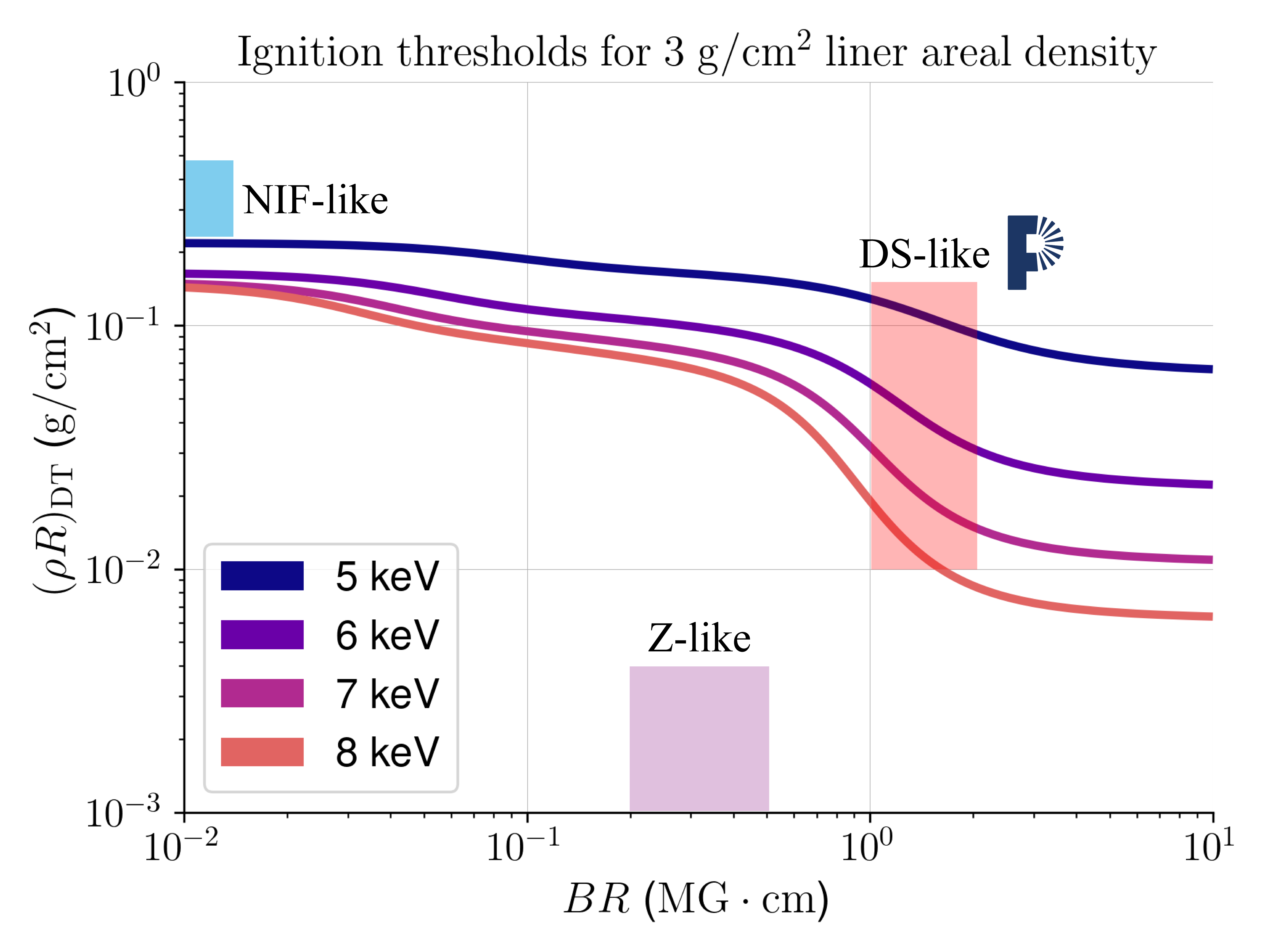}
    \caption{Contours of a generalized Lawson ignition threshold criterion ${\rm GLC}'=1$ (Eq.~\ref{eq:glcmod}) calculated for plasma temperatures ranging from 5 to 8 keV. We assume the DT is tamped by a liner with areal density $(\rho\delta R)_l=3~{\rm g/cm^2}$, a value accessible to the DS. Plasma conditions above and to the right of each curve are capable of robust self-heating and ignition. Colored rectangles represent the approximate parameter regimes distinguishing ignition-scale NIF implosions, contemporary MagLIF experiments on Z, and anticipated fuel conditions attainable on DS. Magnetization with DS-like targets can considerably reduce fuel areal density required for ignition, easing driver requirements compared to what is required for NIF-like targets.}
    \label{fig:glc_nif_z_ds}
\end{figure}

The fuel energy balance conditions represented by Eq.~\ref{eq:DT_powbal} can be integrated into an ignition proximity metric by additionally considering the time scale during which DT fusion occurs relative to the time scale the fuel is hydrodynamically confined by the compressed liner.  Following a canonical formulation of a ``generalized Lawson criterion'' (GLC) widely used by the ICF community \cite{Lindl2015IgnitionMetrics}, we write a generalized expression suitable for analyses of inertially-confined, magnetized fusion plasmas as
\begin{equation}
    {\rm GLC}'\doteq \left(\frac{(\hat{W}_\alpha - \hat{W}_{ci} - \hat{W}_{ce} - \hat{W}_r)\tau_{TN}}{\hat{E}_{\rm DT}}\right)f.
    \label{eq:glcmod}
\end{equation}
The numerical parameter $f\approx 1.2$ has been calibrated to reproduce ignition threshold features observed in simulations of NIF-like ICF implosions\cite{Lindl2015IgnitionMetrics}. We make a common simplifying assumption that the $\hat{W}_{PdV}$ and $W_{\rm end}$ terms become small as the liner ``bounces'' on the high-pressure DT fuel and expands after peak compression. Simulations provide a more detailed, self-consistent picture that continues to account for these energy transport mechanisms.  The parameter $\tau_{TN}$ represents the characteristic fusion burn duration, which is generally proportional to the liner areal density according to $\tau_{TN} \propto (\rho \delta R)_l^{1/2}$, where $\rho_l$ and $\delta R_l$ represent the mean liner mass density and thickness at peak compression.  A deeper discussion of these concepts is provided in Appendix~\ref{sec:ignitionmetric}.  In physical terms, the ignition threshold condition ${\rm GLC}'\approx 1$ can be interpreted as the scenario where the net heating of the DT fuel by fusion reactions is energetically comparable to the energy of the compressed fuel just prior to the onset of fusion burn.

Figure~\ref{fig:glc_nif_z_ds} shows the benefits of magnetizing fusion fuel by plotting the ignition threshold contours of ${\rm GLC}'=1$ as a function of the DT fuel areal density, $(\rho R)_{\rm DT}$ (the `DT' subscript is added for clarity), and the magnetic field-radius product, $BR$, at fusion-relevant plasma temperatures between 5 and 8 keV.  These curves assume the DT fuel is confined by a long (high aspect ratio when compressed) cylindrical liner with total areal density $(\rho\delta R)_l=3~{\rm g/cm^2}$, implying the liner provides the majority of the inertial confinement. This is representative of the liner compression achievable by the DS, cf. Sec.~\ref{sec:scaling_to_Qf}, and represents a roughly $3 \times$ increase in the inertial tamping characteristic of pulser ICF relative to ignition-scale NIF experiments, where the total confining areal densities are closer to $1~{\rm g/cm^2}$.\cite{NIFLawson_2022,NIFTargetGain_2024} While laser-driven schemes typically require heating and ablating $\approx$90 \% of the initial capsule mass to drive high-velocity implosions, leaving $\approx$10 \% to confine the hot spot at ignition time \cite{Lindl1995}, the current drive employed in pulser ICF accelerates the majority of the liner material inward to form the final compressed fuel-liner assembly, leaving very little trailing mass behind.

In inertial fusion schemes, areal density is synonymous with compression and pressure. For cylindrical systems, fuel mass conservation implies $\rho R^2\approx {\rm const}$, so $(\rho R)=(\rho R)_0 (R_0/R)\doteq (\rho R)_0 {\rm CR}$, where ${\rm CR}$ is the fuel convergence ratio. Increasing fuel compression requires increasing the ${\rm CR}$, which requires more drive pressure and can increase the risk of hydrodynamic instabilities. Robust self-heating in unmagnetized DT, represented on the left side of Fig.~\ref{fig:glc_nif_z_ds}, generally occurs for $(\rho R)_{\rm DT}\gtrsim 0.2 ~{\rm g/cm^2}$. This is consistent with the high-pressure, high-compression design space of NIF ignition platforms, involving stagnation pressures $>$300 Gbar.\cite{NIFLawson_2022,NIFTargetGain_2024}

When the DT fuel is magnetized, self-heating can occur at substantially lower values of $(\rho R)_{\rm DT}$. Figure~\ref{fig:glc_nif_z_ds} shows that at sufficiently high temperatures, magnetization reduces the fuel compression and associated pressures needed by more than $10\times$. For $BR$ values $\gtrsim 1.0~{\rm MG\cdot cm}$, the benefits of magnetization saturates as 3.5~MeV alpha particles are largely confined within the magnetized cylindrical fuel assembly. MagLIF experiments on Z have already demonstrated $BR\approx0.4$-$0.5$~MG$\cdot$cm,\cite{SchmitKnapp2014PRL,KnappSchmit2015,Lewis2021BRwithML} depicted schematically by the box at the bottom of Fig.~\ref{fig:glc_nif_z_ds}; scaling from the 20~MA Z facility to 60~MA on the DS could lead to an additional $2.6\times$ increase in $BR$,\cite{Ruiz2023Iscaling} such that the benefits of magnetization are fully realized.  The typical hot spot areal density and magnetic field-radius products occurring in simulations of MagLIF-like implosions on the DS are also illustrated in Fig.~\ref{fig:glc_nif_z_ds}, with examples shown in Sec.~\ref{sec:scaling_to_Qf}.

\subsection{Efficient energy delivery to target}\label{sec:efficiency}
Pulser ICF provides a path to high wall-plug efficiency due to the simplicity of the pulser technology and the high efficiency of (1) stored energy delivered to the target and (2) target absorbed energy coupled to the fuel internal energy. Due to the high combined efficiency of both of these factors, pulser ICF offers a $200\times$ increase in overall efficiency (stored energy coupled to fuel internal energy) relative to laser indirect-drive on the NIF. 

For laser-driven systems, the conversion of electrical energy into coherent laser photons tends to be an inefficient process.  For example, NIF stores 330~MJ of electrical energy in capacitor banks to deliver 2.2~MJ of UV photons to a target, a driver efficiency of $\eta_{\rm D} = 0.007$.  Modern laser architectures using solid state pumps instead of flash lamps promise efficiencies as high as $\eta_{\rm D}=0.1$.  Pacific Fusion's DS stores 80 MJ in the capacitors and transfers $\approx 90$ \% of that energy to an electrical pulse with 100 ns rise time, of which $\approx 44$ MJ is delivered to the insulator stack (a function of machine design choices only, independent of the load) and $\approx 8$ MJ is ultimately delivered to the fusion target area, a driver efficiency of $\eta_{\rm D}=0.1$. The inherent efficiency of the IMG driver technology, which eliminates inefficient pulse-shaping stages used in legacy architectures, provides clear advantages for energy delivery and facility size. We note that a conventional Marx generator delivering a similar power pulse would require $6\times$ the stored energy for comparable energy delivered to the target \cite{pmfe_whitepaper_2024}. 

The efficiency of energy coupled to the target is an additional advantage of pulser-driven ICF. The target, or load, is electrically connected to the driver, so the amount of energy the driver can deliver to the target depends on its electrical characteristics. We define the energy delivered to the load as the energy delivered to the initial volume of the fusion target\cite{Ruiz2023tscaling}. As noted above, of the 80 MJ of stored energy, approximately 8 MJ is delivered to the fusion target volume. The metric of interest for defining the efficiency of coupling to a target is the work done to compress and accelerate the fusion fuel liner relative to the total electrical energy supplied by the driver, often referred to as the hydrodynamic or implosion efficiency.  Pulser ICF is a ``direct drive'' scheme, meaning the pulser directly magnetically accelerates the metal liner that contains the fusion fuel to compress and heat the fuel. We approach this from a circuit perspective, as the motion of the liner changes its inductance, which feeds back on the driver.  

The work done on the liner is characterized by the response of the circuit to the changing inductance of the target.  As the liner implodes, its inductance changes, where the inductance is defined using the expression for a coaxial system of conductors,
\begin{equation}
    L = \frac{\mu_o h}{2 \pi } \ln(\frac{r_c}{r_l}). \label{load_L}
\end{equation}  
Here $r_c$ is the return current radius, assumed constant, and $r_l$ is the radius of the fusion target liner, which decreases with time.  The electrical energy delivered to the liner volume is partitioned into two channels: about half of the energy goes into the magnetic field surrounding the liner, which increases in magnitude and in volume as the energy is delivered. The other half of the energy is mechanical work on the liner. Therefore, the power associated with the mechanical work done on the liner is $P_{\rm mech} = \frac{1}{2}I_{\rm load}^2 \dot{L}_{\rm load}$\cite{mcbride2018primer}, where 
\begin{equation}
\dot{L} = -\frac{\mu_o h}{2 \pi }\frac{\dot{r}_\ell}{r_\ell}
\end{equation}
 is the time derivative of the load inductance.  The mechanical work done to the target is then
\begin{equation}
    E_{\rm mech} = \frac{1}{2}\int_0^{t_{max}} I_{\rm load}^2 \dot{L}_{\rm load} dt = -\frac{1}{2}\int_0^{t_{max}} I_{\rm load}^2 \frac{\mu_o h}{2 \pi }\frac{\dot{r}_\ell}{r_\ell} dt. \label{eq:eliner}
\end{equation}

In the equations above, $t_{max}$ is the time which maximizes the integral, $r_\ell$ is the liner outer radius and $\dot{r}_\ell$ its time derivative. Power is delivered to the liner during the rising edge of the driver power pulse, and the implosion time is matched to the driver rise time. The implosion efficiency of the system is then defined as $\eta_{\rm imp} = E_{\rm mech}/E_{\rm driver}$. After the implosion, the driver continues delivering power to the remnants of the load. This means that much of the ``inefficiency'' in a pulsed power driver is wrapped up in magnetic energy stored in the volume of the transmission lines after the implosion stagnates. In principle, this energy can be partially recaptured or used to enhance confinement late in time with a properly engineered implosion scheme. 

An accurate calculation of the energetics of a liner implosion requires a model that couples the liner implosion to a circuit that represents the DS. The model should include physical processes such as compression of the liner material, ionization, ohmic heating, and other processes. One option is the Semi-Analytic MagLIF Model (SAMM)\cite{McBride_2015}, which captures the physics of a liner implosion in a series of ordinary differential equations. Another option is a radiation hydrodynamics code that uses a resistive magnetohydrodynamics model, such as the FLASH simulations described in Section \ref{sec:simulation}. In 1D FLASH simulations, the mechanical work done on the target is $\approx 4.5$ MJ, so the coupling efficiency of driver stored energy to implosion energy is $\eta_{\rm D} \eta_{\rm imp} = 5.6$ \%. At stagnation, some of the mechanical energy of the imploding shell is converted to internal energy of the fusion fuel shell and hot-spot, with efficiency $\eta_{\rm stag}$. In 1D FLASH simulations, the internal energy of the fusion fuel at stagnation is 1 MJ (of which about half is in the hot spot), corresponding to $\eta_{\rm stag} \approx 22$ \%. Therefore, conversion efficiency of target energy to fusion fuel internal energy is 12 \% (1 MJ fuel internal energy compared to 8 MJ delivered to the target), and the net efficiency of converting driver stored energy to fusion fuel internal energy is 1.25 \% (1 MJ fuel internal energy compared to 80 MJ initially stored). 

For laser indirect drive (LID) and laser direct drive (LDD) ICF schemes, there are several efficiency terms that must be considered to arrive at a reasonable comparison: the fraction of energy absorbed by the fuel-containing capsule, the efficiency of converting absorbed energy to kinetic energy of the imploding shell, and the stagnation efficiency. In LID, most of the laser energy delivered to the target goes into heating the hohlraum wall that emits and contains the soft x-rays that drive the implosion. A small fraction of the energy delivered to the hohlraum target is coupled to the fusion-containing capsule: for the recent NIF ignition experiments, this was $\eta_{\rm hohl} \approx 12$ \% \cite{Kritcher2024}. For LDD, we are concerned with the coupling efficiency of the UV light to the capsule itself, which is typically $\eta_{\rm DD}\sim 80$ \%\cite{Campbell_2017}. Both the LID and LDD schemes accelerate the capsule to high implosion velocity via ablation, and most of the coupled energy is used to heat up and accelerate the ablated mass, which is $\sim90 \%$ of the initial mass of the system. This means that the mass of the remaining liner and fuel shell at peak kinetic energy is quite small compared to the initial mass, and the kinetic energy is similarly small compared to the total energy absorbed by the capsule. The efficiency of the ``rocket'' effect is generally in the range of $7-15\%$.  
Due to the deeper penetration of x-rays into the ablator material compared to laser light, LID has higher rocket efficiency than LDD, which partially compensates for the low efficiency of coupling energy from the hohlraum to the capsule. Ultimately, the implosion efficiency is $\eta_{\rm imp}\approx 2$ \% for LID (LDD is higher, $\eta_{\rm imp}\approx 6$ \%). The total efficiency of converting target absorbed energy to fuel energy is 1 \% ($>5 \times$ lower than for pulser-driven ICF on the DS), and the net efficiency of converting stored energy to fuel energy $\eta_{\rm D} \eta_{\rm hohl} \eta_{\rm imp} \eta_{\rm stag} \approx 6 \times 10^{-5}$, or about $200 \times$ lower than anticipated for DS. A little less than half of the work done on the fuel is converted to hot-spot internal energy, which leads to the onset of ignition and propagating fusion burn.

\begin{figure}
\centering
\includegraphics[width=6in]{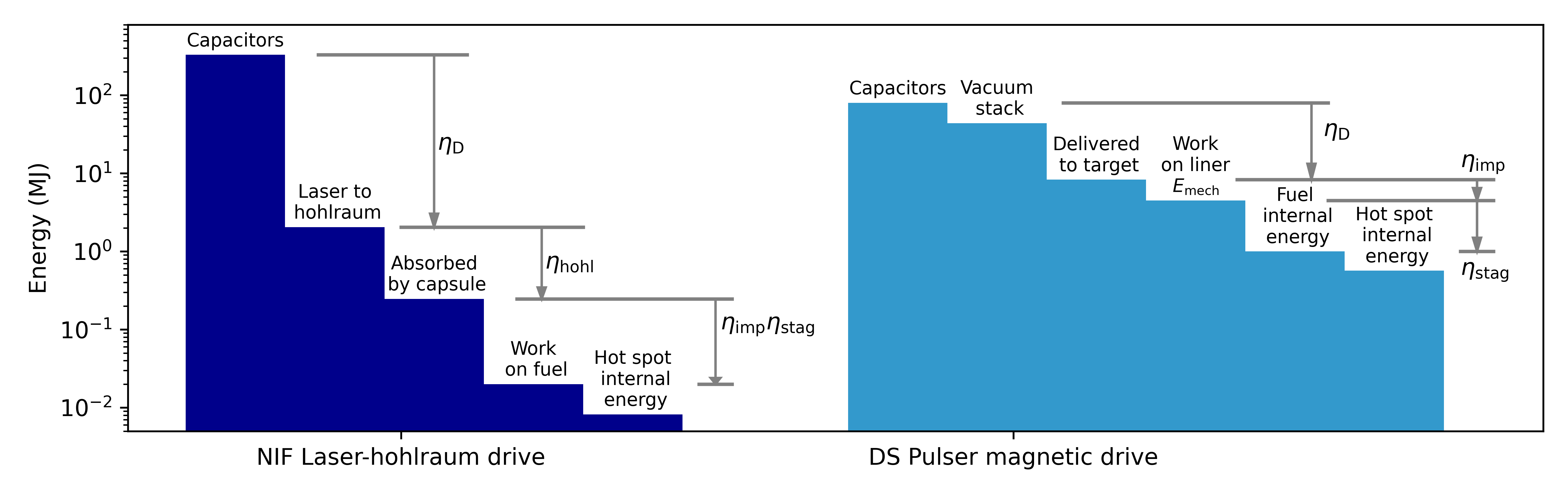}
\caption{\label{fig:LID_efficiency}  Energy flow from driver energy storage to fusion fuel internal energy for laser indirect (hohlraum) drive (LID) ICF on the NIF and pulser driven ICF on the DS. NIF numbers are from published simulations\cite{Kritcher2024}. DS numbers are from simulations described below in Section \ref{sec:simulation}. Labels refer to the various efficiencies $\eta$ described in the text. }
\end{figure}

In summary, pulser-driven ICF offers a total efficiency of converting stored energy to fuel energy of $\eta_{\rm D} \eta_{\rm imp} \eta_{\rm stag} \approx 1.25 \times 10^{-2}$, roughly $200 \times$ higher than LID on the NIF. As a result, we anticipate that DS, despite storing about $4 \times$ less energy than the NIF, will deliver $50 \times$ more energy to the fusion fuel; the impact of this on yield and gain is discussed in the following sections. These efficiencies are compared graphically in Fig. \ref{fig:LID_efficiency}. Finally, we note that even if modern solid state lasers can achieve promised high efficiencies, pulser-driven ICF also remains more efficient than LDD.

\subsection{Simulation capability for ignition and high gain}\label{sec:simulation}
Radiation hydrodynamics simulations have enabled ICF target design on today's leading ICF facilities, including HYDRA's key contributions to obtaining ignition on the NIF \cite{Marinak_sims_ignition2024, Kritcher2024} and the use of HYDRA (e.g., Ref.~\cite{Ruiz2023Iscaling}), ALEGRA (e.g. Ref.~\cite{Knapp_2017}), and LASNEX (e.g. Ref.~\cite{Sinars_2011}) for designing fusion targets at Z. Here, we describe Pacific Fusion's  target design code, present a benchmark against a published scaling study to high-current MagLIF, and show simulations of high-gain fusion targets driven by a representative circuit model for our DS. 

For its target design simulation capabilities, Pacific Fusion is co-developing the FLASH radiation MHD code with the host institution for the code, the Flash Center at the University of Rochester. FLASH \cite{Fryxell2000} is a publicly-available, massively parallel, multi-physics, adaptive mesh refinement (AMR), finite-volume Eulerian hydrodynamics and magneto-hydrodynamics (MHD) code. FLASH has a world-wide user base of more than 4,600 scientists, and more than 1,300 papers have been published using the code to model problems in a wide range of disciplines, including plasma astrophysics \cite{Tzeferacos2018, Bott2021}, combustion, fluid dynamics, high energy density physics (HEDP) \cite{Fatenejad2013, Sauppe2023}, and fusion energy. Over the past decade and under the auspices of the U.S. DOE NNSA, the Flash Center has added extensive High Energy Density Physics (HEDP) and extended-MHD capabilities to FLASH~\cite{Tzeferacos2015}, providing an excellent foundation for Pacific Fusion's target simulation needs. Recent FLASH improvements include multiple state-of-the art hydrodynamic and MHD shock-capturing solvers~\cite{Lee2013}, three-temperature extensions~\cite{Tzeferacos2015} with anisotropic thermal conduction and high-fidelity magnetized heat transport coefficients~\cite{JiHeld2013}, electron-ion thermal equilibration, multi-group radiation diffusion, tabulated multi-species EOS and opacities, laser energy deposition, circuit models, and numerous synthetic diagnostics~\cite{Tzeferacos2017}. 

To establish confidence in FLASH's target design capabilities, Pacific Fusion has verified and validated the code against analytic test problems and data from integrated experiments. Details of this effort are presented in a separate article \cite{ellison_flashvnv_2024}. The experimental benchmarks include platforms for measuring hydrodynamic instability growth rates such as the magneto-Rayleigh-Taylor \cite{Sinars_2010, Sinars_2011} and Richtmyer-Meshkov \cite{Knapp_2020} instabilities. We also benchmarked against a platform for measuring the inertial confinement time of a compressed liquid deuterium fuel \cite{Knapp_2017} and precursor MagLIF experiments that radiographed imploding beryllium liners. Finally, we benchmarked FLASH against fully integrated MagLIF experiments, such as Z shot 2977\cite{Lewis2021BRwithML}, and against the similarity scaling study for scaling MagLIF to high currents\cite{Ruiz2023Iscaling}. Results of these benchmarks can be found in the upcoming publication \cite{ellison_flashvnv_2024}. In all cases, FLASH simulations have been shown to agree well with the experimental data and with the published simulation results from the codes used to design the targets, including LASNEX, HYDRA, and ALEGRA.

\begin{figure}
    \centering
    \includegraphics[width=3.5in]{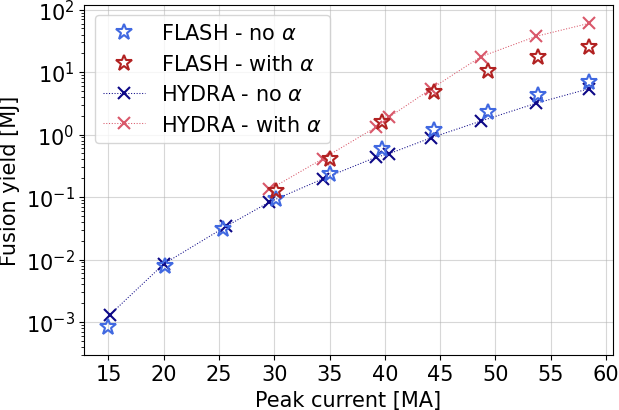}
    \caption{Fusion yield from 2D FLASH simulations of the MagLIF current-scaling study, compared with 2D HYDRA simulations reported in Ref.~\citenum{Ruiz2023Iscaling}. Calculations including alpha heating (``with $\alpha$'') are shown alongside those without alpha heating (``no $\alpha$''). }
    \label{fig:flash_current_scaling}
\end{figure}

Figure~\ref{fig:flash_current_scaling} shows an example of a validation benchmark that also illustrates the exciting fusion regime accessible at high currents. The figure compares FLASH simulations to published HYDRA simulations of a MagLIF target scaled from 15 MA to 60 MA peak current \cite{Ruiz2023Iscaling}. This scaling study was conservative, meaning it attempted to maintain the dynamics of the implosion designs on Z by conserving the values of several dimensionless parameters that describe the system, i.e., similarity scaling. The FLASH simulations were initialized with MagLIF target geometries described in Ref.~\cite{Ruiz2023Iscaling} and coupled to a lumped-element external circuit model representing a pulsed power driver. Native FLASH does not possess models for fusion reactions of hydrogenic isotopes, so we have extended our internal version of FLASH to include relevant fusion reactions and to couple the resulting energy to the material via either a local-deposition model or an alpha-energy diffusion model. Figure ~\ref{fig:flash_current_scaling} shows that FLASH is able to accurately model the fusion yield of these MagLIF targets, representing the culmination of many physical processes. The scaling trend is captured extremely well, validating that FLASH is solving the relevant physical processes to match the theory established in Ref.~\cite{Ruiz2023Iscaling} and the HYDRA simulations from the same Reference. Moreover, the fusion gain factors and alpha amplification factors are similar to the HYDRA simulations, indicating that ignition occurs at similar values of current drive in the two codes. We note that these simulations were in no way ``calibrated'' to the HYDRA simulations but were merely run with the same problem definition and the inclusion of the relevant physical processes. At high currents with alpha deposition on, HYDRA simulations indicate slightly higher yield than FLASH; this difference is an ongoing area of development.

Figure~\ref{fig:flash_current_scaling} clearly shows the onset of self-heating and ignition described in Section~\ref{sec:theory}: the simulated fusion yield from calculations that include alpha heating (``with $\alpha$'') diverge from calculations without alpha heating (``no $\alpha$'') at $I_\text{max} \geq 40$ MA, with yield $Y_{DT} \geq 60$ MJ at  $I_{max} = 60$ MA.  The main result of the similarity scaling study is that MagLIF performance on Z today scales conservatively to ignition (as quantified by the Generalized Lawson Criteron) at peak driver current $I_\text{max} \geq 50$ MA\cite{Ruiz2023Iscaling}. The similarity-scaled designs are conservative in both senses of the word: the scaling conserves the key dimensionless parameters that describe the system, and it is baselined by performance on the Z facility today, with no major innovations in target design and technology assumed. 

Target designs can be improved beyond similarity scaled versions of MagLIF on Z, which can further increase gain and reduce driver cost. One promising approach involves adding a cryogenic DT fuel liner to the inner surface of the imploding cylinder~\cite{Slutz2012HighGain, Slutz2016Scaling}, as is done in spherical NIF capsules. The DT ice liner serves multiple purposes: in addition to shielding the fuel hot-spot from the metallic liner, which could mix into the fuel and increase radiation losses, the fuel serves as a reservoir of mass for the hot-spot, some of which is squeezed out the top and bottom of the cylindrical implosion. The igniting hot-spot can drive a fusion burn wave into the dense fuel liner, dramatically increasing the fusion yield compared to gas-fuel-only MagLIF targets. 

We present FLASH simulations of such a prototype cryogenic target concept driven by a circuit model of the DS pulser. Simulation-based target design will be a major ongoing effort at Pacific Fusion; here we seek to show a proof-of-concept illustration that the drive energy is sufficient for achieving ignition and facility gain with a target appropriately impedance-matched to the driver. In the FLASH simulations of the DS-driven targets, the circuit model represents the voltage drive provided by the DS pulser and includes impedances and losses of the key components of the system, including losses in the vacuum transmission lines. The circuit loss models are calibrated against particle-in-cell kinetic simulations (described in Section~\ref{sec:target_area}).  

\begin{figure}
    \centering
    \includegraphics[width=0.9\linewidth]{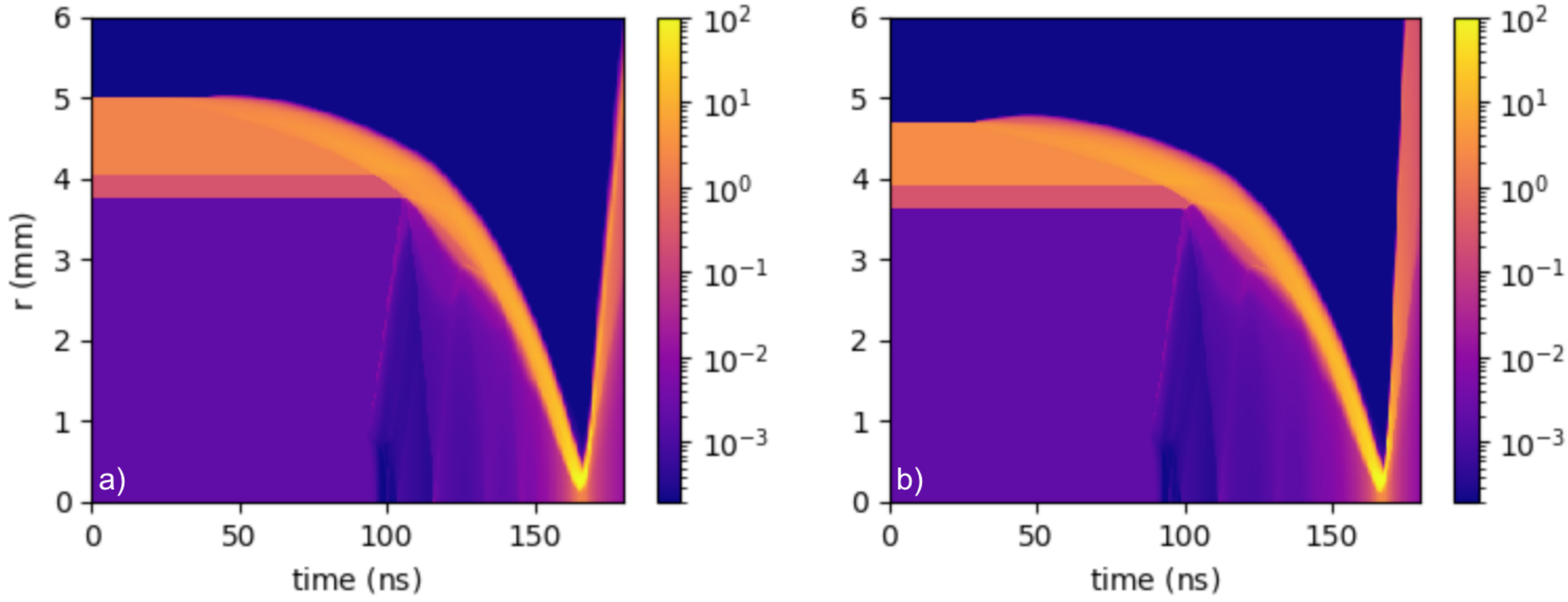}
    \caption{FLASH simulations of 6 mm tall MagLIF targets designed to match the DS pulser. The pulser delivers $I_{max} = 56$ MA and $E_{liner} \approx 8.3$ MJ of electrical energy to each design. Both targets have 300 \textmu m thick DT ice liners, $B_z = 15$ T, and 18 kJ preheat energy: (a) Beryllium liner design generates 109 MJ yield (b) Aluminum liner design generates 380 MJ yield.}
    \label{fig:flash_ds3_designs}
\end{figure}

Figure \ref{fig:flash_ds3_designs} shows streak plots of density for two MagLIF targets designed to match the DS pulser. The beryllium liner design (Fig. \ref{fig:flash_ds3_designs}a) uses a thinner liner than the scaled designs from Fig. \ref{fig:flash_current_scaling}, and is based on recent high-performing MagLIF experiments using thinner liners \cite{HarveyThompson2024}. The aluminum liner design (Fig. \ref{fig:flash_ds3_designs}b) is appropriately scaled from the beryllium design. Both designs are 6 mm tall and are driven to $I_{max} = 56$ MA. Both example designs are part of an extensive computational study of MagLIF performance achievable on the DS. The examples shown here have 300 \textmu m thick DT ice liners, $B_z = 15$ T, and 18 kJ preheat energy and both achieve facility gain. The beryllium liner generates 109 MJ of fusion yield for a facility gain of $Q_f = 1.36$ and the aluminum liner generates 380 MJ for a facility gain of $Q_f = 4.75$.

\begin{figure}
    \centering
    \includegraphics[width=0.9\linewidth]{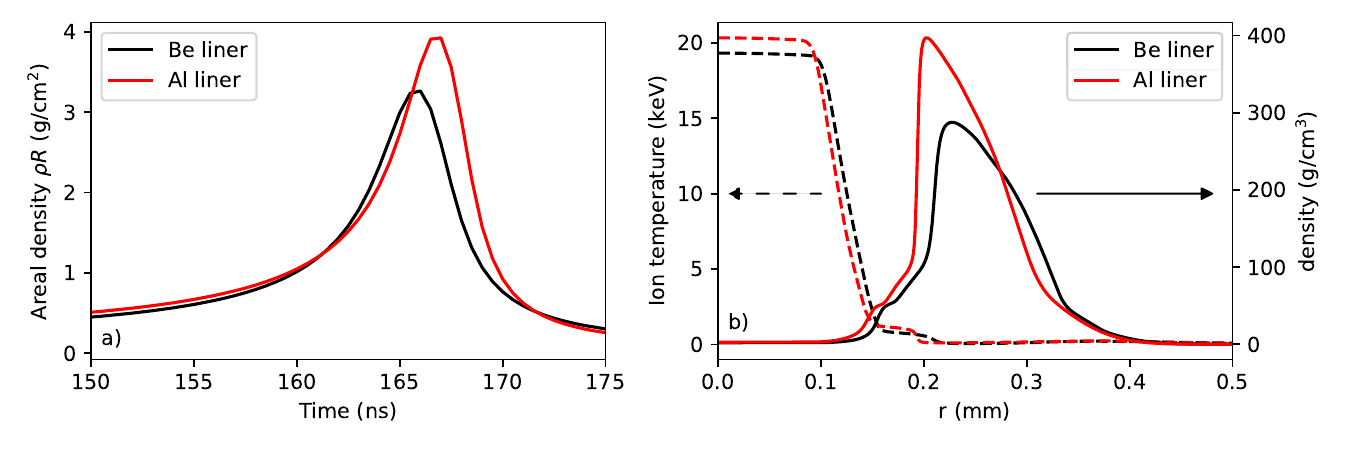}
    \caption{Characteristic plasma conditions of the designs shown in Fig. \ref{fig:flash_ds3_designs}: (a) Liner areal density ($(\rho\delta R)_l$) vs. time (b) Ion temperature (dash) and mass density (solid) vs. radius at the time of peak $(\rho\delta R)_l$. The higher $(\rho\delta R)_l$ of the aluminum liner design results in higher temperature, density, and total yield.}
    \label{fig:flash_ds3_curves}
\end{figure}

In summary, Pacific Fusion has a high-fidelity radiation hydrodynamics simulation capability that has been validated against a range of integrated experiments at Z. This code has reproduced the key results of the MagLIF current-scaling study of Ref.~\cite{Ruiz2023Iscaling}, which indicates that conservatively scaled MagLIF targets are well into the ignition regime at the 60 MA scale. Pacific Fusion is currently using FLASH to design MagLIF targets for the DS pulser that can achieve $Q_f > 1$. In the next section, we use FLASH simulations to discuss scaling of MagLIF to facility gain and how this scaling compares to laser-driven ICF. 

\subsection{Scaling to facility gain}\label{sec:scaling_to_Qf}

In the inertial regime, the requirements to heat and compress fusion fuel to ignition conditions are independent of the compression scheme\cite{Hurricane2023RMP}. Ignition and target gain were achieved on the NIF laser by accelerating a DT fusion fuel shell to high velocity ($v \geq 400$ km/s) while tailoring the compressibility of the implosion and increasing the size of the implosion system to provide sufficient confinement. 
Magnetically-driven inertial fusion systems are typically scaled differently from laser-driven systems. It is instructive to compare how these two different systems scale to understand the technical basis for pursuing pulser-driven ICF. 

\subsubsection{Scaling laser-driven ICF}\label{sec:laser_scaling}
Hydro-equivalent or hydrodynamic scaling is often used to assess how laser-driven fusion systems scale to larger facility drivers\cite{NoraBetti2014}. In hydro-equivalent scaling, the fundamental characteristics of the implosion remain constant as the system is enlarged. The implosion velocity and ablation pressure are held constant, leading to constant values of the stagnation pressure, density, and temperature, in the absence of alpha-heating. Hydro equivalence is achieved by scaling geometric quantities (lengths) and temporal quantities (timescales) together:
\begin{eqnarray*}
r = sr_0 \\
t = st_0,
\end{eqnarray*}
where $r_0$ and $t_0$ are the characteristic length- and time-scales of the reference implosion.  Thus, geometric areas scale as $s^2$, and intensity on the capsule surface is kept constant by scaling the laser power $P_L$ by the same factor $s^2$. The laser energy $E_L$, proportional to $P_L \times t$, scales as $E_{laser} \propto s^3$. Alternatively, the scaling can be written as $r \propto E_{laser}^{1/3}$. Density is constant, so the scaled system mass also increases by $s^3$.  Ignoring the relationship between $E_{laser}$ and hohlraum efficiency, which is relatively weak for NIF-size and larger hohlraums, this scaling applies to both laser-direct-drive and laser-indirect-drive schemes. 

As an implosion design is hydrodynamically scaled to larger drivers, it passes through three performance regimes: the pre-ignition regime at low capsule absorbed energy $E_{cap}$, where the yield scales weakly with energy; the boot-strap regime, where the yield rapidly increases due to the onset of self-heating; and the robustly-burning regime, where the yield depends only on the DT fuel mass and the inertial burn-up fraction $\phi$, defined as
\begin{equation}
Y_{DT} = \phi m_{fuel} e_{DT}. \label{phi_def} 
\end{equation}
Here, $e_{DT} = 3.4 \times 10^{11}$ J/g is the Q-value of the DT reaction (17.59 MeV) per mass. The burn-up fraction for spherical inertial fusion systems can be derived as a function of the DT fuel areal density $\rho R$,
\begin{equation*}
\phi = \frac{\rho R}{\rho R + 6},
\end{equation*} 
where $\rho R$ has units of g/cm$^2$\cite{Lindl1995}. Since hydro-equivalent scaling maintains a constant density $\rho$, the areal density scales only as $r$, $\rho R \propto E_L^{1/3}$, and for $\rho R \ll 6$, $\phi \propto E_L^{1/3}$. This means that when hydrodynamically scaling a robustly-burning design, the DT yield and target gain $Q_{target}$ scale as
\begin{eqnarray*}
    Y_{DT} &\propto& E_L^{4/3} \\ \label{eq:y_el_scaling}
    Q_{target} = \frac{Y_{DT}}{E_L} &\propto& E_L^{1/3} \label{eq:Q_el_scaling}
\end{eqnarray*}
This unfavorable scaling as an implosion transitions from boot-strapping to robust ignition means, for example, that hydrodynamically scaling the recent NIF experiments with $Q_{target}\approx2.4$ at $E_L$ = 2.05 MJ to $Q_{target}\approx10$ would require $E_L\approx 10$ MJ\cite{Landen2024}. 

\subsubsection{Scaling pulser-driven ICF}\label{sec:pmf_scaling}
In pulser ICF, magnetically-driven liners are continuously driven throughout the power pulse by the Lorentz force. For certain target designs, pre-heat energy and magnetic field are injected into the fuel cavity before the implosion stagnates, preconditioning the hot spot to reach ignition temperatures. In the MagLIF concept, for example, the pre-heat energy is injected using a laser, and the fuel is magnetized by external Helmholtz coils. This approach has been studied extensively but is limited by the capabilities of the Z Facility at Sandia National Laboratories, notably that Z cannot drive more than $\sim20$ MA through a MagLIF target. 

To simplify the engineering of and achieve low costs in future pulser IFE power plants, we are developing target designs that eliminate the need for a preheat laser and external Helmholtz coils. For example, one can pre-heat and magnetize the fuel by using a current pre-pulse that uses a small fraction of the energy used to compress the target. Alternatively, higher pressure implosion concepts can reduce or entirely eliminate the need for pre-heat and/or magnetization \cite{Lindemuth2024}.
 
Previous studies have investigated similarity scaling of MagLIF to larger (higher current) drivers than Z \cite{SchmitRuiz2020,Ruiz2023Theory,Ruiz2023Iscaling}. When the amount of current delivered to the target is increased while maintaining the $\approx 100$ ns rise time of the current pulse, the drive pressure and stagnation pressure also increase. As a result, the fusion yield scales as $Y_{DT} \propto I_{max}^6$\cite{Ruiz2023Iscaling}, even before accounting for plasma self-heating from alpha particles. When the target height is held constant with $I_{max}$, the energy delivered to the liner scales as $E_{liner} \propto I^2$ (see Eq. \ref{eq:eliner}). Thus, yield and gain scale as
\begin{eqnarray*}
    Y_{DT} &\propto& E_{liner}^3 \\
    Q_{target} = \frac{Y_{DT}}{E_{liner}} &\propto& E_{liner}^2,
\end{eqnarray*}
even before self-heating is taken into account. This gain scaling is much more favorable --- growing faster with energy delivered to liner/target --- than the hydro-equivalent scaling that applies to laser ICF. Combined with the low cost and high efficiency of pulser technology, this scaling is a major reason Pacific Fusion is developing pulser IFE on its DS. 

\subsubsection{Comparing scaling on the NIF to Pacific Fusion's DS}
Combining our understanding of inertial fusion theory, driver efficiency, and scaling laws enables us to compare our goal of $Q_f > 1$ on Pacific Fusion's DS to the recent achievement of ignition on the NIF. Ignition and target gain $Q_{target} > 1$ on the NIF demonstrate that fusion gain is possible in inertial systems. Building a system that can reach $Q_f > 1$ requires efficiently driving larger fuel masses than those used in NIF targets. Pulser-driven ICF is ideally suited to meet this challenge, due to its high efficiency and because it can achieve $Q_f > 1$ by generating fusion fuel conditions that have already been demonstrated on the NIF, as discussed below.  

\begin{figure}
    \centering
    \includegraphics[width=0.65\linewidth]{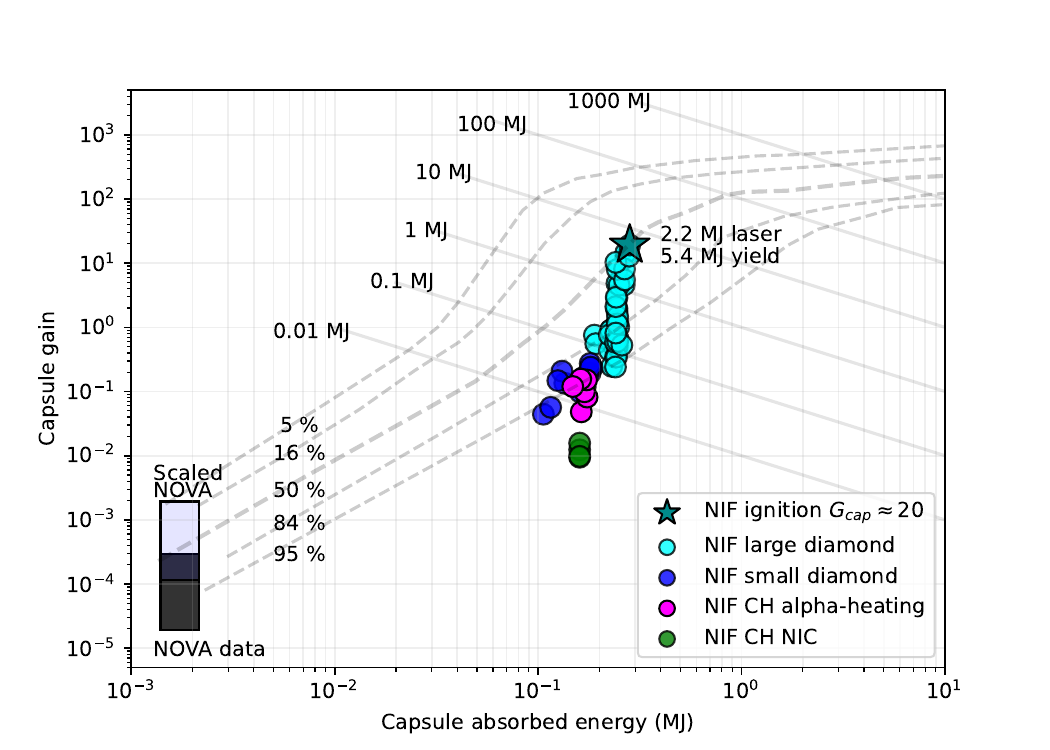} 
    \caption{Ignition and gain scaling plot for laser-indirect-drive ICF, based on the 1990 study\cite{Mead_IAEA_1990}. The gain curves (dashed gray) were generated from LASNEX simulations with more (high confidence) or fewer (low confidence) sources of degradation. The bars in the lower left represent unpublished deuterium implosions from the NOVA laser and the same data scaled up for DT fuel. Circles are NIF DT implosion data from various capsule designs. The dark cyan star is the target gain $Q_{target} \approx 2$ ($Y \approx 5$ MJ) result from February 2024.} \label{fig:mead_laser}
\end{figure}

The basics of fusion ignition and gain were understood decades before they were demonstrated on the NIF. Hydro-equivalent scaling was used to generate a set of ``gain curves'' for laser-indirect-drive ICF by Mead \emph{et al.}, in 1990\cite{Mead_IAEA_1990}. Figure \ref{fig:mead_laser} shows the gain curves (dashed gray) reproduced from the original report, with the x-axis converted from laser energy $E_{laser}$ to capsule absorbed energy $E_{cap}$. These curves were generated by running the ICF design code LASNEX\cite{Zimmerman1975} with various levels of degradation to the implosion. The curves are expressed as confidence bands, with higher probabilities corresponding to more degraded implosions and lower probabilities corresponding to more optimal implosions. For example, the 5 \% confidence curve corresponds to near-optimal, highly compressed, perfectly symmetrical implosions with no mix. The authors concluded there was a 5 \% probability of achieving this performance level given the state of ICF knowledge in 1990. The curves clearly show the three performance regimes described in Sec. \ref{sec:laser_scaling} (pre-ignition, boot-strapping, and robust burn).

Note that the x-axis of Fig. \ref{fig:mead_laser} has units of MJ, whereas the y-axis is dimensionless. Yield = gain $\times$ energy, so lines of constant fusion energy output appear as straight lines (with negative slope) on the plot. The subset of NIF implosion data shown in Fig. \ref{fig:mead_laser} reveals the prescience of the 1990 report: Early plastic (CH) shell implosions from the National Ignition Campaign (NIC, green) fall below the 95 \% confidence band. This makes sense, as calculating the poor performance of these implosions required 3D simulations that incorporated degradation sources that were unanticipated in 1990\cite{ClarkHero3D_PoP2015}. Improving the hydrodynamic stability of these implosions led to a $\approx 10 \times$ improvement in gain (magenta)\cite{Hurricane2014}. Switching to diamond capsules led to higher performance than plastic at approximately the same scale (blue)\cite{LePape2018}. Larger diamond capsule designs (cyan), scaled up in size from the smaller diamond designs\cite{Kritcher2018}, achieved the largest improvement in performance and gain, reaching ignition and target gain $Q_{target} > 2$ \cite{NIFTargetGain_2024,Kritcher2024}. The increase in target performance at nearly constant $E_{cap}$ resulted from improving target design and fabrication to reduce degradation of the implosion by ablator mix. 

\begin{figure}
    \centering
    \includegraphics[width=0.65\linewidth]{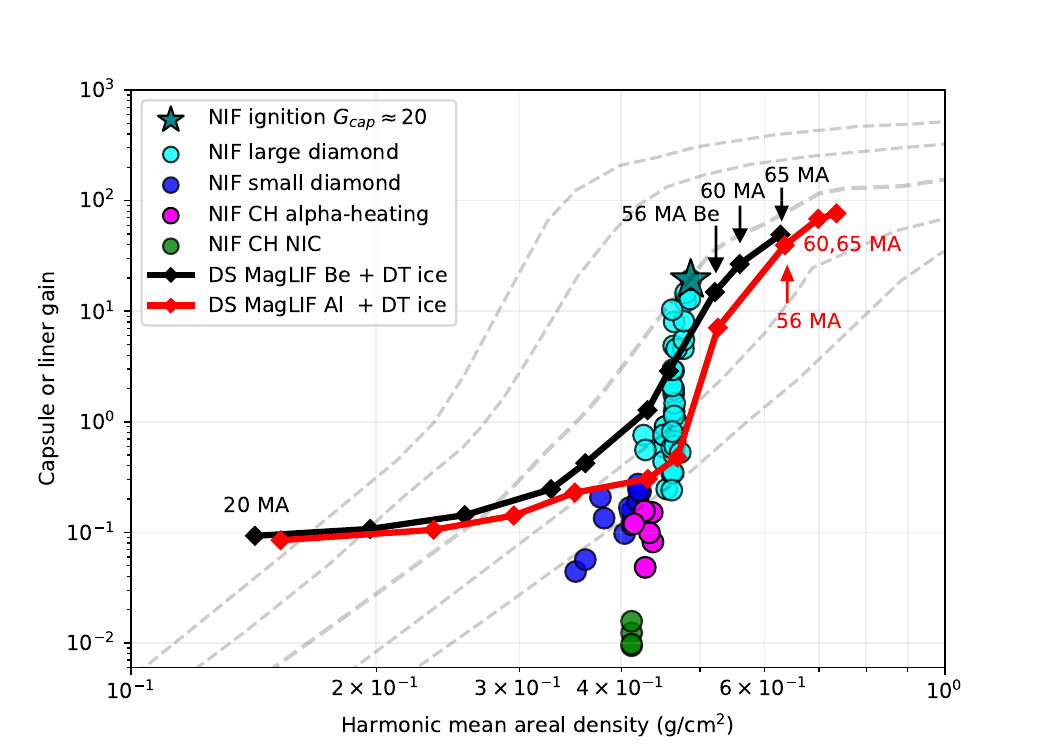}
    \caption{Ignition and gain scaling plot for laser-driven ICF and pulser-driven ICF, plotted against the harmonic mean areal density $= \sqrt{\rho R_{HS}~(\rho\delta R)_l}$, an important term in the Generalized Lawson Criterion (Eq. \ref{eq:glc_scale}). The red and black lines are target gain estimates from a current-scaling study of the two DT ice layer MagLIF designs described in Fig. \ref{fig:flash_ds3_designs} and Table \ref{table:NIF_Dynos_gains}. Only the 20 MA points are accessible on Z today; larger currents will be accessed on the DS. Both designs start to ignite at $I_{max} \geq 50$ MA (the last 4 points of each line), where the MagLIF liners achieve similar areal density to igniting NIF capsules. At $I_{max} \geq 60$ MA, the MagLIF designs reach larger target gain $Q_{target}$ and higher overall yield than the NIF ignition shot.} \label{fig:mead_pmf}
    \label{fig:mead_both}
\end{figure}

The highest performing NIF shot to-date (dark cyan star in Fig. \ref{fig:mead_laser}) lies on the 50 \% confidence gain curve. Higher-performing, lower-confidence curves correspond to implosions with more compression and larger $\rho R$. It remains to be seen if further advances in target design can lead to additional performance gains or whether future NIF experiments will ``roll over'' and follow the 50 \% gain curve into the robustly-burning regime. Numerous design improvements are needed for existing ignition designs on the NIF to scale favorably to $Q_F>1$ on a future laser facility\cite{Landen2024}. 

In Fig. \ref{fig:mead_both}, we visualize pulser-driven systems in the same space as NIF implosions. For this purpose, we transform the x-axis from absorbed energy $E_{cap}$ to areal density $\rho R$. Specifically, we use the harmonic-mean $\rho R$ described in Eq. \ref{eq:glc_scale} of appendix \ref{sec:ignitionmetric}, $\sqrt{\rho R_{HS}~(\rho\delta R)_l}$. This expression shows that an inertial fusion assembly can ignite when the product of fuel areal density and liner areal density is sufficiently large. Since the requirements for ignition and propagating burn are independent of the driver, MagLIF-like ICF designs can be plotted in this same space alongside NIF data. For NIF points, we use published values of $\rho R_{HS}$ and $(\rho\delta R)_l = \rho R_{total} - \rho R_{HS}$\cite{Hurricane2016NaturePhysics,Patel2020HotSpot,Kritcher2024}. For experiments without published data, the $\rho R$ was scaled as $\rho R \propto E_{cap}^{(1/3)}$, per hydro-equivalent scaling. The range of $\rho R$ spanned by the NIF experiments is quite small, so the uncertainty in the estimated $\rho R$ does not change the overall conclusions of the plot. 

The red and black lines are from FLASH simulations of a current-scaling study of MagLIF targets with DT ice layers. These simulations are based on the two ignition designs described in Section \ref{sec:simulation} and Figs. \ref{fig:flash_ds3_designs}-\ref{fig:flash_ds3_curves}. The two designs, with $I_{max} = 56$ MA, were similarity-scaled to span a range of drive current $I_{max} = 20 - 65$ MA using the scaling prescriptions presented by D.~E. Ruiz \cite{Ruiz2023Iscaling,RuizIFSA2023}. The 20 MA Z-scale points generate $\sqrt{\rho R_{HS}~(\rho\delta R)_l} \approx 0.15$ g/cm$^3$, far below what is needed for ignition. As the drive current increases, the MagLIF designs increase in areal density and gain slowly. From $I_{max} \approx 50 - 65$ MA, the curves increase rapidly in areal density and target gain. The 56 MA Be-liner design point is just to the right of the NIF ignition star, whereas the 56 MA Al-liner point reaches higher areal density and gain at the same current. 

We can compare the efficiency of the DS driver/fuel-lined MagLIF combination to a recent NIF ignition experiment. There are several useful metrics for fusion system efficiency. The target gain $Q_{target}$ and the facility gain $Q_f$  quantify the overall driver-target system efficiency, whereas the fuel gain $Q_{fuel}$ and burn-up fraction are useful for assessing the efficiency of the burn required to generate large fusion yields. 

For a recent igniting implosion on the NIF using 2.05 MJ of laser energy, the total work done on the DT fuel was approximately 20 kJ, or 1 \% of the laser energy delivered to the hohlraum. Due to incomplete conversion of kinetic to internal energy and other losses, 8.2 kJ of the 20 kJ were converted to fuel internal energy\cite{Kritcher2024}. We extrapolate this to 9-10 kJ fuel internal energy in the absence of alpha heating for a 2.2 MJ NIF shot. Recall from Eq. \ref{phi_def} that the burn-up fraction $\phi$ is the ratio of the total nuclear energy output to the output that would be generated by reacting 100 \% of the fuel. Recent NIF ignition shots driven by 2.05 MJ with fuel mass 0.22 mg DT released 3.2 MJ and 3.9 MJ, corresponding to $\phi = 4.3$ \% and $\phi = 5.3$ \%, respectively~\cite{Landen2024}. Assuming the same fuel mass, the highest performing NIF ignition shot, which yielded 5.2 MJ at 2.2 MJ laser energy, burned up $\phi \approx 7$ \% of its DT fuel. 

We compare this to the 56 MA MagLIF-like target designs for the DS described in Section \ref{sec:simulation} and Figs. \ref{fig:flash_ds3_designs}-\ref{fig:flash_ds3_curves}. The electrical energy delivered to the load (target), given by Eq. \ref{eq:eliner}, depends on both the pulser rise-time and the target height $h$, so we wish to minimize both to maximize $Q_f$; this is why the designs in Fig. \ref{fig:flash_ds3_designs} are 6 mm tall. Table \ref{table:NIF_Dynos_gains} compares simulated performance metrics for these two DS targets to the highest performing NIF ignition experiment to-date. Table \ref{table:NIF_Dynos_gains} shows that reaching $Q_f>1$ with MagLIF on the DS requires fuel gain and burn-up fraction levels that are already demonstrated on NIF. Facility gain $Q_f > 1$ is achievable on the DS due to the high efficiency of energy delivery to the targets, which enables compression and heating of fuel masses nearly $50 \times$ larger than NIF ignition targets. 

\begin{table}
  \centering
    \caption{Comparison of performance metrics for NIF ignition and simulated DS targets. The Be DS and Al DS targets, driven with a peak current of 56 MA,  achieve $Q_f>1$ in 1D simulations.}\label{table:NIF_Dynos_gains}
  \begin{tabular}{| l | c | c | c |}
\hline
\textbf{Metric}  & \textbf{NIF Ignition} & \textbf{Be DS target} & \textbf{Al DS target} \\
\hline
 Energy stored in capacitors & 330 MJ & 80 MJ  & 80 MJ \\
 Energy delivered to target  & 2.2 MJ & 8.3 MJ & 8.4 MJ \\
 DT fuel mass  & 0.22 mg & $\approx 10.6$ mg & $\approx 10.2$ mg  \\
 Nuclear energy yield  & 5.2 MJ & 109 MJ & 380 MJ \\
 $Q_{target} = Y_{TN}$ / $E_{driver}$  & $\approx$ 2.4  & $\approx 13$ & $\approx 45$ \\
 $Q_{fuel} = Y_{TN}$ / $E_{int,DT}$  & $\approx$ 260  & $\approx$100 & $\approx$350 \\
 DT burn-up fraction $\phi$  & 7 \% & 3 \% & 10 \% \\
 Facility gain $Q_f$ & 0.016 & 1.36  & 4.75 \\
\hline
\end{tabular} 
\end{table}

In summary, the MagLIF ICF concept provides a conservative scaling path to $Q_{target} > 1$. With the addition of a cryogenic DT fuel layer, MagLIF scales to $Q_f > 1$ on Pacific Fusion's DS with fuel gain $Q_{fuel}$ and burnup fraction $\phi$ comparable to those already achieved on the NIF. Facility gain can be increased further by driving to larger peak current. For example, A 65 MA aluminum liner target scaled from the design shown in Table \ref{table:NIF_Dynos_gains} and Fig. \ref{fig:mead_both} generates 816 MJ of yield, corresponding to facility gain $Q_f \gtrsim 7$, sufficient for a fusion power plant. In addition to driving to higher peak currents, additional improvements to target design can further increase gain.

\subsection{Demonstrated performance with thermal fusion reactions } \label{sec:thermal}
The history of magnetically-driven pulsed fusion systems contains examples where the predominant mechanism of fusion production was non-thermal beam-driven (that is, highly energetic fuel ion beams, resulting from instabilities, that drove fusion by collisions with other fuel) rather than generated by a thermal distribution of ions.  This can be seen in the early days of quasi-static Z-pinches, where sausage- and kink-mode instabilities tended to break apart the plasma column, producing large transient voltages across the column\cite{Haines_2011}, accelerating ions to high energy ($\sim 1$~ MeV), resulting in non-thermal neutron production. Similar behavior has been observed in a variety  of dynamic Z-pinch configurations, e.g. the dense plasma focus (DPF)\cite{krishnan_2012, Klir_PPCF_2012, Schmidt_PoP_2014} and gas puff Z-pinch\cite{Klir_PRL_2014, coverdale_2007, velikovich_2007, Welch_PoP_2010, Appelbe_PoP_2015, Offerman_PRL_2016}. Neutrons produced by non-thermal means tend to not scale favorably to ignition and gain.  A common factor in all of these systems is that the driver current is present within the fuel itself.  This means that fuel ions will directly experience the transient electric fields resulting from MHD and kinetic instabilities.  

Fortunately, pulser ICF concepts such as MagLIF do not suffer from this problem. In MagLIF implosions, a thick metal liner implosion is used to compress the fusion fuel, preventing the penetration of the driver current into the fuel and the creation of energetic ion beams. All observations of density, temperature, and neutron spectra in MagLIF experiments are consistent with fully thermal nuclear burn processes, exhibiting none of the signatures of non-thermal ion populations, e.g. high energy neutrons and anisotropic neutron emission\cite{Knapp_PoP_2013}. Gomez \textit{et al.} showed that the neutron spectra observed along the axial and radial directions were roughly equivalent\cite{Gomez_PRL_2014, Hahn_2016, Yager-Elorriaga_2022}. When fitting the core of the spectrum with a thermal model over a wide range of experiments, no excess of neutrons in the high energy tail was observed. An excess of neutrons is observed in the low energy portion of the spectrum, but this is consistent with neutron scattering. Furthermore, the yields measured along different directions all agree within the measurement uncertainty\cite{Gomez_PRL_2014}. In short, the neutron emission indicates no need to invoke a non-thermal explanation for any of the observable features. A Maxwellian ion distribution is perfectly adequate to explain the observed neutron signatures.

Other evidence from the wide range of experiments conducted at Sandia National Laboratories supports this claim. Knapp \textit{et al.} demonstrated a Bayesian data assimilation approach utilizing a thermal plasma model that is able to match all observables (neutron and x-ray) simultaneously over a wide range of experiments\cite{Knapp_2022}. The spectral characteristics of both neutrons and x-rays are used to fit multiple diagnostic signatures simultaneously, assuming the ions and electrons are in thermal equilibrium. Gomez \textit{et al.} showed that the target yield scaled as expected with inferred ion temperature over a range of $\sim 1-3$~keV\cite{Gomez_PRL_2020}. Furthermore, Knapp \textit{et al.} and Harvey-Thomspon \textit{et al.} demonstrated that mix of non-fuel material into the stagnation column degrades performance as one would expect for a thermal plasma\cite{Knapp_PoP_2019, Harvey_PoP_2018}. In a beam-target configuration, mix at the few percent level will only degrade performance in proportion to the dilution of the target ions.  However, the trends in MagLIF data are consistent with the reduction in performance expected due to the enhanced radiation losses and resultant cooling of the fusion plasma. The evidence overwhelmingly points to the thermal nature of fusion production in the MagLIF configuration, lending strong support to the scaling of these concepts to ignition and gain.

\section{Pulser Design}\label{sec:pulserdesign}

\subsection{Demonstration System (DS) Overview}\label{ssec:dsoverview}

\begin{figure*}
    \centering
    \includegraphics[width=6.5in]{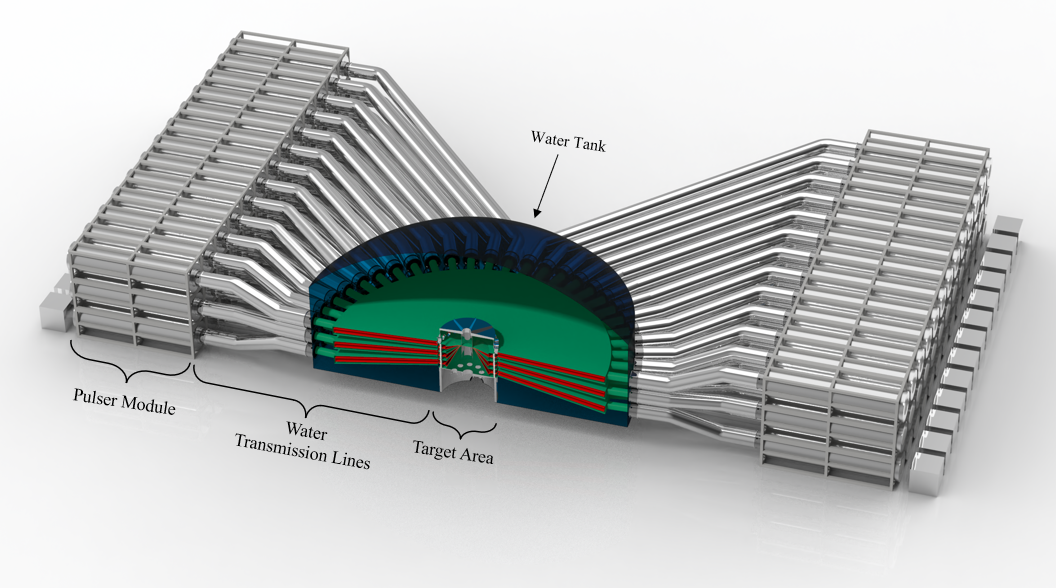}
    \caption{Conceptual schematic cross section showing half of Pacific Fusion's DS. The total dimensions of the full system are $\sim$73m x 80m.}
    \label{fig:Dynos}
\end{figure*}

Pacific Fusion’s DS is designed with the goal of demonstrating fusion performance required for a commercially viable fusion energy system. The DS stores approximately 80 megajoules of electrical energy and will deliver >60 MA over $\sim$100 nanosecond timescales to a fusion target. Figure \ref{fig:Dynos} shows the layout of the DS pulser, which is broken up into three sections: the pulser modules, the water transmission lines, and the target area. The DS consists of three discrete levels of pulser modules with an impedance-matched-Marx-generator (IMG) architecture. Power is delivered from the modules to the target area via water-insulated coaxial transmission lines and tri-plate impedance transformers. The water chamber surrounding the target chamber and the tri-plates will be filled with deionized water providing the necessary neutron shielding -- at least 5 meters -- to ensure capture of all generated radiation. The target area consists of a water-vacuum insulator stack, the target diagnostic suite, and a vacuum area containing a set of magnetically insulation transmission lines (MITLs), detailed in Section \ref{sec:fulldsmodeling}, a triple post-hole convolute to combine the levels, and a final inner MITL that delivers the power to a cassette containing the fusion target. The target itself consists of the fusion fuel surrounded by the liner, and the electrical current delivered through the target generates the Lorentz force that rapidly compresses the fuel. 

\subsection{Pulser Modules: IMGs}

\begin{figure}
    \centering
    \includegraphics[width=4.5in]{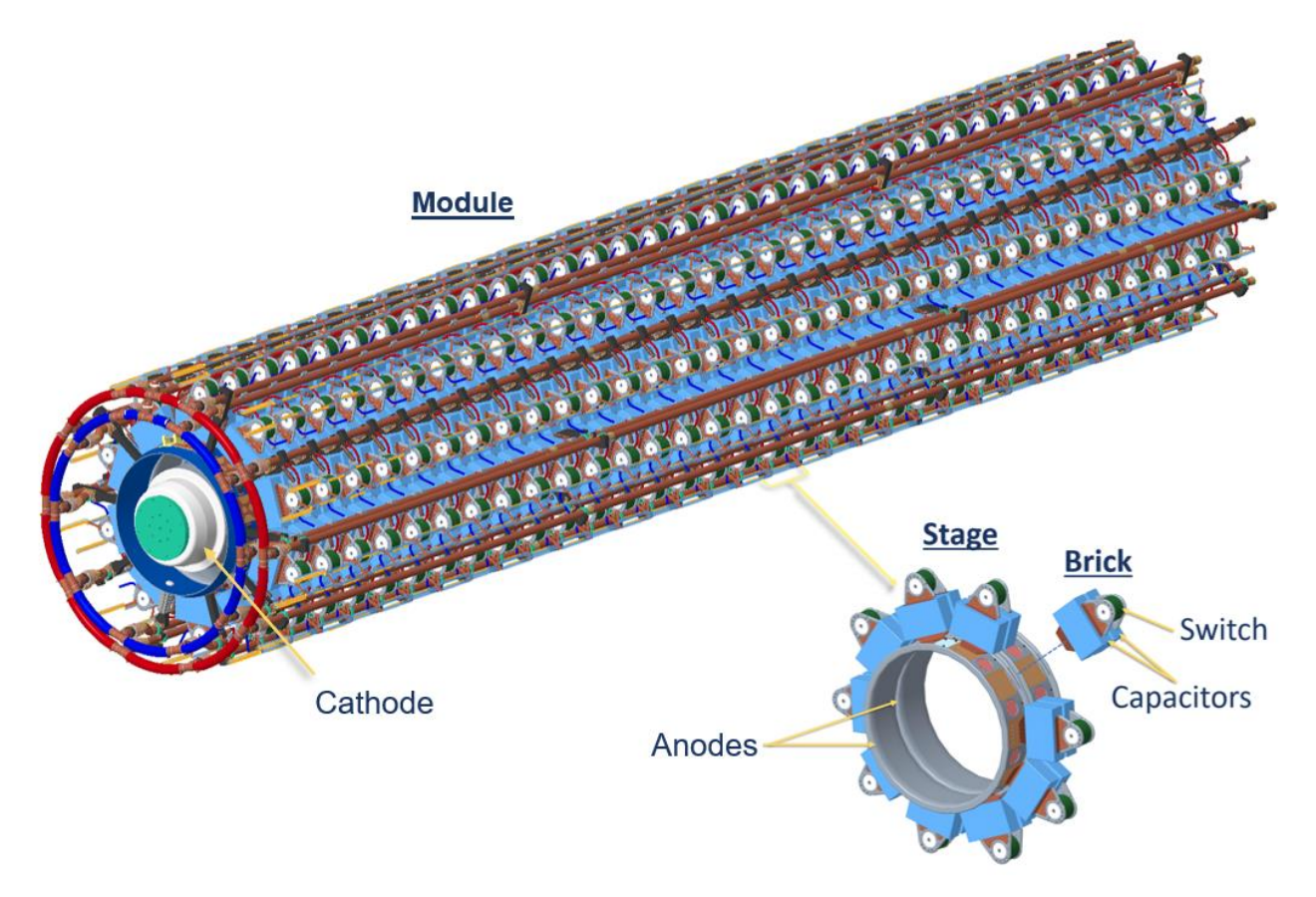}
    \caption{IMG Module and components. The module diameter is approximately 1.9m.}
    \label{fig:IMG_Module}
\end{figure}

The DS pulser uses an IMG architecture\cite{stygar2017impedance, lechien2023sirius}. Figure \ref{fig:IMG_Module} shows an IMG module and its components. The building block of an IMG is called a brick, which contains two capacitors charged at a given voltage differential in series with a low impedance spark gap switch. When the switch is closed, the energy from the capacitors is discharged into the stage, which consists of multiple bricks discharged simultaneously in parallel. The voltage of a given stage is that of the brick charging voltage, and the energy discharged is the sum of the stored energy in all capacitors. Stages are added in series to create a pulser module, with the output voltage of module being the sum of the stage voltages. Pacific Fusion’s baseline IMG architecture consists of 10 bricks per stage and 32 stages per module. The IMG has a bipolar configuration, with each brick charged to +/- 100 kV in series with a triggered spark gap switch. With a capacitance of 160 nF, the total energy stored per capacitor is 800 J, resulting in a total energy storage of 512 kJ per module.  

The full DS pulser shown in Figure \ref{fig:Dynos}  contains 156 modules in parallel, stored in canisters filled with dielectric oil.  The modules collectively  store up to $\sim$ 80 MJ of energy, and the pulser can deliver > 350 TW to the water section.

\subsection{Water Transmission Lines}

Current is delivered to the target chamber through the coaxial pulse tube and tri-plate transmission lines using deionized water as the dielectric, as seen in Figure \ref{fig:Dynos}. The vacuum target chamber at the center of the pulser contains the magnetically insulated transmission lines that delivers the current to the ICF target. Due to the low capacitance and low inductance of the system, pulse compression is not needed for pulses as short as 100 ns. Longer pulses can be created by varying the triggering times of the modules. 

\subsection{Target Area}\label{sec:target_area}

\begin{figure}
    \centering
    \includegraphics[width=6.5in]{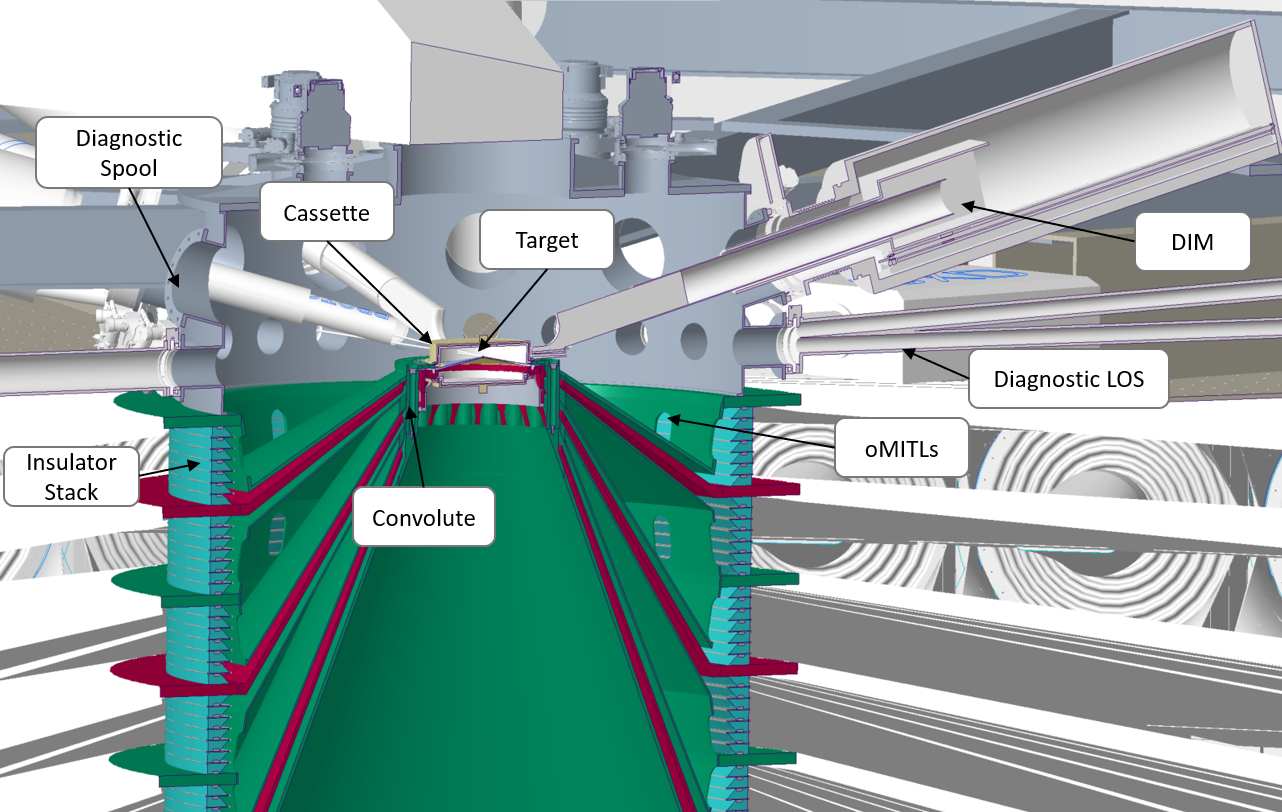}
    \caption{Target Area Cross Section of the DS Target Area: The design of the outer MITL (oMITL) is similar to that used on the Z machine but with 6 conical transmission lines consisting of four split anodes and three solid cathodes leading to the convolute. For scale, the insulator stack has a radius of 3.2m.}
    \label{fig:Target Area}
\end{figure}

Pacific Fusion breaks down the Target Area into five subsystems: vacuum power flow system, inner-MITL/cassette, chamber support systems, target diagnostics, and hardware used for blast management. The vacuum power flow system consists of the insulator stack, the outer-MITLs, a triple post-hole-convolute, and the inner-MITL (iMITL), detailed more in Section \ref{sec:fulldsmodeling}. It is responsible for delivering energy from the water tri-plates through the vacuum MITL levels into a single iMITL. The iMITL transfers the current to the target. The cassette is a small portable chamber that houses the final portion of iMITL, the target and support utilities, including the interfaces to the cryogenic system. The target chamber, which is made up of the insulator stack and the diagnostic spool, maintains vacuum during operation. The support systems are responsible for maintaining the target chamber vacuum, proving the structural load path and dynamic response, and installation and alignment of components into the system during the build. The target diagnostics include nuclear, x-ray and optical diagnostics used to inform and iterate on our experiments. Each diagnostic is mounted to a port on the diagnostic spool, giving the diagnostic its own line of sight to the target. Diagnostic ports include 200mm diameter equatorial line of sight ports and 400mm diameter 15 deg line of sight ports (Figure \ref{fig:Target Area}). 

The cassette concept is designed for flexibility, with the cassette replaced after every shot. This allows Pacific Fusion to iterate and optimize the target design over time. The cassette is assembled offline from the main system. During the build process, the target and iMITLs are meticulously measured to ensure precision alignment. Once the target is ready, the cassette is brought down to vacuum and the cryo layering process can begin, if desired. Several cassettes can be built at a time to be ready for shots. This allows for high-tempo shot operations, including high-yield shots. The cassette design, along with other strategic design decisions, allows the DS to achieve at least 1 shot per day while maintaining the flexibility to use solid fuel and multiple target types. 

Another advantage of the DS design is the water tank. The target chamber and tri-plates are contained by a water tank with 5 m of water sitting over the vacuum target chamber. This provides sufficient neutron shielding to ensure that the desired background radiation levels are achieved outside the system. 

\subsection{Full DS Modeling}\label{sec:fulldsmodeling}
\begin{figure}
    \centering
    \includegraphics[width=6.5in]{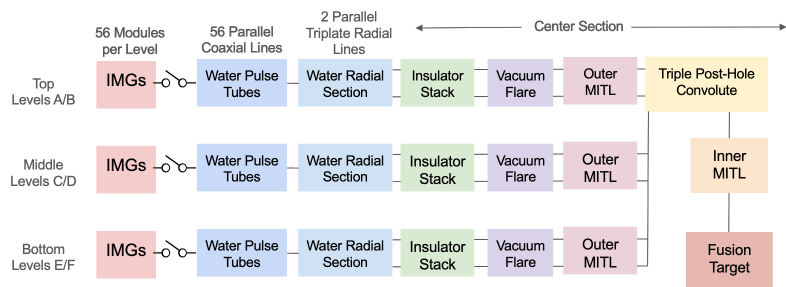}
    \caption{Block diagram of the full DS: This shows major subsystems required for circuit modeling and electromagnetic simulations.}
    \label{fig:BlockC}
\end{figure}

The entire DS has been modeled using the circuit code CASTLE (Circuit Analysis and Simulation with Transmission-Line Emphasis) \cite{Allen2023}, and the model includes all components down to the module level. Every subsystem (Figure \ref{fig:BlockC}) has subsequently been modeled using 2D/3D full electromagnetic-only (EM) or electromagnetic-with-plasma simulations using the Particle-In-Cell code CHICAGO \textregistered \cite{welch2020hybrid} to validate the circuit representations and to determine the transport efficiency of the vacuum section. These CHICAGO \textregistered simulations include the full circuit, with individual components removed from the circuit. The input/outputs of removed components are coupled as boundary conditions to the circuit simulation. The circuit model is incorporated in a proprietary optimizer that balances multiple constraints, including the specific component designs and spatial constraints such as the layout of the facility.

\subsubsection{Power Delivery in Vacuum Section}
The vacuum section of the DS is subject to the highest electric and magnetic fields originating in the pulser, so it is the most likely region for current to be diverted en route to the fusion target. Efficient transport of current in this regime requires the use of Magnetically Insulated Transmission Lines (MITLs) \cite{Creedon1975,Creedon1977,VanDevender1979,DiCapua1983,Mendel1992,Mendel1996,Ottinger2006} to carry the current from the water-vacuum interface to the load.  In the DS, this is accomplished through six outer MITLs which are connected together through a triple-post-hole-convolute. The convolute combines the currents and sends them to the inner MITL, which delivers power to the fusion target.

The water-vacuum interface is an insulator stack of discrete dielectric rings similar to the Z design \cite{Stygar2005} (but scaled to a radius of 3.2 m) and has been simulated in full-scale dynamic CHICAGO \textregistered EM simulations. The vacuum stack modeling includes detailed electrode shaping on the water side and the vacuum side to control the electric field profile across the stack, keeping average electric field (V/d) across an individual insulator ring to below 160 kV/cm.  The full spatiotemporal dynamics of the applied electric field from CHICAGO \textregistered simulations have been used in a statistical flashover model \cite{Stygar2004}, predicting a full stack flashover of any vacuum level at < 0.2\% when using a conservative $\gamma_{SM}$ \cite{Stygar2005}. With a full charge voltage on the modules, the pulser will deliver $\approx 310$ TW of peak power to the insulator stack.

Full EM, 2D/3D kinetic PIC simulations using CHICAGO \textregistered were used to model the entire system, starting at the water-vacuum interface and extending to the liner surface.  The vacuum section is spatially too large to efficiently model in a single simulation, so the vacuum section was broken into 3 main components: the inner MITL, the triple post-hole convolute\cite{stygar2015conceptual}, and the outer MITL/insulator stack.  The simulations spatially overlap so that the convolute simulation includes the last $\approx 25$\% of the outer MITLs and the first $\approx 10$\% of the inner MITL, ensuring continuity of conditions between simulations. The simulations of the outer MITLs and convolute are resolved to 250 \textmu m spatially with timesteps that resolve the electron cyclotron frequency.  The convolute has a larger radius than simply scaling up from the current Z design \cite{Bennett2019transport,Gomez2017}, to minimize heating of the convolute and to increase its survivability. The increased survivability comes at a cost of early-time inefficiency relative to a scaled Z, as the magnetic insulation takes some time to be established in the convolute. The convolute is intended to remain in place between shots, so it should be a low source of surface contaminants, as it is not regularly exposed to ambient air.  

The convolute simulations use similar models and algorithmic approaches to prior state-of-art MITL simulations \cite{Bennett2019transport,welch2020hybrid,welch2019hybrid,welch2009hybrid,Bennett_milestone:2023,tummel2022flow,Rose2015} and include space-charge-limited emission \cite{LangmuirSCL} of electrons for electric field stresses exceeding 240 kV/cm, ion emission\cite{Rose2008,Blaugrund1977} for surfaces heated above 600 K, and 1D transport models for particle heating, thermal and magnetic diffusion attached to each electrode surface element. 

The outer MITL design is similar to the Z design \cite{Stygar2009} but with 6 conical transmission lines (Figure \ref{fig:Target Area}) consisting of four split (hollow) anodes and three solid cathodes leading to the convolute. Since the convolute will be at a relatively large radius, these transmission lines can maintain constant impedance up to the post-hole region in the convolute and still have sufficiently large A-K gaps to minimize losses. In the current design, the outer MITLs have an average impedance of $\approx 2.3 ~\Omega$.

\subsubsection{Driver coupling efficiency}

The convolute simulations are attached to a low inductance, low mass inner MITL simulation. This simulation has a variable resolution from  25 $\mu$m to 125 $\mu$m spatially with an adaptive time-step set to $\frac{8}{\omega_c}$, where ${\omega_c}$ is the electron cyclotron frequency and uses an implicit field solver \cite{welch2020hybrid} to handle high magnetic fields and plasma densities. The target and inner MITL are contained and replaced with a new cassette every shot. The target cassette allows most of the MITL system to remain under vacuum and free of contaminants. These simulations include the physics present in prior convolute simulations and additional electrode plasma sources from the presence of surface contaminants\cite{welch2019hybrid}. These plasmas have previously been shown to be able to cross the A-K gap via a Hall-like mechanism \cite{Bennett2019transport,Bennett2021transport,Bennett_milestone2023}, leading to the generation of additional current pathways that can divert current from reaching the liner surface. The inner MITL simulations also include electrons and ions flowing from the convolute into the inner MITL region.

\begin{figure}
    \centering
    \includegraphics[width=4.5in]{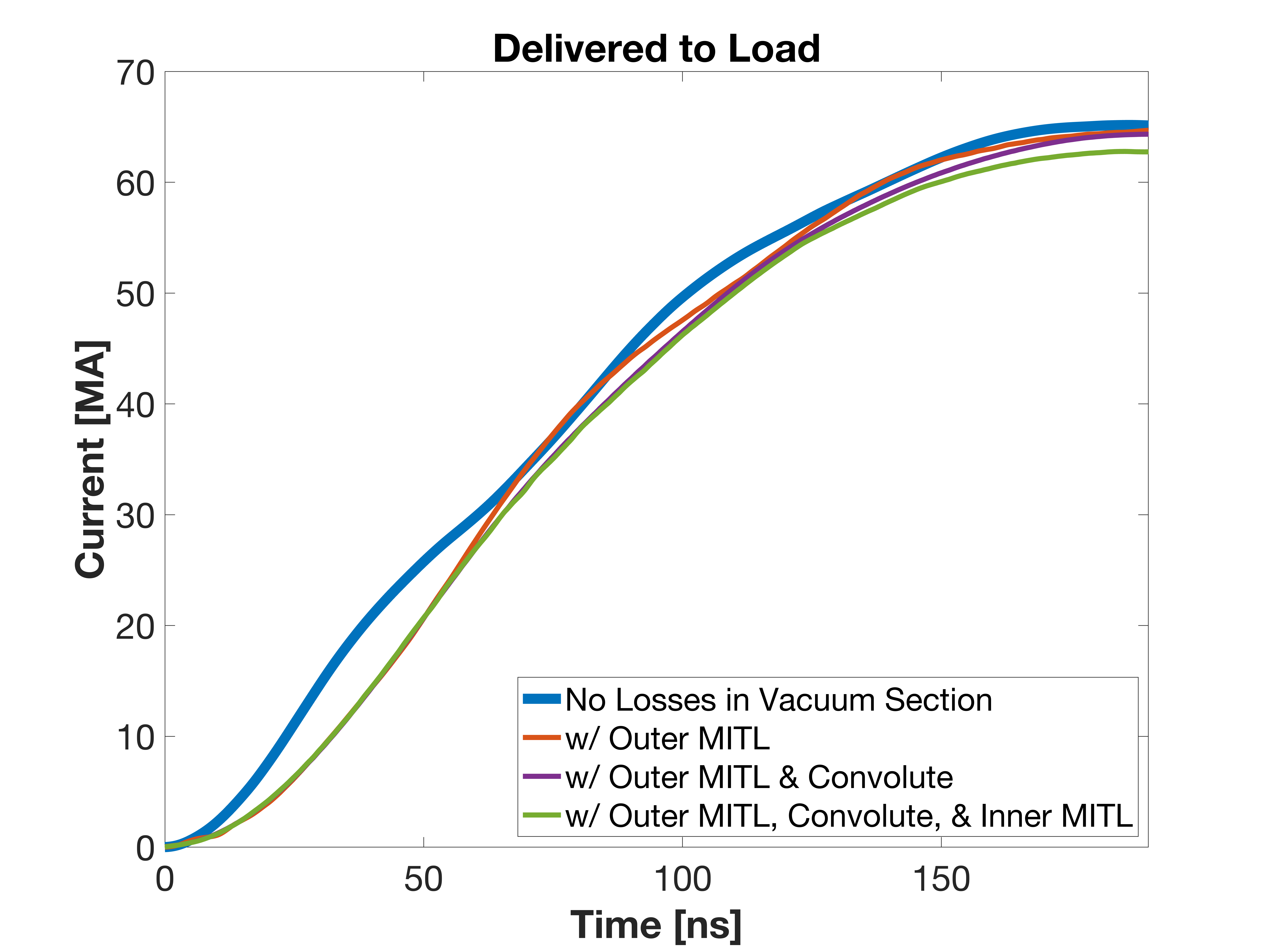}
    \caption{Current delivered to the liner surface for calculations with increasing vacuum components.  Zero time is taken to be when the first kA arrives for the ideal circuit case.}
    \label{fig:CurrentD}
\end{figure}

Figure \ref{fig:CurrentD} shows the net current delivery to the load region. Simulated current pulses are shown for the ideal circuit, with plasmas only in the outer MITLs; plasma sources in both the outer MITLs and the convolute; and plasmas in the entire vacuum section. In these simulations, the load has a fixed impedance of 27 m$\Omega$. At early times, less than 60 ns in Figure \ref{fig:CurrentD}, some current is diverted from reaching the load until magnetic insulation is established in each level. Pulses on the top, middle, and bottom levels arrive at the outer MITL at different times to take into account the differences in the length of their respective outer MITLs. The different levels magnetically insulate at different points along the current pulse. At peak current, the DS delivers about 62.8 MA, 115 TW, and 10 MJ to the load region, coupling \emph{96}\% of the current delivered to the insulator stack to the final load. Further optimization of the vacuum region is underway to further increase current delivery, through a combination of reducing the system inductance and improving current delivery through the final components. 

Detailed interactions between the inner MITL region and the target are also important considerations and an active area of research. Physics considerations and modeling methodologies for assessing this interaction are outlined in Appendix~\ref{sec:pulser-target_coupling}.

\subsection{Pulse Shaping}

\begin{figure}
    \centering
    \includegraphics[width=6.5in]{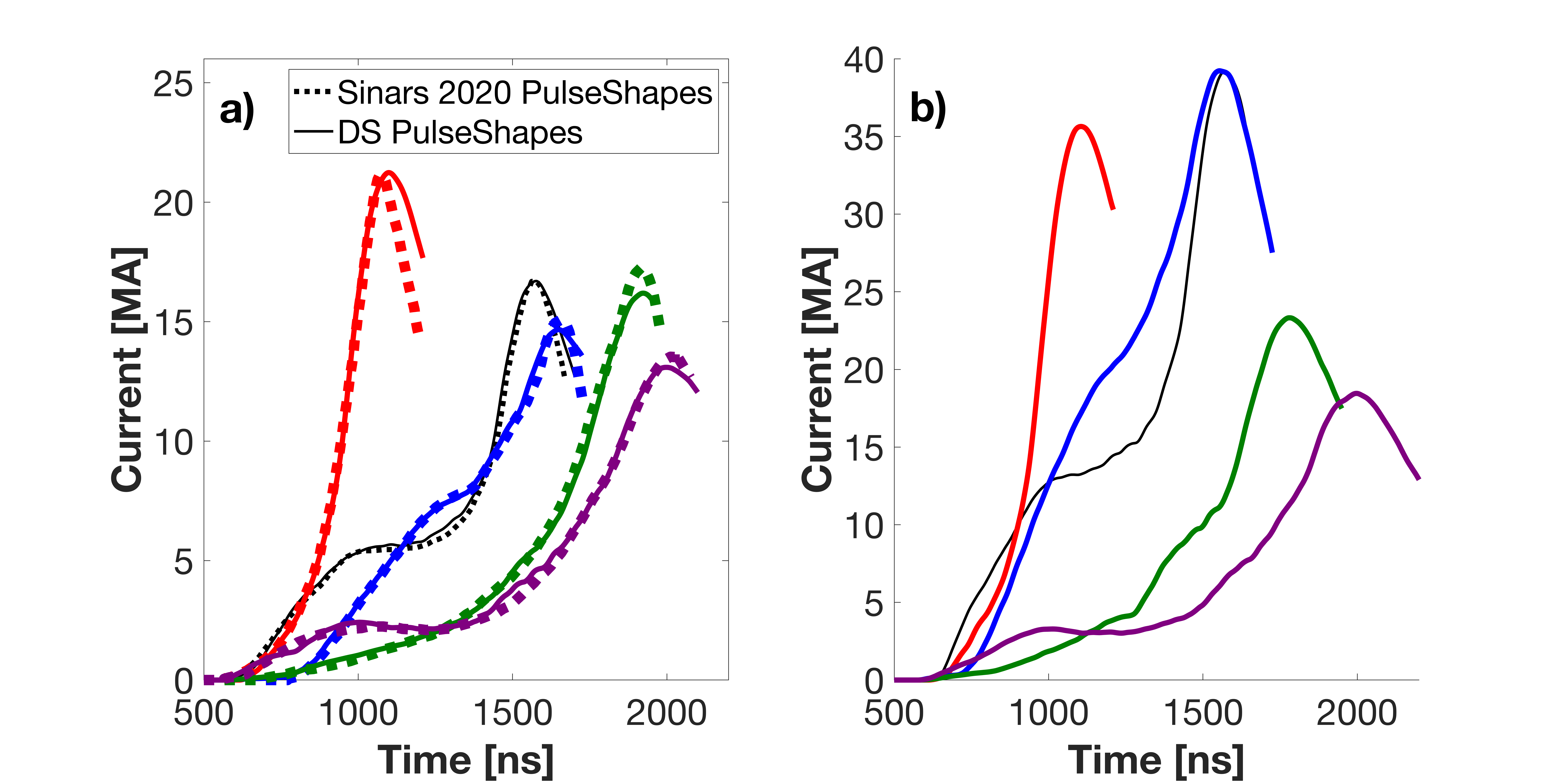}
    \caption{(a) Comparison of pulse shapes presented in Ref.~\cite{sinars2020review} for the Z machine (dashed lines) and those achievable on the DS with only changing module timing and charge voltage (solid lines). (b) Same DS pulse shapes when utilizing as much of the stored energy as possible.}
    \label{fig:PulseShape}
\end{figure}

The modular design of the DS, along with water insulated transmission lines, allows for two levels of pulse-shaping.  The individual modules can be independently triggered and are transit-time isolated by the pulse tubes at an individual level for $\sim 715$ ns from the outermost to innermost set of modules. The top, middle, and bottom levels are further isolated by an additional $\sim 750$ ns. Longer pulses outside this range could be generated in a manner similar to how long pulses are generated on the Z-Machine (see e.g. Ref.~\cite{sinars2020review}), through the insertion of water switches in-between the module outputs and the water impedance transformer for further isolation when needed. Panel (a) of Fig.~\ref{fig:PulseShape} shows a comparison between the pulse shapes from Ref.~\cite{sinars2020review} and equivalent pulse shapes achievable on the DS (in a full circuit model of the pulser) by only varying the initial charge voltage and switch times while utilizing less than $\sim$ 30\% of the total stored energy available. Panel (b) of Fig.~\ref{fig:PulseShape} shows these same profiles with many of the modules charged to full voltage, leading to a $2.75\times$ increase in peak current.

\subsection{Diagnostic Suite}\label{sec:diagnostics}
The primary goal of the DS is to achieve net facility gain. We expect that achieving this goal will require rapidly evolving the target design based upon understanding built from experiments. This process will require that each DS experiment be appropriately diagnosed to understand its performance and to characterize the degree to which the experiments agree with, or differ from, computational models. Additional diagnostics will be required for targeted physics measurements on dedicated experiments, for example to study driver-target coupling or hydrodynamic instabilities. As on the NIF, this required the development of a full suite of diagnostics\cite{kilkenny2016national} to measure key observables.

\begin{figure}
    \centering
    \includegraphics[width=6.5in]{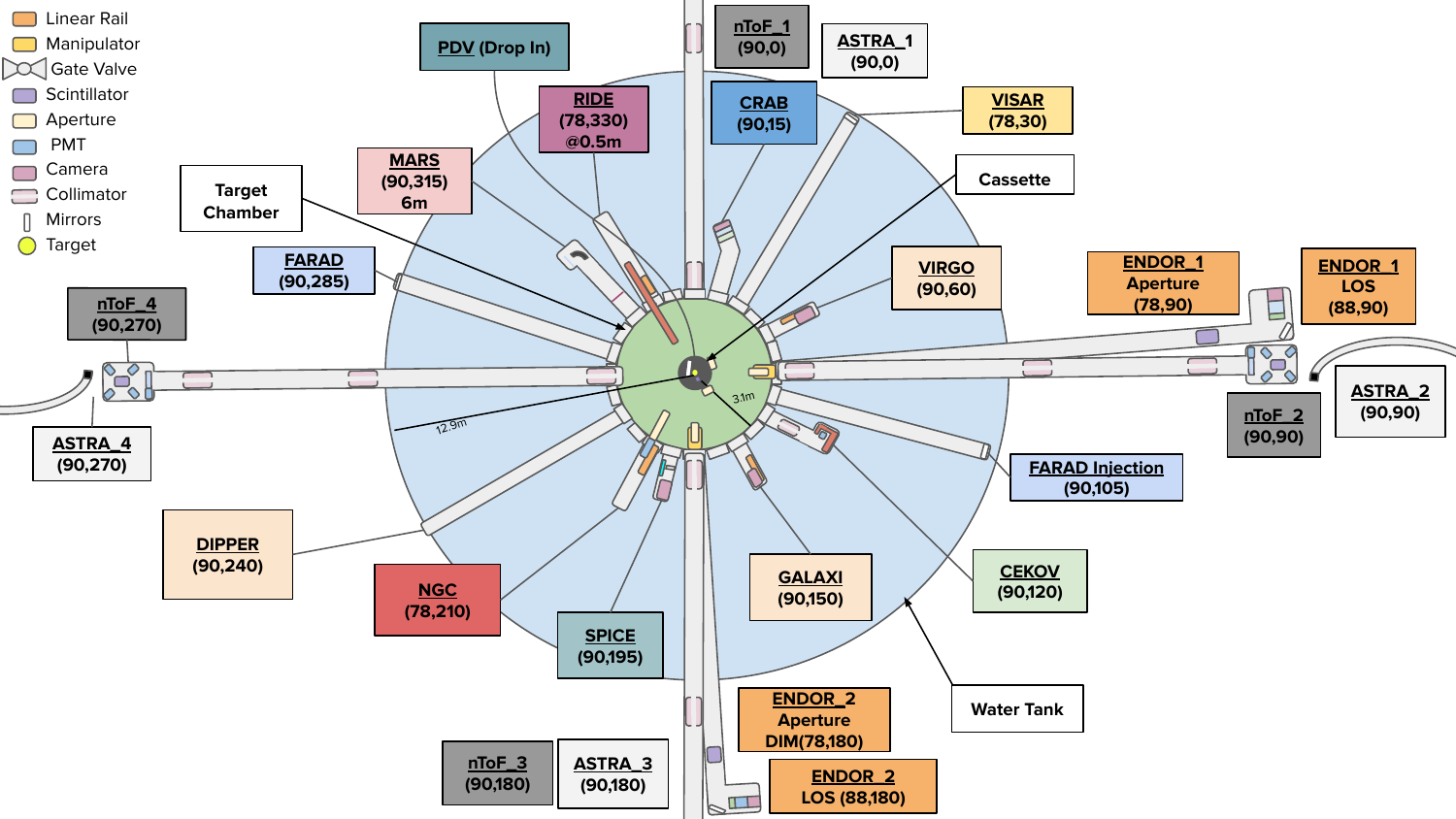}
    \caption{Conceptual plan for the DS diagnostic suite and layout. Diagnostics are noted by their distance from TCC and angle in facility coordinates ($\theta$, $\phi$) where $\theta$ is the polar angle and $\phi$ is the azimuthal.}
    \label{fig:diagnostic_layout}
\end{figure}

Our initial baseline diagnostic plan is shown schematically in Fig. \ref{fig:diagnostic_layout}. The water chamber is on the order of 13 meters in radius. Six long collimated lines of sight are planned for 6 neutron time of flight (NTOF) and 2 Energetic Neutron Detectors Observing with Resolution (ENDOR) for neutron imaging; four of the NTOFs are arranged in the equatorial plane and shown in Fig. \ref{fig:diagnostic_layout}. These are positioned so that the NTOFs are arranged in opposing pairs that are mutually orthogonal to best diagnose spectral moments (e.g. as in \cite{hartouni2023evidence}). The 2 ENDOR are orthogonal to optimize for 3-D reconstruction \cite{volegov2015three}. Positioning of these instruments far from Target Chamber Center (TCC), and therefore outside of the water chamber, is necessary to achieve the desired spectral resolution on NTOF through the time of flight dispersion, and the desired spatial resolution for ENDOR. All other diagnostics are either within the water chamber or have optical relays to a side laboratory space. This approach maximizes the inherent shielding of the water chamber.

Completing the nuclear diagnostic suite are a DT $\gamma$-ray reaction history measured by a Compton Electron cherenKOV (CEKOV) detector\cite{herrmann2010diagnosing}, two DD nuclear burn history measurements based on a Recoil Ion DEtector \cite{fry2023geant4} (RIDE) and on (n,$\gamma$) Cherenkov detection (NGC), a MAgnetic Recoil Spectrometer (MARS)\cite{gatu2022high} for neutron yield and spectrometry, and Activation Samples for Thermonuclear Reaction Analysis (ASTRA) \cite{bleuel2012neutron} in the equatorial NTOF lines of sight.

The DS x-ray diagnostic suite will include an emitted power measurement using a Calibrated Radiometry and Bolometry (CRAB) diagnostic composed of a filtered streak camera (e.g. similar to the SPIDER system on NIF\cite{khan2012measuring}), which provides the highest temporal resolution, complemented by filtered diodes\cite{chandler1999filtered}; two equatorial orthogonal x-ray cameras for self-emission imaging, Gated Active Long Aspect X-ray Imager (GALAXI), and radiography, Versatile Imaging for RadioGraphic Observations (VIRGO); and an x-ray SPectroscopic Instrument for Collisional Excitation (SPICE), potentially with several crystals for key energy ranges. A line of sight is reserved opposite VIRGO for a source, the Dynamic Irradiation with Photons for Producing Energetic Radiographs (DIPPER). For optical diagnostics, we plan on an optical transmission probe to enable FAraday Rotatation and Deflectometry (FARAD) in low-density near-target plasmas, a line VISAR\cite{datte2020design} for drive characterization and materials studies, and fiber-coupled photon Doppler velocimetry (PDV).

We continue to improve upon this conceptual diagnostic plan for DS, and we are actively developing detailed designs for individual systems, to be detailed in future publications. The diagnostics ports not shown in Fig. \ref{fig:diagnostic_layout} are either used for chamber infrastructure or are available as spares. Approximately 10 ports are potentially available for additional diagnostics; Pacific Fusion is actively evaluating additional measurements beyond our baseline suite.

The bulk of the diagnostic suite is focused on the core mission of demonstrating facility gain. This requires that we are able to track the performance of a variety of ICF targets by inferring the Generalized Lawson Criterion (GLC) $\chi$ for each implosion experiment as well as obtaining enough information to ascertain the root cause of observed failure modes and performance limiting effects. The potential performance limiters are numerous, including symmetry, end losses, mix, and inefficient flux compression. Each has different observable consequences.  We are developing a model-driven Bayesian approach to infer performance from the ensemble of diagnostic observables. This methodology has been demonstrated on MagLIF implosions on the Z machine using a static quasi-2D model to infer burn-averaged quantities and $\chi$ across a wide range of experiments\cite{Knapp_2022}. Additional examples and extensions to this approach are discussed in Ref. \cite{Knapp_2023}. This approach allows for the reconstruction of the target performance and extraction of unobserved dynamics, with uncertainties, from the underlying model. This formalism will be used in the near-term to aid in the design of the diagnostic suite by optimizing parameters of the individual instruments to obtain maximally informative observations\cite{Knapp_JPP_2023}. This quantitative optimization enables us to weigh the consequences of different design choices for the diagnostics in the context of the entire suite of observations, enabling tradeoffs to be considered that can have dramatic impacts on cost, schedule, and performance.  

\subsection{Facility cost comparison} \label{sec:faccost}
In previous sections, we have described the design and parameters of Pacific Fusion's DS. Here, we briefly describe its cost relative to the NIF. 

Our present detailed estimate puts the cost to build the 80 MJ DS at slightly over $\$6/J$, or approximately $\$500M$. This estimate is comparable to the inflation-adjusted per-Joule cost of the ZR project --- which completely overhauled and rebuilt the Z facility into its present state --- of $\$4/J$ in 2006 dollars or $\$6.3/J$ in 2025 dollars \cite{FPA2010}. The capital cost to build the NIF was roughly $\$3.5B$ in 2012, inflation-adjusted to about $\$4.9B$ in 2025. In short, the DS costs $\sim10\times$ less to build than the NIF.

\section{Path to commercial power}\label{sec:reactor}

Many aspects of the DS will translate directly to the pulser required for subsequent commercial power plants, which will have: a similar system size and layout; near-identical module architecture; similar scale chambers; and the same sort of targets. The DS may achieve facility gains well beyond its design goal of $Q_f>1$. However, commercial pulser IFE systems differ in several respects; they must last for decades of cycling, be cycled rapidly, and efficiently capture fusion output and breed tritium. Specifically, they require three additional technical developments:
\begin{itemize}
    \item{\textbf{Component lifetime.} Development of pulser components that allow for repetitive operation with long cycle life and infrequent, inexpensive maintenance.}
    \item{\textbf{Fusion chamber.} Development of a fusion chamber that allows for high-gain shots, survives repetitive stresses arising from shots, and can be rapidly cleared and prepared for subsequent shots.}
    \item{\textbf{Tritium breeding.} Development of blankets and first solid walls that are compatible with a durable, rep-ratable fusion chamber and that permit a practical tritium breeding ratio and efficient fusion energy capture.}
\end{itemize}
We will discuss these topics in more depth in subsequent publications, but give an overview of the main issues here.

\subsection{Component lifetime}\label{sec:test}

Operating a pulser IFE power plant will require shooting targets at repetition rates between 0.1-10 Hz, compared to the approximately single shot per day for current leading scientific facilities (NIF, Sandia Z machine). The pulser therefore must operate at repetition rates 10\textsuperscript{4}-10\textsuperscript{6}-fold higher than present large-scale facilities at an affordable capital cost, with component lifetimes averaging 1 billion shots. Currently, the two components within the IMG pulser modules that do not typically meet this lifetime requirement at operating parameters are the high-voltage capacitors and spark gap switches. Improving the lifetime of these key components can be achieved by two approaches: (1) reducing operating conditions (e.g., lower voltage and stored energy per component) by using more components in parallel and (2) innovative component design. 

An example of the first approach would be to reduce operating voltages for capacitors by 3-fold, from 100 kV to 35 kV, which would be expected to increase capacitor lifetime from $\approx10^5$ shots to $\approx10^9$ shots, at the expense of $9\times$ less energy stored per capacitor~\cite{CSI}. This results in the need for 9-fold more capacitors to be able to store and discharge the same amount of total energy.  Lower voltages and capacitance also reduce the charge transferred by the switches, which could result in a substantial increase in switch lifetime. 

The second approach would involve developing innovative designs to increase the lifetime of capacitors and high voltage switches. For capacitors, for example, this includes developing dielectric polymers with improved dielectric constants and breakdown strengths, and capacitor designs that reduce electric field stresses to decelerate capacitor aging. 
For switches, improved spark gap switch designs can reduce the charge transferred per unit area through increased multi-channeling, reducing ablation and therefore increasing switch lifetime; alternate switch technologies (e.g., solid state) as well as in-situ refurbishment for traditional spark-gap switches are other avenues for increasing switch lifetimes. 

Pacific Fusion is pursuing both above approaches to component improvement --- using modern non-destructive evaluation methods, accelerated testing, and parallelization --- to develop optimized components that meet commercial requirements and enable affordable, firm fusion power.

\subsection{Fusion chamber}
To achieve high power output in a pulser IFE power plant, one needs to rapidly cycle the fusion chamber, i.e. drive fusion targets in rapid succession, preparing the chamber after each shot. To do so, one must ensure that replaceable electrodes can be ejected without damaging the walls, that the post-shot impulse does not damage the wall, and that high vacuum can be rapidly restored. We discuss these issues below.
\subsubsection{Replaceable electrode disassembly}
\begin{figure}
    \centering
    \includegraphics[width=6in]{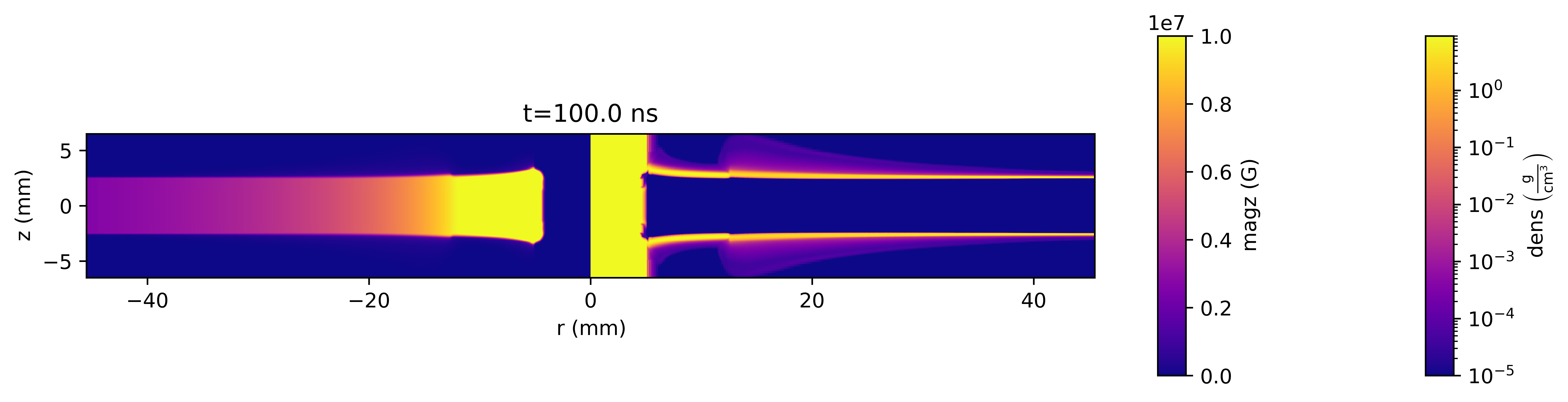}
    \caption{Low-mass inner MITLs are advantageous for low cost, high repetition systems. A 100$\mu$m MITL with a 200$\mu$m, 1.25mm radius washer placed on the transmission lines is one promising such configuration. This has a 2.49nH inductance using a 100ns linear ramp to 60MA, 13.7\% greater than large-mass, theoretically immovable MITLs. \cite{mcbride2018primer}.}
    \label{fig:low_mass_iMITLs}
\end{figure}

In past MagLIF and other pulsed power experiments, the destructible target mass has included large replaceable electrodes (inner MITLs, or iMITLs) that fracture from the shot blast and the magnetic pressures temporarily constrained by them. This generates a considerable amount of debris and requires a large amount of mass to be inserted and evacuated. One option we are pursuing is low-mass iMITLs that minimize the amount of material consumed and recycled. The criteria for a successful iMITL are that it must carry the surface current supplied by the pulsed power system and must not deform significantly during the shot as to not impose too large of an impedance penalty on the system \cite{slutz2003low}. The skin depth of copper is just under 50$\mu$m during these timescales, setting a small minimum thickness necessary for carrying the current.  One promising iMITL option is shown Fig. \ref{fig:low_mass_iMITLs}.  

\subsubsection{Post-shot vacuum chamber impulse response}
Pulser ICF systems require extreme temperatures and pressures over a very short period of time. The magnetic energy driving the fusion target system is released as the imploding target ``bounces'' and expands. The alpha particle energy from an igniting high-gain target, which can be many times the driving energy, is absorbed by the target and near-target material over a few nanoseconds. The resulting rapid energy release and material expansion imparts loads on target-facing structures. These loads are broadly the same as other inertial confinement fusion facilities and can be categorized as the hydrodynamic blast, the radiative x-ray emissions and resulting ablative shocks, neutron damage, and any debris or shrapnel~\cite{masters2016debris, masters2014evaluation, yager2021optical}. We are using a combination of large-scale, late-time radiation magnetohydrodynamic simulations and high strain rate, shock EOS, structural dynamics simulations to predict the loads on the structure and analyze the short term effects.  We are extending this out to implicit dynamics and fatigue simulations to determine the long-term effects on the structure. The flexibility of our approach allows us to selectively steer any debris and/or locally protect any solid materials that could face damage. 

\begin{figure}
    \centering
    \includegraphics[width=2.16in]{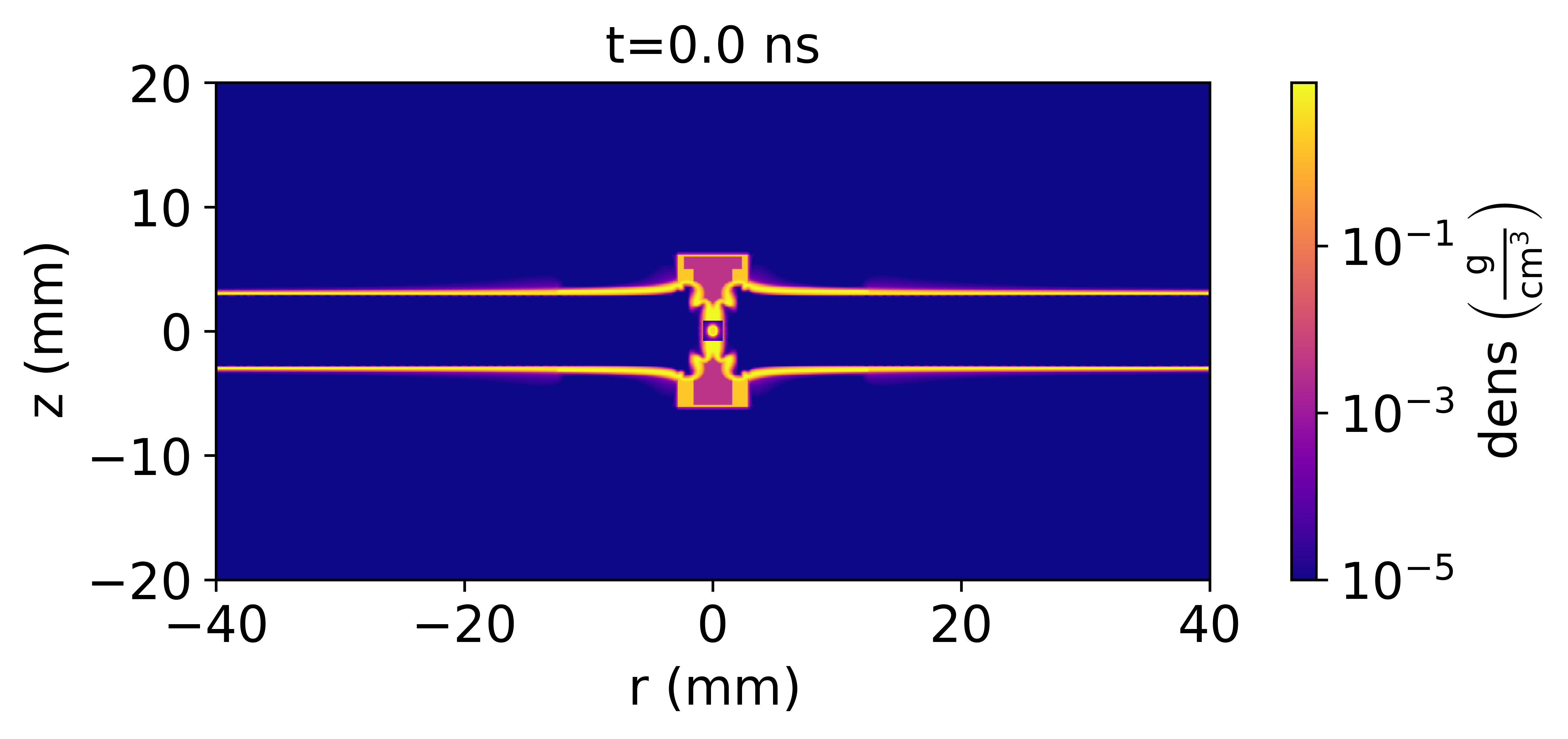}
    \includegraphics[width=2.16in]{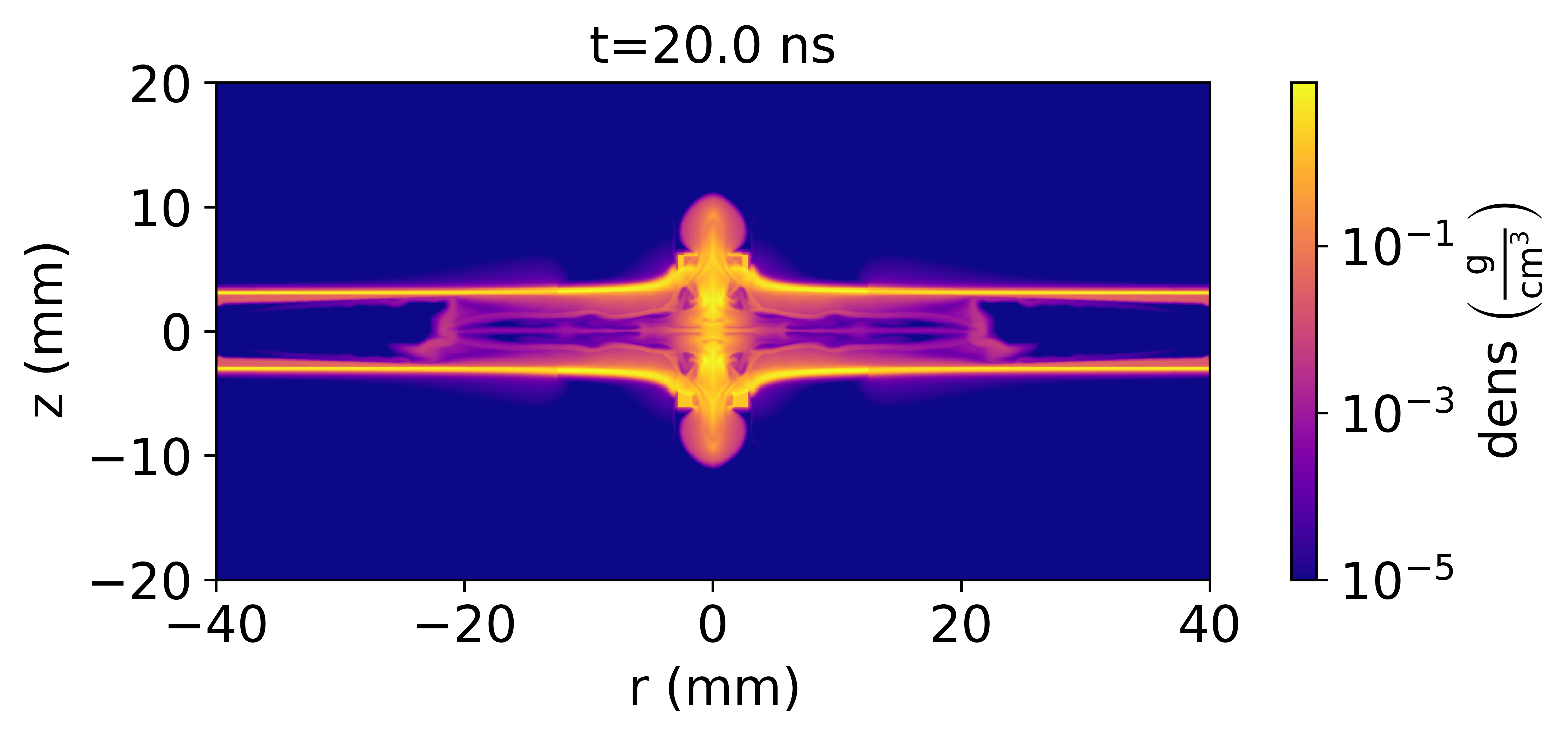}
    \includegraphics[width=2.16in]{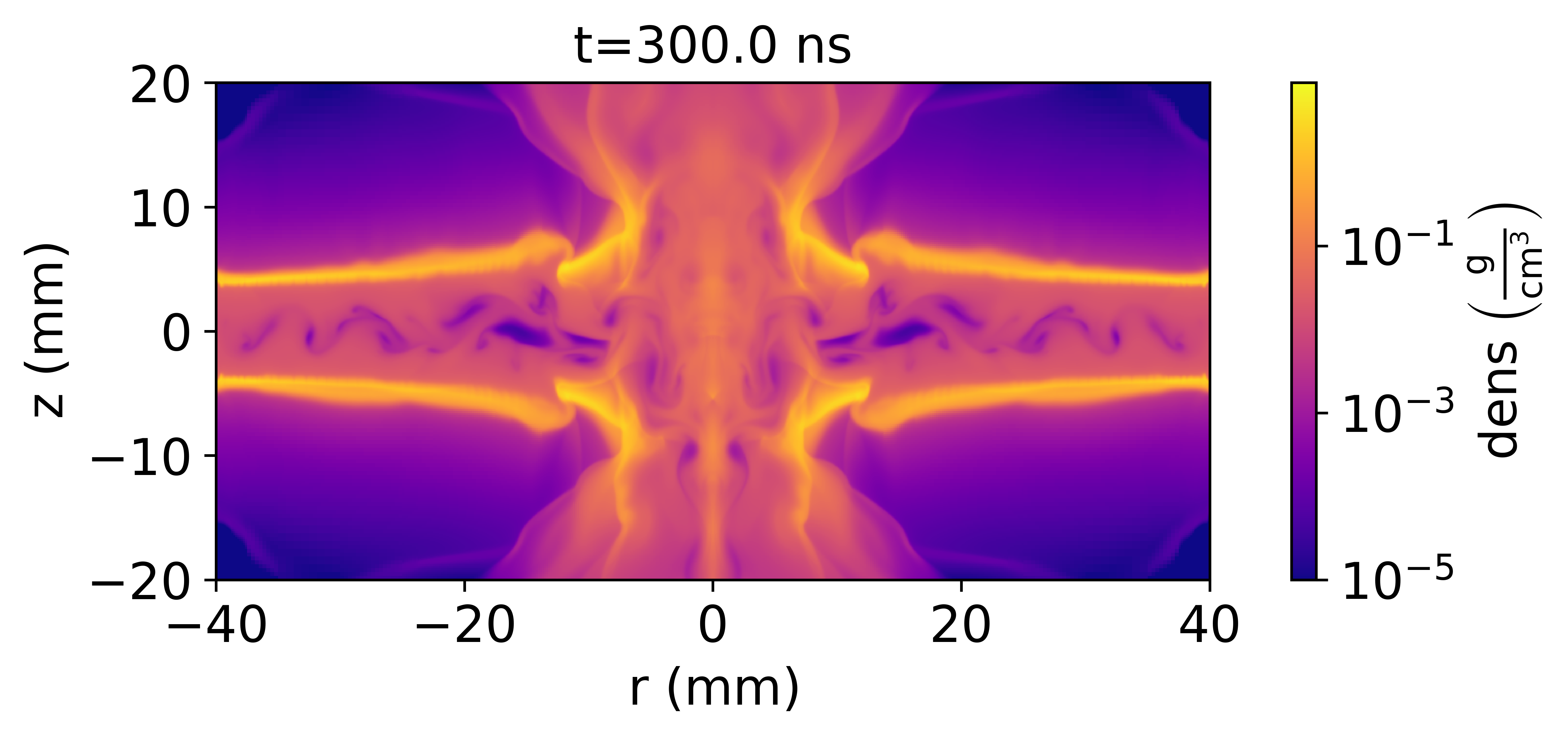}
    \includegraphics[width=2in]{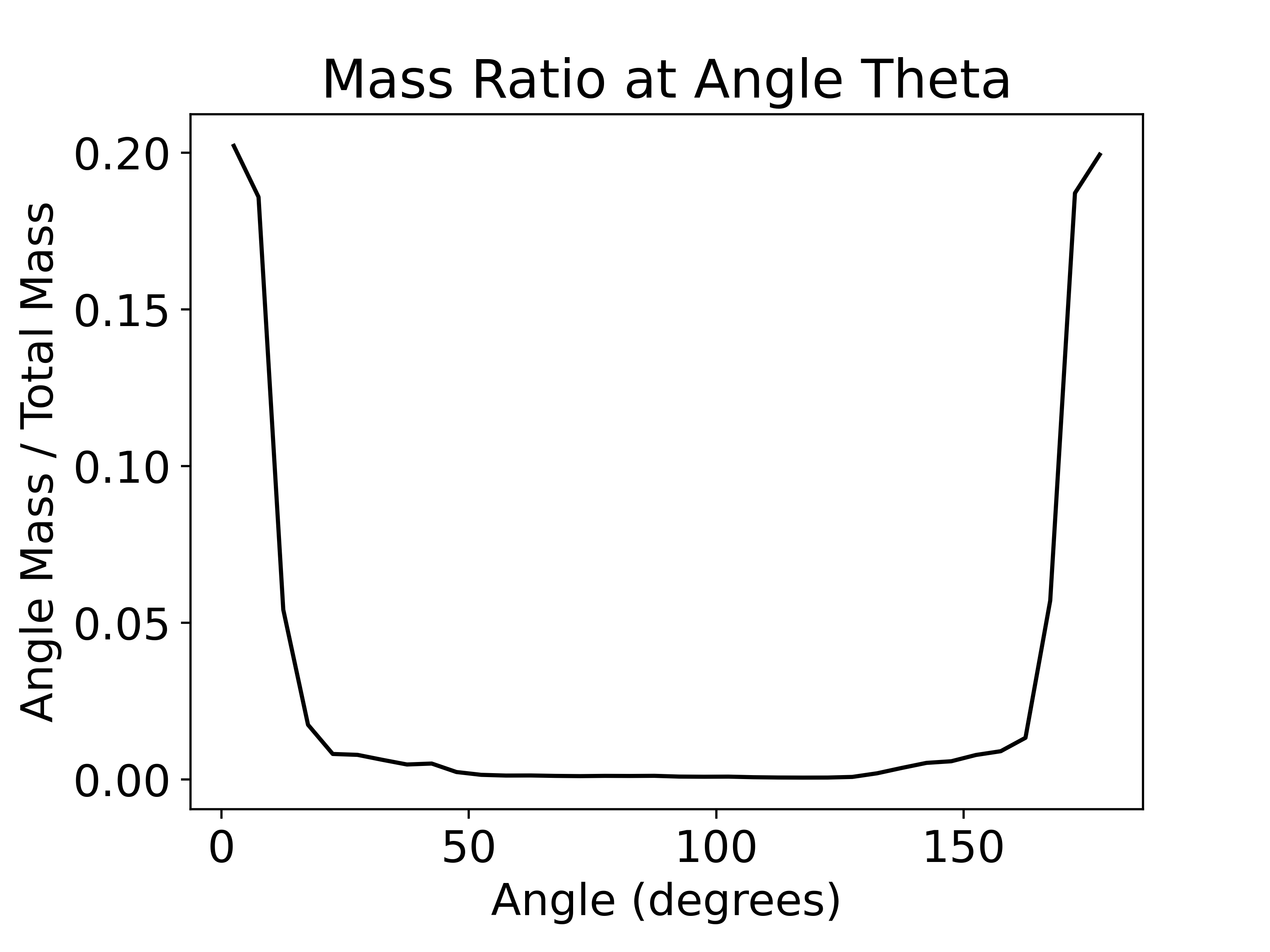}
    \includegraphics[width=1.95in]{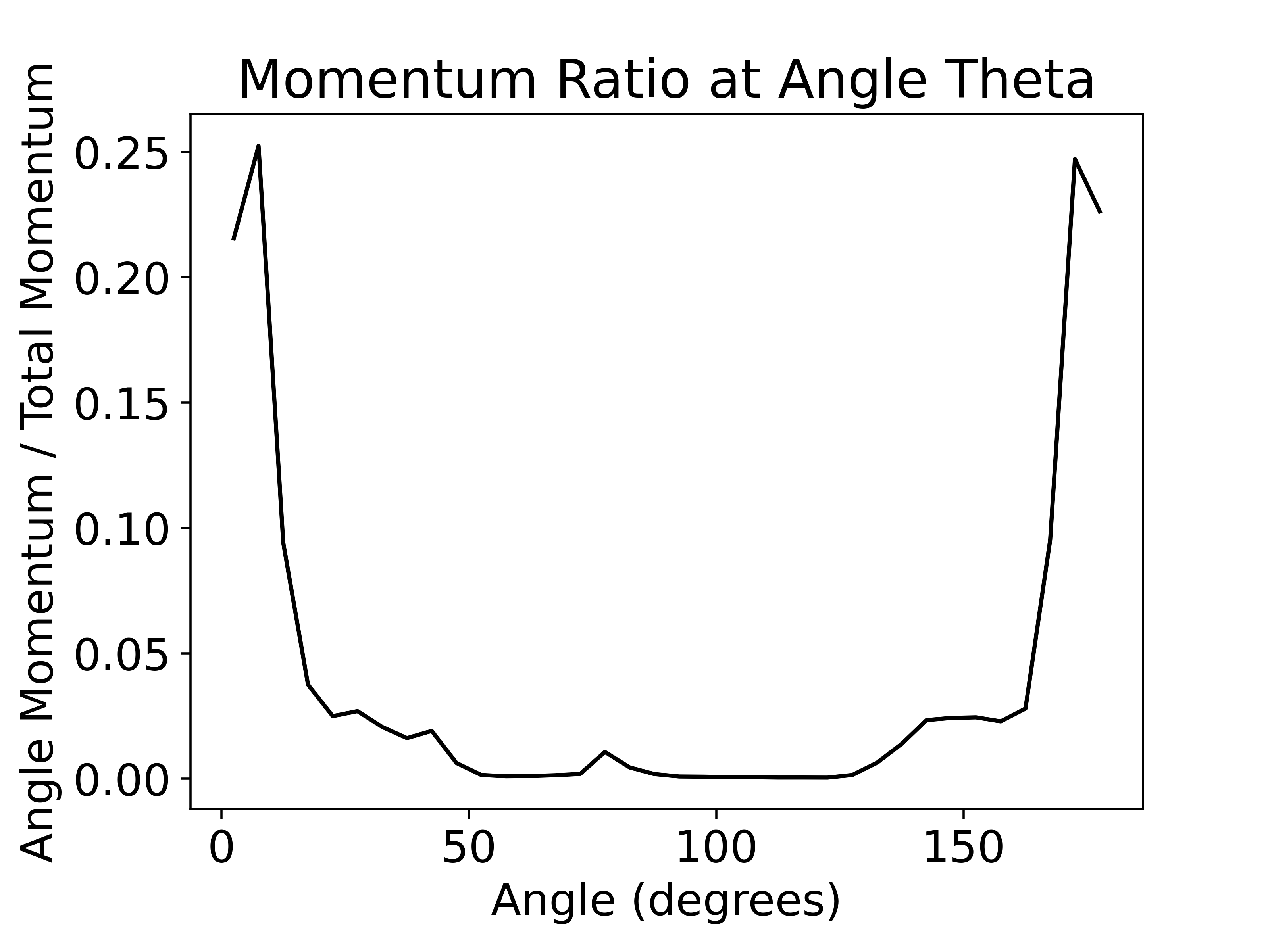}
    \includegraphics[width=2in]{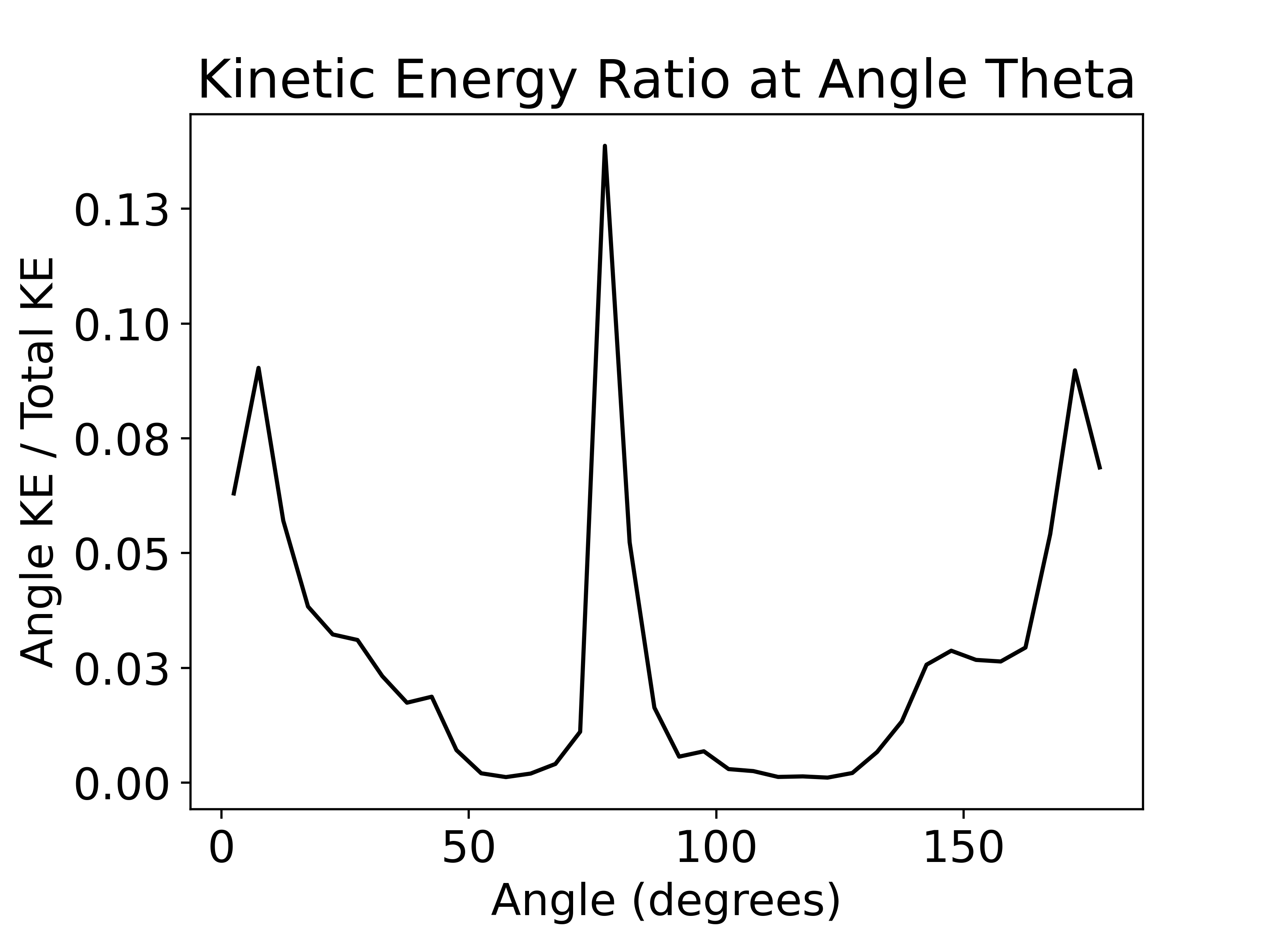}
    \caption{Late-time radiation magnetohydrodynamic simulations with the resulting distribution of mass, momentum, and energy ejected from the explosion. The simulations shown ran to 2.4$\mu$s past the fusion ignition time and tracked material leaving the simulation boundaries. Zero degrees is at the north pole and 180 degrees at the south pole showing 93\% of the mass, 85\% of the momentum, and 55\% of the kinetic energy traveling within 25 degrees of the poles.}
    \label{fig:post_shot}
\end{figure}

For example, magnetic pressure pushes our target liner inward but also pushes the iMITLs axially during the implosion. The residual magnetic field resists the radial expansion of the rebounding liner and the result is that the radial momentum is minimized and the target primarily expands axially, orthogonal to the driving hardware.  The mass, momentum, and kinetic energy distribution for the hydrodynamic blast can be seen in (Fig. \ref{fig:post_shot}). For the few pieces of hardware that are exposed to damaging shocks, we are developing consumable shielding strategies to mitigate the effects.

\begin{figure}[htb]
    \centering
    \includegraphics[width=5in]{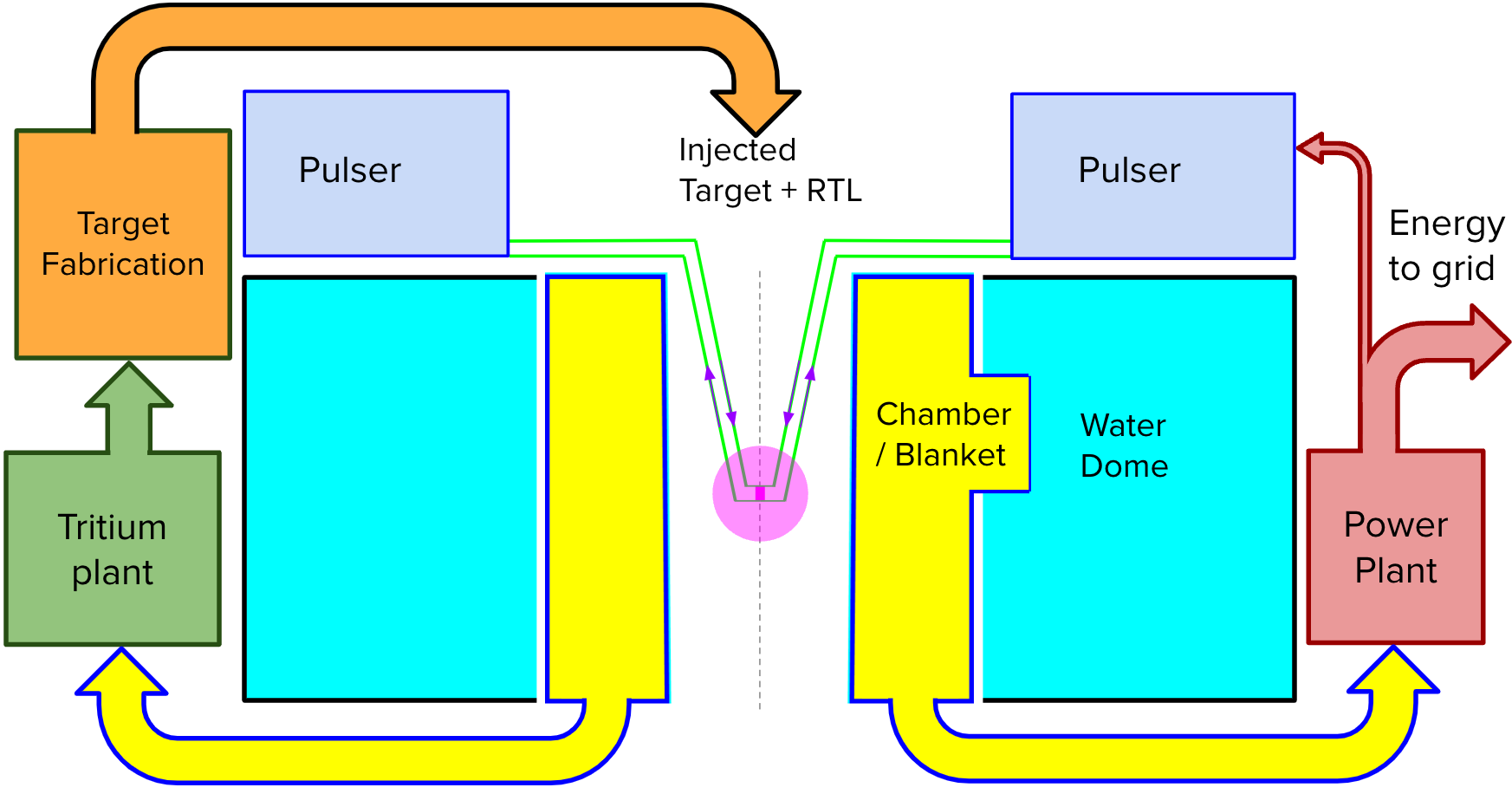}
    \caption{Pulser-driven IFE system schematic, illustrating incorporation of injected targets, recyclable transmission lines, and liquid or solid first wall / breeder blanket.}
    \label{fig:reactor_schematic}
\end{figure}

\subsubsection{Chamber pumping and reloading}
While many challenges for a commercial fusion system are shared among pulsed fusion concepts, considerations that are unique to pulser-driven ICF systems include $\sim 1$ Hz repetitive injection of inner MITL sections in a manner that is compatible with MITL surface condition and vacuum requirements for current delivery, blanket neutronics and tritium breeding characteristics, energy recovery and heat exchange, and radiation shielding of critical system components and the overall facility shielding envelope. For each cycle, the chamber must be prepared for the next shot. This includes inserting a new target, inserting new iMITLs, and re-establishing the vacuum state in between the MITLs. The iMITLs are split between (i) low-mass consumable transmission line (CTL) \cite{slutz2003low} or recyclable transmission line (RTL) \cite{de2003recyclable, sharpe2007recyclable} sections and (ii) permanent MITLs that are shielded by the tritium breeding material and are meant to last for a large number of shots. The target and consumable iMITL insertion can be performed mechanically via target exchange hardware. This can be done in a flexible way given the fact that the energy is transmitted along the gap between the MITLs and doesn't have any significant line-of-sight requirements. The volume between the MITLs must be rapidly pumped down to a high vacuum in preparation for the next shot. Several strategies are being pursued to do so, such as sweeping the chamber with a getter after each shot; more details will be shared in subsequent publications.

\subsection{Tritium breeding}
\begin{figure}[htb]
    \centering
    \includegraphics[width=1.6in]{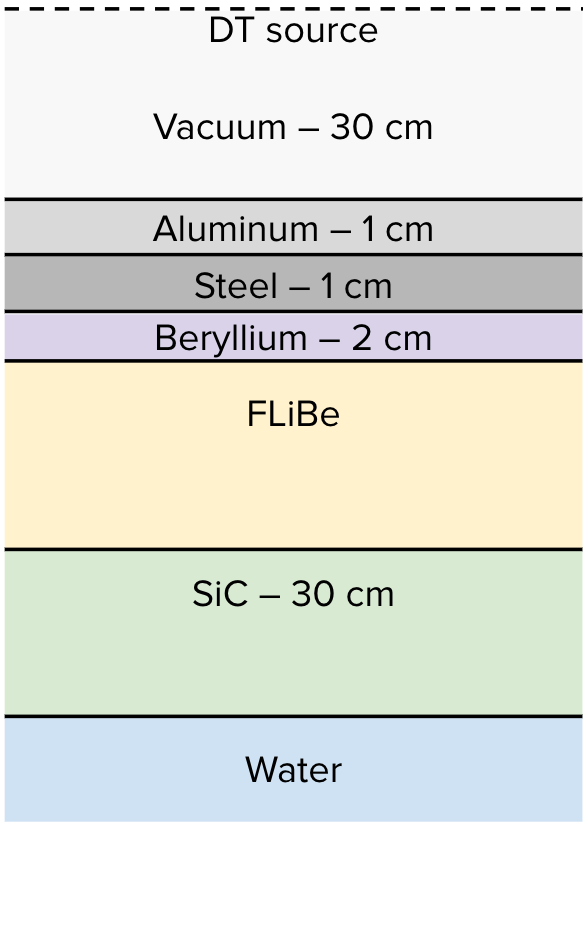}
    \includegraphics[width=4.0in]{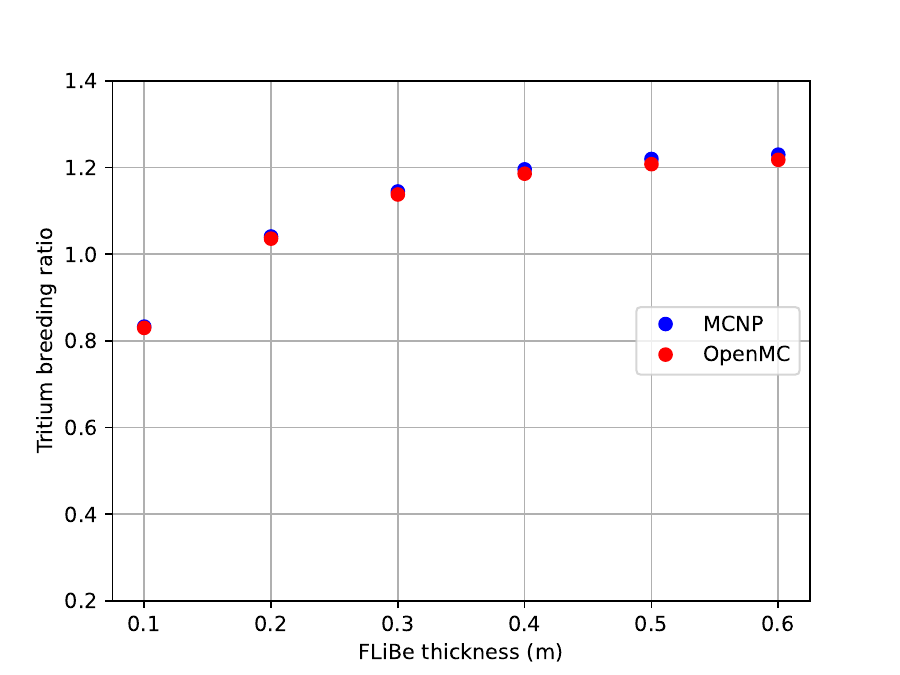}
    \caption{Radiation Monte-Carlo calculation of pulser-IFE-relevant chamber neutronics, blanket thickness, and tritium breeding ratio}
    \label{fig:nuclear_engineering}
\end{figure}

Achievement of a tritium breeding ratio appreciably exceeding unity within a stable target chamber first wall is a necessary condition for a feasible fusion power plant. A pulsed-power-MITL-driven system is compatible with lithium-containing blanket and shielding technologies that meet this requirement via neutron reactions with lithium-6, using configurations similar to those shown in the HYLIFE \cite{blink1985high, moir1991hylife, ogando2024preliminary} and Z-IFE \cite{olson2004z, anderson2006z, meier2004analyses, olson2005development} reactor design studies, (cf. Fig.~\ref{fig:reactor_schematic} and reference~\citenum{anderson2006z}). Neutronics analysis in MCNP for notional 1D target chamber and blanket configurations confirms that sufficient tritium breeding ratio can be achieved with appropriate radial thickness of a FLiBe molten salt blanket, c.f. Fig.~\ref{fig:nuclear_engineering}. Other blanket materials (e.g., PbLi) and configurations are compatible with our approach as well.

In contrast to laser-based IFE concepts --- which require line of sight between the final optics and target -- the pulsed power MITL-based energy coupling is less constraining for blanket configurations. For example, power may be delivered via a stepped conical feed to injected targets in such a way that radiation line of sight from the target is fully occluded. Thus the target can be effectively fully surrounded by blanket material, moderating water, and structural / shielding materials at the time of the repetitive fusion energy release. The surface-conformal energy delivery of the MITLs is also advantageous for approaches to the mitigation of many first wall issues, allowing for the protection of the target chamber first wall from dynamic blast loading\cite{abbott2005dynamic} and prompt x-ray and gamma ray emissions via thick injected liquid blanket material and/or aerosol/gas cushions, and shielding of chamber structural steels to within material DPA limits for the lifetime of the system.

More details of a pulser IFE power plant, building upon the schematic in Fig~\ref{fig:reactor_schematic}, will be discussed in future publications.

\section{Utility to the broader fusion community}\label{ssec:Utility}

In aggregate, the high scientific maturity, high efficiency, and relatively low cost of pulser ICF makes Pacific Fusion's DS a valuable complement to the other flagship fusion facilities in the US's strategic path to commercializing fusion energy. The National Academy of Sciences report \textit{Bringing Fusion to the U.S. Grid} recommended (i) pursuing a Fusion Pilot Plant in the 2035-2040 time frame and (ii) investing in concept-agnostic fusion technology developments \cite{NationalAcademySci_Fusion_2021}. The DS contributes significantly to both of these objectives. The schedule for delivering the DS and achieving facility gain would allow pilot-plant scale facilities to be developed well before this targeted time frame. 

As discussed in this paper, we believe pulser IFE has substantial prospects for innovations and cost reductions in delivering fusion power. In terms of technology developments, there are many community-wide benefits to pursuing these objectives, including improved pulsed power technology, the development of high flux neutron sources, improved plasma diagnostics, improved simulation capabilities, and improved understanding of burning plasma dynamics in MHD regimes. 

The DS also advances the feasibility of high-yield fusion for national defense. High-yield (>100 MJ) inertial fusion is a critical national defense tool because it is the only way to create the extreme conditions essential for ensuring the effectiveness and survivability of key defense capabilities. Although the National Ignition Facility’s recent achievement of $\sim5$ MJ was a major milestone, it falls well short of the yields needed to replicate threat-level environments and obtain the data required for comprehensive validation of these systems. Meanwhile, adversaries are racing to achieve 100+ MJ fusion by 2030, potentially two decades before the current U.S. goal of 2050 \cite{NNSA24Enterprise}. Pulser-driven inertial fusion offers the United States a timely opportunity to maintain its leadership in advanced defense technologies.

The DS delivers a low-cost capability that can couple $\sim10$ MJ to the fusion target to generate many tens or hundreds of MJ of fusion energy in adjacent parameter regimes to those accessed by laser-driven ICF. High-gain fusion platforms fielded at the DS can therefore advance the NNSA's mission, and collaborations on plasma diagnostics, target design, simulations, and fundamental plasma physics can benefit all participants. 

One effective way to advance such efforts is through the public-private partnership model. Public-private partnerships are becoming increasingly important to the US's fusion strategy, and were highlighted as a key recommendation in the National Academy of Sciences report \cite{NationalAcademySci_Fusion_2021}. The US DOE has launched several funding programs focused on private companies, including the INFUSE network and the Milestone-Based Fusion Development Program. It has also published formal Requests for Information on \textit{Cost-Sharing Partnerships with the Private Sector in Fusion Energy} and, recently, on \textit{Pulsed Power Public/Private Partnerships}. Pacific Fusion has already procured significant private funding for building its DS, creating an opportunity for US DOE to leverage these investments by partnering to further advance the broader fusion community's objectives.

\section{Conclusion}\label{sec:Conclusion}

Pulser-driven inertial fusion can enable affordable, abundant commercial fusion power and also close the gap to high-yield (>100 MJ) capabilities. By directly delivering tens of MJ of electrical energy to magnetized fusion targets at high efficiency, pulser ICF offers a near-term path to achieving $Q_f>1$ — an achievement that would mark the decisive turning point in fusion energy development. Pacific Fusion's DS, capable of exceeding 60 MA, shows how new pulsed-power technologies can be scaled and optimized for affordable, manageable, practical, and scalable (AMPS) fusion. Critically, it represents a $1000\times$ increase in fusion capabilities by producing $100\times$ the facility-level gain at about 1/10 the cost of the National Ignition Facility, accelerating progress toward both commercial power and high-yield fusion readiness.

\section{Acknowledgments}\label{sec:Acknowledgments}
The authors would like to thank the following individuals for their contributions: Andrew Benson, Kurt Boehm, Jonathan Carroll-Nellenback, Philip Datte, Michael Farrell, Edward C. Hansen, Amber Hennessy, Yannick de Jong, Ian Scott, Petros Tzeferacos, Daan van Vugt, and Dale Welch.

The following article has been submitted to Phys. Plasmas. After it is published, it will be found at: \\
https://publishing.aip.org/resources/librarians/products/journals/  \\ \\

\appendix

\textbf{\Large{Appendix}}

\section{An ignition metric for pulser ICF}\label{sec:ignitionmetric}

In this section, we illustrate the core concepts behind our generalized Lawson criterion (GLC') suitable for assessing the proximity of pulser-driven ICF concepts to ignition, cf. Eq.~\ref{eq:glcmod}. Beginning with the hot spot energy balance expression given by Eq.~\ref{eq:DT_powbal}, the power terms on the right hand side scale with the compressed fusion fuel conditions as:\cite{SchmitRuiz2020}
\begin{align}
    ({\rm alpha~deposition})\to \hat{W}_\alpha & \propto \eta_\alpha (\rho R)^2 s(T), \label{eq:p_a_repeat} \\
    ({\rm compression/expansion})\to \hat{W}_{PdV} & \propto (\rho R) T U_{\rm liner}, \label{eq:p_pdv} \\
    ({\rm bremsstrahlung~radiation})\to \hat{W}_r & \propto (\rho R)^2 T^{1/2}, \label{eq:p_r} \\
    ({\rm electron~conduction})\to \hat{W}_{ce} & \propto T^{7/2} g_e(x_e), \label{eq:p_ce} \\
    ({\rm ion~conduction})\to \hat{W}_{ci} & \propto T^{7/2} g_i(x_i), \label{eq:p_ci} \\
    ({\rm end~losses})\to \hat{W}_{\rm end} & \propto T^{3/2}\rho R^2/h \propto T^{3/2}(\rho R)(R/h). \label{eq:p_end}
\end{align}
The energies and powers in the above equations are expressed per unit length of the cylindrical fuel assembly, denoted by the hat notation. The hot spot mass density is $\rho$, the liner velocity is $U_{\rm liner}$ (where negative values correspond to fuel compression), and the radius and length of the DT cylinder are $R$ and $h$, respectively. We assume electrons and ions are in thermal equilibrium ($T_e=T_i\doteq T$).

The degree of magnetic inhibition of thermal conduction from the hot fuel to the cooler liner wall is captured by the functions $g_{e,i}(x_{e,i})$, where the electron and ion \textit{magnetization} are represented by the Hall parameters $x_{e,i}\propto T^{3/2}B/\rho$. In the strongly magnetized limit, $g(x)\propto 1/x^2$. When $B\to 0$, $g\to 1$. In the intermediate regime where a species is marginally magnetized, i.e., $x\sim\mathcal{O}(1)$, the transport regime is ``Bohm-like'' and $g\propto 1/x$.  In MagLIF, where the magnetic field entrained in the fuel is sufficient to strongly suppress electron heat conduction but weak enough that the plasma thermal pressure exceeds the local magnetic pressure, the ions are often in this Bohm-like regime.

\begin{figure}
    \centering
    \includegraphics[width=0.8\textwidth]{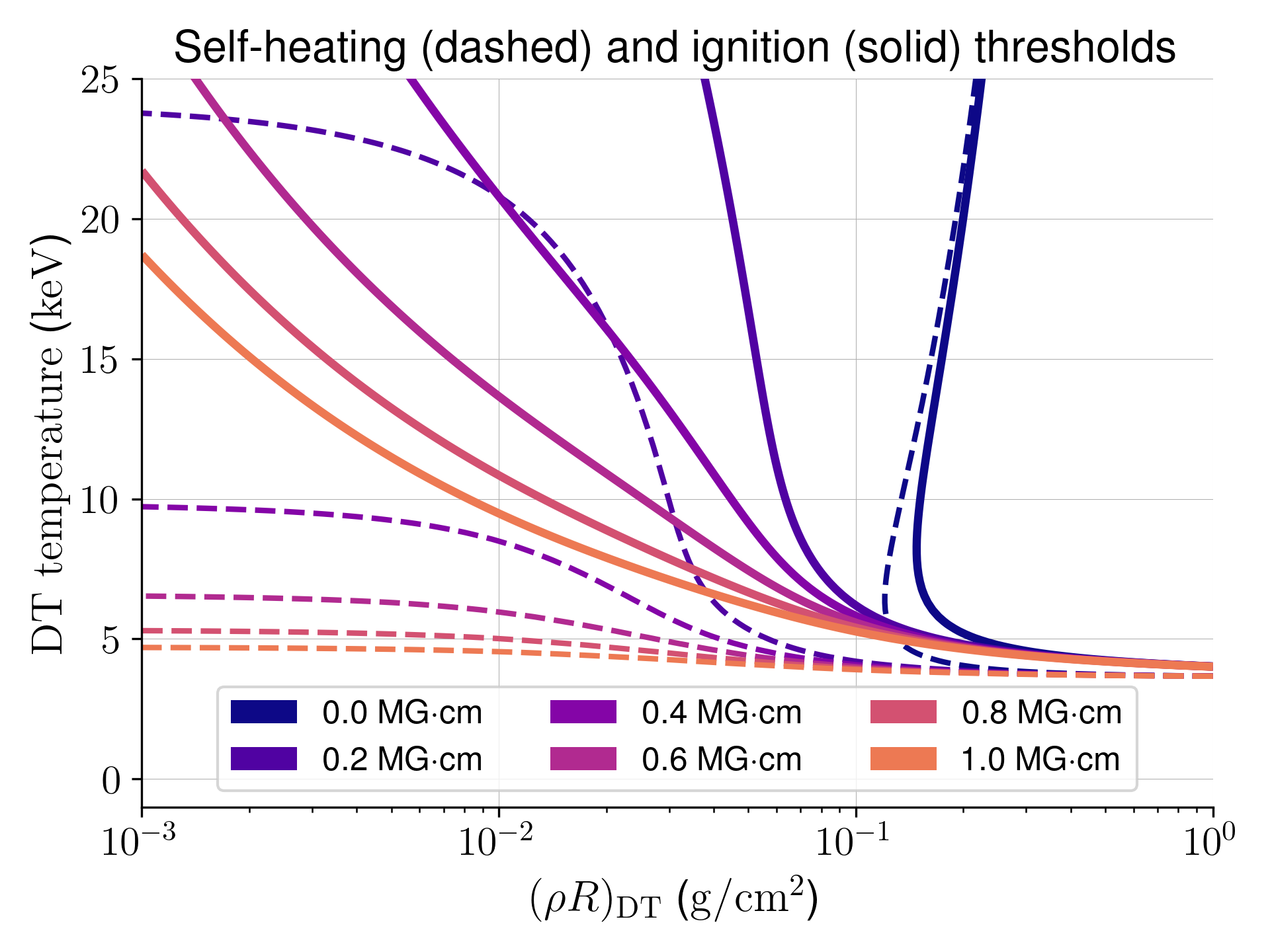}
    \caption{Self-heating threshold contours (dashed lines) for magnetized DT plasma spanning magnetic-field-radius products $BR$ from 0 (darkest) to 1.0 MG$\cdot$cm (lightest) in 0.2 MG$\cdot$cm increments. All plasma configurations with $(\rho R)_{\rm DT}$ and $T_{\rm DT}$ residing above and to the right of the respective dashed contour undergo instantaneous positive DT heating in the absence of plasma compression or expansion, i.e., at stagnation. Solid lines are contours of a generalized Lawson ignition threshold criterion ${\rm GLC}'=1$ (Eq.~\ref{eq:glcmod}) spanning the same range of $BR$ values, assuming the DT is tamped by a liner with areal density $(\rho\delta R)_l=3~{\rm g/cm^2}$, conditions accessible to the DS. Note that while the Lawson threshold to ignite is more stringent than the threshold for instantaneous self-heating, alpha heating can help the assembly cross over the ignition threshold.}
    \label{fig:selfheating}
\end{figure}

Figure~\ref{fig:selfheating} illustrates the expanded space of self-heating conditions afforded by pulser ICF when the DT fuel is magnetized.  Here, we plot contours of neutral power balance within the DT fuel $\hat{W}_\alpha=\hat{W}_{ce}+\hat{W}_{ci}+\hat{W}_r$ at peak compression (stagnation), where we make the common simplifying assumptions that $\hat{W}_{PdV}$ and $\hat{W}_{\rm end}$ are momentarily ignorable. We assume a fuel radius of 100~$\mu$m, the approximate scale at peak compression of simulated MagLIF-like implosions matched to the DS. Each dashed contour corresponds to a different amount of axial magnetic flux entrained in the fuel, expressed in terms of the magnetic-field-radius product, $BR$, which is proportional to the ratio of the fuel radius to the 3.5~MeV alpha-particle gyroradius and thus quantifies the degree of confinement of the alpha particles. The value $BR \approx 0.5~{\rm MG\cdot cm}$ corresponds to the case where the alpha gyroradius equals the fuel radius and alphas are marginally confined within the DT column \cite{KnappSchmit2015}. Combinations of $T$ and $(\rho R)_{\rm DT}$ above and to the right of the solid contours are capable of self-heating. The dashed lines in Fig.~\ref{fig:selfheating} demonstrate that increasing $BR$ enables self-heating at decreasingly small fuel areal densities, with the impact saturating for $BR>1~{\rm MG\cdot cm}$.

In addition to achieving net fusion-driven self-heating, an igniting target must also remain confined long enough for the cumulative effect of self-heating to significantly increase the internal energy of the DT fuel. This ignition requirement is captured in the generalized Lawson criterion, or GLC. We adopt the version used in Ref.~\cite{Lindl2015IgnitionMetrics}, Eq.~(9):
\begin{equation}
    {\rm GLC}\doteq \frac{\hat{E}_\alpha/\tau_{TN}}{\hat{E}_{\rm DT}/\tau_E}
        = \left(\frac{\hat{E}_\alpha}{\hat{E}_{\rm DT}}\right)f
        = \left(\frac{\hat{Y}_{\rm DT}/5}{\hat{E}_{\rm DT}}\right)f
        = \left(\frac{\hat{Y}_{\rm DT}}{\hat{M}_{\rm DT}}\right)\left(\frac{m_p}{6k_B T_{\rm DT}}\right)f
    \label{eq:glc}
\end{equation}
Here, $\hat{E}_\alpha$ is the alpha particle energy generated by fusion, which is 1/5 the total DT yield, $\hat{Y}_{\rm DT}$. (This expression assumes 100\% of the alpha-particle energy is deposited in the hot spot, though we revise this assumption in our modified GLC below.) The characteristic burn duration and energy confinement times are given by $\tau_{TN}$ and $\tau_E$, respectively. Using the value $f=1.2$ for the ratio of the two confinement times reproduces the ignition threshold behavior seen in simulations of NIF-like ICF implosions when ${\rm GLC}\approx 1.0$~\cite{Lindl2015IgnitionMetrics}.

We find from the GLC given by Eq.~\ref{eq:glc} that ignition in the lower-$(\rho R)_{\rm DT}$ pulser ICF design space requires a large areal density of the confining liner, $(\rho\delta R)_l$. Following Ref.~\cite{Betti2010IgnitionMetrics}, we represent the liner as a thin cylindrical shell of linear mass density $\hat{M}_l$, whose interaction with the hot DT fuel can be approximated by the equation of motion,
\begin{equation}
    \hat{M}_l\ddot{R}_l=2\pi R_l P_{\rm DT}.
    \label{eq:liner_eom}
\end{equation}
In Eq.~\ref{eq:liner_eom}, the external drive pressure is ignored relative to the much larger fuel pressure $P_{\rm DT}$ near stagnation. By dimensional analysis, the hydrodynamic confinement time $\tau_E$ is well approximated by $\tau_E\approx (\hat{M}_l/2\pi P_{\rm DT})^{1/2}\approx [(\rho R\delta R)_l/P_{\rm DT}]^{1/2}$, where the liner density $\rho_l$, radius $R_l$ (assumed equal to the fuel outer radius $R_{\rm DT}=R$), and thickness $\delta R_l$ are evaluated at stagnation.  From Eqs.~\ref{eq:p_a} and \ref{eq:glc}, we find $\hat{Y}_{\rm DT}\approx \hat{W}_\alpha \tau_{TN} = \hat{W}_\alpha \tau_E/f$, while the DT linear mass density $\hat{M}_{\rm DT}=\pi R^2 \rho_{\rm DT}$. Inserting these relationships into Eq.~\ref{eq:glc} leads to an important scaling relation applicable to all inertial fusion approaches:
\begin{equation}
    {\rm GLC}\propto \left[\frac{s(T)}{T^{3/2}}\right]\left[ (\rho R)_{\rm DT}(\rho\delta R)_l\right]^{1/2}
    \label{eq:glc_scale}
\end{equation}
In Eq.~\ref{eq:glc_scale}, the GLC is proportional to the harmonic mean of the fuel (hot-spot) and liner areal densities; thus, to ignite at low $(\rho R)_{\rm DT}$, the liner areal density $(\rho\delta R)_l$ must be increased accordingly. 
Fortunately, pulser ICF is ideally suited to compressing liners to large values of $(\rho\delta R)_l$.  MagLIF experiments on the Z facility have already demonstrated $(\rho\delta R)_l\approx 1~{\rm g/cm^2}$\cite{Knapp_2022}.  The liner areal density increases approximately linearly with peak current,\cite{SchmitRuiz2020} so scaling up present-day MagLIF experiments from $I_{max} \approx 20$~MA to $I_{max} \approx 60$~MA should triple $(\rho\delta R)_l$, putting such designs in a favorable ignition space.  

In Figs.~\ref{fig:glc_nif_z_ds} and \ref{fig:selfheating}, we plot a variation of the GLC representing the \textit{net} heating of the DT fuel by accounting for many of the sources and sinks described by Eq.~\ref{eq:DT_powbal}.  Specifically, we replace $\hat{E}_\alpha\doteq\hat{W}_\alpha \tau_{TN}$ with $\hat{E}_{\rm net}\doteq (\hat{W}_\alpha - \hat{W}_{ci} - \hat{W}_{ce} - \hat{W}_r)\tau_{TN}$, such that
\begin{equation}
    {\rm GLC}'\doteq \left(\frac{(\hat{W}_\alpha - \hat{W}_{ci} - \hat{W}_{ce} - \hat{W}_r)\tau_{TN}}{\hat{E}_{\rm DT}}\right)f.
    \label{eq:glcmod_repeat}
\end{equation}
This expression combines the self-heating and confinement requirements for ignition and can only be positive in regions where instantaneous net self-heating occurs at stagnation.  We plot contours representing the ignition threshold criterion ${\rm GLC}'=1$ as solid lines in Figs.~\ref{fig:glc_nif_z_ds} and \ref{fig:selfheating} for the specific case $(\rho\delta R)_l=3~{\rm g/cm^2}$, demonstrating the large design space opened up by fuel magnetization. Note that for boundary cases above the dashed curves but below the solid ${\rm GLC}'\approx 1$ curves in Fig.~\ref{fig:selfheating}, self-heating can push the fuel above the ignition boundary.

\section{Pulser-target coupling physics}
\label{sec:pulser-target_coupling}
Accurate modeling of the evolution of plasma within the inner MITL and load region requires inclusion of physics beyond that usually included in radiation MHD models (PIC or Hall physics as just discussed); however, some simple MHD concepts are often useful in understanding how the evolved plasma influences current delivery to or within the target. The MHD-like picture can capture gross features of the plasma flow when the particle dynamics are dominated by $E \times B$ drift. This applies when both ions and electrons are strongly magnetized, i.e, gyro-radii for both species are small compared to the AK gap and the $E \times B$ drift dominates over polarization drift (when the gyro-period is short compared to the rate of change of the flow) as well as gradient B and curvature drifts (again in the limit of small gyro-orbits). These conditions typically apply within the inner MITL and target region during the main current rise but are violated early in time.  

If the conductivity of the plasma is sufficient (above a few eV in electron temperature) then the MHD concept of ``frozen flux'' is useful: the concept that magnetic field and plasma are approximately ``tied'' together. In a system with a purely azimuthal field, ignoring the comparatively weak axial field that might be used to magnetize a MagLIF target, the magnetic field and plasma can be treated as a collection of  magnetic flux loops of infinitesimal area $\delta$A, with magnetic field strength B, and magnetic flux B$\delta$A. The plasma mass entrained within the loop is $2\pi R \rho \delta A$, where R is the cylindrical radius. If the magnetic flux and mass are tied together, then the flux/mass of the loop is proportional to $B/(\rho R)$, which is a constant of the motion for the MHD-like flow. Since the magnetic field, ignoring small current losses, varies as $R^{-1}$, this indicates that the inward (target-directed) plasma flow will tend to concentrate the density as ~ $R^{-2}$. The simple flux loop picture offers another insight: If the plasma in the inner MITL is accelerated toward the target due to its increasing inductance and associated electric field increase, then the flux loops must increase their kinetic energy. It might be expected that the change in kinetic energy of the loops would be limited by the available magnetic energy in the loops, i.e. $M v^2 / 2 < B^2/2\mu_0 \cdot V$, where $v$ is the $E \times B$ velocity and V is the volume of the loop = $2\pi R \delta A$. This condition is equivalent to $v < v_\textrm{Alfven}$, where $v_\textrm{Alfven} = \frac{B}{\sqrt{\mu_0 \rho}}$ is the classic Alfv\'en velocity for wave propagation in magnetized plasmas. Since $B$ varies as $R^{-1}$ and $\rho$ varies as $R^{-2}$ there is a tendency for the power flow region to have a characteristic (time-dependent) Alfv\'en velocity. In practice, the presence of plasma where the increase in flow velocity is more than a few percent of the Alfv\'en velocity leads to noticeable magnetic energy loss, which manifests as current loss to the load. 

An idealized version of load and plasma evolution helps to understand this effect. If we imagine an initial state with a static cylindrical liner and a plasma-filled MITL with density varying as $R^{-2}$ and magnetic field varying as $R^{-1}$, i.e., a constant Alfv\'en velocity, then once the liner begins to move, a rarefaction wave is launched outward into the MITL plasma. In a low beta (beta = plasma pressure/ magnetic pressure) magnetized plasma, the speed of the magneto-acoustic rarefaction is $v_\textrm{Alfven}$. The magnetic field and magnetic pressure at the surface of the liner are less than the vacuum values due to the rarefaction, implying less than 100 percent current delivery to the load.  On the Z accelerator, the magnitude of these effects depends on the nature of the load and the condition of electrode surfaces that contribute plasma into the MITL. Low velocity loads such as MagLIF may have less sensitivity than high velocity like gas puffs or wire arrays, although high inductance loads can have a high MITL electric field and $E \times B$ velocity since $dL/dt$ is proportional to both the liner velocity and target height. In any case, attention to vacuum conditioning and electrode contaminants that contribute to MITL plasma are important considerations, although these effects can be captured in the PIC models of the MITL. The effect of plasma on the target itself is usually included in radiation MHD target simulations as a boundary condition or density floor. The plasma can cause current redistribution within the target volume as well as the MITL. At the target scale, the MHD model is usually a better approximation than in the MITL.  The density floors in MHD simulations of Z experiments have typically been chosen empirically, although they are often in the range predicted by PIC. Self-consistent linking of PIC modeling of plasma evolution to radiation MHD models is a goal for improved fidelity modeling.

\bibliographystyle{unsrt}
\bibliography{main}

\begin{thebibliography}{100}

\bibitem{energy_2023}
Statistical review of world energy 2023 72nd edition.
\newblock Technical report, Energy Institute, 2023.

\bibitem{ARC_2015}
B.N. Sorbom, J.~Ball, T.R. Palmer, F.J. Mangiarotti, J.M. Sierchio, P.~Bonoli, C.~Kasten, D.A. Sutherland, H.S. Barnard, C.B. Haakonsen, J.~Goh, C.~Sung, and D.G. Whyte.
\newblock Arc: A compact, high-field, fusion nuclear science facility and demonstration power plant with demountable magnets.
\newblock {\em Fusion Engineering and Design}, 100:378--405, 2015.

\bibitem{NIFLawson_2022}
H.~Abu-Shawareb et~al.
\newblock Lawson criterion for ignition exceeded in an inertial fusion experiment.
\newblock {\em Phys. Rev. Lett.}, 129:075001, Aug 2022.

\bibitem{NIFTargetGain_2024}
H.~Abu-Shawareb et~al.
\newblock Achievement of target gain larger than unity in an inertial fusion experiment.
\newblock {\em Phys. Rev. Lett.}, 132:065102, Feb 2024.

\bibitem{Lawson_1957}
J~D Lawson.
\newblock Some criteria for a power producing thermonuclear reactor.
\newblock {\em Proceedings of the Physical Society (London). Section B}, 70(1):6, Jan 1957.

\bibitem{SfA}
Science for America.
\newblock {\em New Opportunities in Fusion Power}.
\newblock 2023.

\bibitem{JASON}
JASON Report.
\newblock {\em Prospects for Low Cost Fusion Development}.
\newblock 2018.

\bibitem{lindemuth2009fundamental}
Irvin~R Lindemuth and Richard~E Siemon.
\newblock The fundamental parameter space of controlled thermonuclear fusion.
\newblock {\em American Journal of Physics}, 77(5):407--416, 2009.

\bibitem{jet1992fusion}
Jet Team et~al.
\newblock Fusion energy production from a deuterium-tritium plasma in the jet tokamak.
\newblock {\em Nuclear Fusion}, 32(2):187, 1992.

\bibitem{strachan1994fusion}
JD~Strachan, H~Adler, P~Alling, C~Ancher, H~Anderson, JL~Anderson, D~Ashcroft, Cris~W Barnes, G~Barnes, S~Batha, et~al.
\newblock Fusion power production from {TFTR} plasmas fueled with deuterium and tritium.
\newblock {\em Physical Review Letters}, 72(22):3526, 1994.

\bibitem{holtkamp2007overview}
Norbert Holtkamp, ITER~Project Team, et~al.
\newblock An overview of the iter project.
\newblock {\em Fusion Engineering and Design}, 82(5-14):427--434, 2007.

\bibitem{miller2004national}
George~H Miller, Edward~I Moses, and Craig~R Wuest.
\newblock The national ignition facility.
\newblock {\em Optical Engineering}, 43(12):2841--2853, 2004.

\bibitem{boehly1997initial}
TR~Boehly, DL~Brown, RS~Craxton, RL~Keck, JP~Knauer, JH~Kelly, TJ~Kessler, SA~Kumpan, SJ~Loucks, SA~Letzring, et~al.
\newblock Initial performance results of the {OMEGA} laser system.
\newblock {\em Optics Communications}, 133(1-6):495--506, 1997.

\bibitem{fleurot2005laser}
Noel Fleurot, Claude Cavailler, and JL~Bourgade.
\newblock The {L}aser {M}egajoule ({LMJ}) {P}roject dedicated to inertial confinement fusion: Development and construction status.
\newblock {\em Fusion Engineering and Design}, 74(1-4):147--154, 2005.

\bibitem{pmfe_whitepaper_2024}
C.L. Ellison, V.~Garcia, M.~Gomez, G.~P. Grim, J.~H. Hammer, C.~A. Jennings, P.~Knapp1, K.~R. LeChien, N.~B. Meezan, R.~Peterson, A.~C. Reyes, A.~Steiner, W.~A. Stygar, P.~Tzeferacos, D.~Welch, and A.~B. Zylstra.
\newblock Opportunities in pulsed magnetic fusion energy.
\newblock {\em In Preparation}, 2024.

\bibitem{wurzel2021}
Samuel~E. Wurzel and Scott~C. Hsu.
\newblock {Progress toward fusion energy breakeven and gain as measured against the Lawson criterion}.
\newblock {\em Physics of Plasmas}, 29(6):062103, 06 2022.

\bibitem{sinars2020review}
DB~Sinars et~al.
\newblock {Review of pulsed power-driven high energy density physics research on Z at Sandia}.
\newblock {\em Phys. Plasmas}, 27(7), 2020.

\bibitem{slutz2010pulsed}
SA~Slutz, MC~Herrmann, RA~Vesey, AB~Sefkow, DB~Sinars, DC~Rovang, KJ~Peterson, and ME~Cuneo.
\newblock Pulsed-power-driven cylindrical liner implosions of laser preheated fuel magnetized with an axial field.
\newblock {\em Physics of Plasmas}, 17(5), 2010.

\bibitem{Slutz2012HighGain}
S.~A. Slutz and R.~A. Vesey.
\newblock High-gain magnetized inertial fusion.
\newblock {\em Phys. Rev. Lett.}, 108:025003, Jan 2012.

\bibitem{Knapp_2022}
P.~F. Knapp, M.~E. Glinsky, M.~A. Schaeuble, C.~A. Jennings, M.~Evans, J.~Gunning, T.~J. Awe, G.~A. Chandler, M.~Geissel, M.~R. Gomez, K.~D. Hahn, S.~B. Hansen, E.~C. Harding, A.~J. Harvey-Thompson, S.~Humane, B.~T. Klein, M.~Mangan, T.~Nagayama, A.~J. Porwitzky, D.~E. Ruiz, P.~F. Schmit, S.~A. Slutz, I.~C. Smith, M.~R. Weis, D.~A. Yager-Elorriaga, D.~J. Ampleford, K.~Beckwith, T.~R. Mattsson, K.~J. Peterson, and D.~B. Sinars.
\newblock {Estimation of stagnation performance metrics in magnetized liner inertial fusion experiments using Bayesian data assimilation}.
\newblock {\em Physics of Plasmas}, 29(5):052711, 05 2022.

\bibitem{lewis2024developments}
William Lewis, Marc-Andre Schaeuble, Jeffrey Fein, Thomas Awe, Gordon Chandler, Matthias Geissel, Matthew Gomez, Stephanie Hansen, Adam Harvey-Thompson, Hannah Hasson, et~al.
\newblock Developments in assessing the generalized lawson criterion in magnetized liner inertial fusion.
\newblock {\em Bulletin of the American Physical Society}, 2024.

\bibitem{JET_power}
Power supply.
\newblock \url{https://web.archive.org/web/20160105221442/https://www.euro-fusion.org/fusion/jet-tech/jets-flywheels/}.

\bibitem{stygar2017impedance}
WA~Stygar et~al.
\newblock Impedance-matched marx generators.
\newblock {\em Physical Review Accelerators and Beams}, 20(4):040402, 2017.

\bibitem{Christopherson2019}
A.~R. Christopherson, R.~Betti, and J.~D. Lindl.
\newblock Thermonuclear ignition and the onset of propagating burn in inertial fusion implosions.
\newblock {\em Phys. Rev. E}, 99:021201, Feb 2019.

\bibitem{Basko2000}
M.M. Basko, A.J. Kemp, and J.~Meyer ter Vehn.
\newblock Ignition conditions for magnetized target fusion in cylindrical geometry.
\newblock {\em Nuclear Fusion}, 40(1):59, 2000.

\bibitem{Slutz2016Scaling}
S.~A. Slutz et~al.
\newblock {Scaling magnetized liner inertial fusion on Z and future pulsed-power accelerators}.
\newblock {\em Phys. Plasmas}, 23(2):022702, 02 2016.

\bibitem{SchmitRuiz2020}
P.~F. Schmit and D.~E. Ruiz.
\newblock {A conservative approach to scaling magneto-inertial fusion concepts to larger pulsed-power drivers}.
\newblock {\em Physics of Plasmas}, 27(6):062707, 06 2020.

\bibitem{Ruiz2023Iscaling}
D.~E. Ruiz et~al.
\newblock {Exploring the parameter space of MagLIF implosions using similarity scaling. II. Current scaling}.
\newblock {\em Phys. Plasmas}, 30(3):032708, 2023.

\bibitem{Ruiz2023Theory}
D.~E. Ruiz, P.~F. Schmit, D.~A. Yager-Elorriaga, C.~A. Jennings, and K.~Beckwith.
\newblock {Exploring the parameter space of MagLIF implosions using similarity scaling. I. Theoretical framework}.
\newblock {\em Physics of Plasmas}, 30(3):032707, 03 2023.

\bibitem{Lindl2015IgnitionMetrics}
J.~D. Lindl, S.~W. Haan, O.~L. Landen, A.~R. Christopherson, and R.~Betti.
\newblock {Progress toward a self-consistent set of 1D ignition capsule metrics in ICF}.
\newblock {\em Physics of Plasmas}, 25(12):122704, 12 2018.

\bibitem{Lindl1995}
John Lindl.
\newblock {Development of the indirect-drive approach to inertial confinement fusion and the target physics basis for ignition and gain}.
\newblock {\em Phys. Plasmas}, 2(11):3933--4024, 1995.

\bibitem{SchmitKnapp2014PRL}
P.~F. Schmit, P.~F. Knapp, S.~B. Hansen, M.~R. Gomez, K.~D. Hahn, D.~B. Sinars, K.~J. Peterson, S.~A. Slutz, A.~B. Sefkow, T.~J. Awe, E.~Harding, C.~A. Jennings, G.~A. Chandler, G.~W. Cooper, M.~E. Cuneo, M.~Geissel, A.~J. Harvey-Thompson, M.~C. Herrmann, M.~H. Hess, O.~Johns, D.~C. Lamppa, M.~R. Martin, R.~D. McBride, J.~L. Porter, G.~K. Robertson, G.~A. Rochau, D.~C. Rovang, C.~L. Ruiz, M.~E. Savage, I.~C. Smith, W.~A. Stygar, and R.~A. Vesey.
\newblock Understanding fuel magnetization and mix using secondary nuclear reactions in magneto-inertial fusion.
\newblock {\em Phys. Rev. Lett.}, 113:155004, Oct 2014.

\bibitem{KnappSchmit2015}
P.~F. Knapp, P.~F. Schmit, S.~B. Hansen, M.~R. Gomez, K.~D. Hahn, D.~B. Sinars, K.~J. Peterson, S.~A. Slutz, A.~B. Sefkow, T.~J. Awe, E.~Harding, C.~A. Jennings, M.~P. Desjarlais, G.~A. Chandler, G.~W. Cooper, M.~E. Cuneo, M.~Geissel, A.~J. Harvey-Thompson, J.~L. Porter, G.~A. Rochau, D.~C. Rovang, C.~L. Ruiz, M.~E. Savage, I.~C. Smith, W.~A. Stygar, and M.~C. Herrmann.
\newblock {Effects of magnetization on fusion product trapping and secondary neutron spectraa)}.
\newblock {\em Physics of Plasmas}, 22(5):056312, 05 2015.

\bibitem{Lewis2021BRwithML}
William~E. Lewis, Patrick~F. Knapp, Stephen~A. Slutz, Paul~F. Schmit, Gordon~A. Chandler, Matthew~R. Gomez, Adam~J. Harvey-Thompson, Michael~A. Mangan, David~J. Ampleford, and Kristian Beckwith.
\newblock {Deep-learning-enabled Bayesian inference of fuel magnetization in magnetized liner inertial fusion}.
\newblock {\em Physics of Plasmas}, 28(9):092701, 09 2021.

\bibitem{Ruiz2023tscaling}
D.~E. Ruiz, P.~F. Schmit, M.~R. Weis, K.~J. Peterson, and M.~K. Matzen.
\newblock Exploring the parameter space of maglif implosions using similarity scaling. iii. rise-time scaling.
\newblock {\em Physics of Plasmas}, 30(3):032709, 2023.

\bibitem{mcbride2018primer}
RD~McBride, WA~Stygar, ME~Cuneo, DB~Sinars, MG~Mazarakis, JJ~Leckbee, ME~Savage, BT~Hutsel, JD~Douglass, ML~Kiefer, et~al.
\newblock A primer on pulsed power and linear transformer drivers for high energy density physics applications.
\newblock {\em IEEE Transactions on Plasma Science}, 46(11):3928--3967, 2018.

\bibitem{McBride_2015}
Ryan~D. McBride and Stephen~A. Slutz.
\newblock {A semi-analytic model of magnetized liner inertial fusion}.
\newblock {\em Physics of Plasmas}, 22(5):052708, 05 2015.

\bibitem{Kritcher2024}
A.~L. Kritcher et~al.
\newblock {Design of the first fusion experiment to achieve target energy gain $G>1$}.
\newblock {\em Phys. Rev. E}, 109:025204, Feb 2024.

\bibitem{Campbell_2017}
E.M. Campbell, V.N. Goncharov, T.C. Sangster, S.P. Regan, P.B. Radha, R.~Betti, J.F. Myatt, D.H. Froula, M.J. Rosenberg, I.V. Igumenshchev, W.~Seka, A.A. Solodov, A.V. Maximov, J.A. Marozas, T.J.B. Collins, D.~Turnbull, F.J. Marshall, A.~Shvydky, J.P. Knauer, R.L. McCrory, A.B. Sefkow, M.~Hohenberger, P.A. Michel, T.~Chapman, L.~Masse, C.~Goyon, S.~Ross, J.W. Bates, M.~Karasik, J.~Oh, J.~Weaver, A.J. Schmitt, K.~Obenschain, S.P. Obenschain, S.~Reyes, and B.~Van~Wonterghem.
\newblock {Laser-direct-drive program: Promise, challenge, and path forward}.
\newblock {\em Matter and Radiation at Extremes}, 2(2):37--54, 03 2017.

\bibitem{Marinak_sims_ignition2024}
M.~M. Marinak, G.~B. Zimmerman, T.~Chapman, G.~D. Kerbel, M.~V. Patel, J.~M. Koning, S.~M. Sepke, B.~Chang, C.~R. Schroeder, J.~A. Harte, D.~S. Bailey, L.~A. Taylor, S.~H. Langer, M.~A. Belyaev, D.~S. Clark, J.~Gaffney, B.~A. Hammel, D.~E. Hinkel, A.~L. Kritcher, J.~L. Milovich, H.~F. Robey, and C.~R. Weber.
\newblock {How numerical simulations helped to achieve breakeven on the NIF}.
\newblock {\em Physics of Plasmas}, 31(7):070501, 07 2024.

\bibitem{Knapp_2017}
P.~F. Knapp et~al.
\newblock {Direct measurement of the inertial confinement time in a magnetically driven implosion}.
\newblock {\em Phys. Plasmas}, 24(4):042708, 04 2017.

\bibitem{Sinars_2011}
D.~B. Sinars, S.~A. Slutz, M.~C. Herrmann, R.~D. McBride, M.~E. Cuneo, C.~A. Jennings, J.~P. Chittenden, A.~L. Velikovich, K.~J. Peterson, R.~A. Vesey, C.~Nakhleh, E.~M. Waisman, B.~E. Blue, K.~Killebrew, D.~Schroen, K.~Tomlinson, A.~D. Edens, M.~R. Lopez, I.~C. Smith, J.~Shores, V.~Bigman, G.~R. Bennett, B.~W. Atherton, M.~Savage, W.~A. Stygar, G.~T. Leifeste, and J.~L. Porter.
\newblock {Measurements of magneto-Rayleigh–Taylor instability growth during the implosion of initially solid metal liners}.
\newblock {\em Phys. Plasmas}, 18(5):056301, 04 2011.

\bibitem{Fryxell2000}
B.~Fryxell et~al.
\newblock {FLASH}: An adaptive mesh hydrodynamics code for modeling astrophysical thermonuclear flashes.
\newblock {\em Astrophys. J. Suppl. Ser.}, 131(1):273--334, nov 2000.

\bibitem{Tzeferacos2018}
P.~Tzeferacos et~al.
\newblock Laboratory evidence of dynamo amplification of magnetic fields in a turbulent plasma.
\newblock {\em Nat. Commun.}, 9(1):1--8, 2018.

\bibitem{Bott2021}
Archie F.~A. Bott, Petros Tzeferacos, Laura Chen, Charlotte A.~J. Palmer, Alexandra Rigby, Anthony~R. Bell, Robert Bingham, Andrew Birkel, Carlo Graziani, Dustin~H. Froula, et~al.
\newblock Time-resolved turbulent dynamo in a laser plasma.
\newblock {\em Proc. Natl. Acad. Sci. U. S. A.}, 118(11), 2021.

\bibitem{Fatenejad2013}
Milad Fatenejad, B.~Fryxell, J.~Wohlbier, E.~Myra, D.~Lamb, C.~Fryer, and C.~Graziani.
\newblock Collaborative comparison of simulation codes for high-energy-density physics applications.
\newblock {\em High Energy Density Phys.}, 9(1):63--66, 2013.

\bibitem{Sauppe2023}
J.~P. Sauppe, Y.~Lu, P.~Tzeferacos, A.~C. Reyes, S.~Palaniyappan, K.~A. Flippo, S.~Li, and J.~L. Kline.
\newblock {On the importance of three-dimensional modeling for high-energy-density physics experiments}.
\newblock {\em Phys. Plasmas}, 30(6):062707, 06 2023.

\bibitem{Tzeferacos2015}
P.~Tzeferacos et~al.
\newblock {FLASH MHD simulations of experiments that study shock-generated magnetic fields}.
\newblock {\em High Energy Density Phys.}, 17, Part A:24--31, 2015.

\bibitem{Lee2013}
Dongwook {Lee}.
\newblock A solution accurate, efficient and stable unsplit staggered mesh scheme for three dimensional magnetohydrodynamics.
\newblock {\em J. Computat. Phys.}, 243:269--292, jun 2013.

\bibitem{JiHeld2013}
Jeong-Young Ji and Eric~D. Held.
\newblock Closure and transport theory for high-collisionality electron-ion plasmas.
\newblock {\em Phys. Plasmas}, 20(4):042114, 2013.

\bibitem{Tzeferacos2017}
P~Tzeferacos et~al.
\newblock Numerical modeling of laser-driven experiments aiming to demonstrate magnetic field amplification via turbulent dynamo.
\newblock {\em Phys. Plasmas}, 24(4):041404, 2017.

\bibitem{ellison_flashvnv_2024}
C.~L. Ellison et~al.
\newblock Verification and validation of flash for pulsed magnetic fusion target design.
\newblock {\em In Preparation}, 2024.

\bibitem{Sinars_2010}
D.~B. Sinars et~al.
\newblock {Measurements of Magneto-Rayleigh-Taylor Instability Growth during the Implosion of Initially Solid Al Tubes Driven by the 20-MA, 100-ns Z Facility}.
\newblock {\em Phys. Rev. Lett.}, 105:185001, Oct 2010.

\bibitem{Knapp_2020}
P.~F. Knapp, M.~R. Martin, D.~Yager-Elorriaga, A.~J. Porwitzky, F.~W. Doss, G.~A. Shipley, C.~A. Jennings, D.~E. Ruiz, T.~Byvank, C.~C. Kuranz, C.~E. Myers, D.~H. Dolan, K.~Cochrane, M.~Schollmeier, I.~C. Smith, T.~R. Mattsson, B.~M. Jones, K.~Peterson, J.~Schwarz, R.~D. McBride, D.~G. Flicker, and D.~B. Sinars.
\newblock {A novel, magnetically driven convergent Richtmyer–Meshkov platform}.
\newblock {\em Phys. Plasmas}, 27(9):092707, 09 2020.

\bibitem{HarveyThompson2024}
A.~J. Harvey-Thompson, M.~E. Geissel, W.~E. Lewis, D.~A. Yager-Elorriaga, M.~R. Weis, C.~A. Jennings, J.~R. Fein, D.~J. Ampleford, M.~R. Gomez, E.~C. Harding, S.~B. Hansen, D.~E. Bliss, G.~A. Chandler, G.~S. Dunham, E.~S. Field, B.~R. Galloway, M.~Glinsky, K.~D. Hahn, P.~F. Knapp, D.~C. Lamppa, L.~M. Lucero, M.~A. Mangan, R.~R. Paguio, L.~Perea, K.~J. Peterson, J.~L. Porter, P.~K. Rambo, G.~K. Robertson, G.~A. Rochau, D.~E. Ruiz, C.~L. Ruiz, M.~A. Schaeuble, J.~Schwarz, J.~E. Shores, D.~B. Sinars, S.~A. Slutz, G.~E. Smith, I.~C. Smith, C.~S. Speas, K.~Whittemore, and E.~P. Yu.
\newblock {Development of a high performance MagLIF target platform using high aspect ratio coated liners and low-mix laser preheat}.
\newblock {\em Physics of Plasmas}, 31(7):072711, 07 2024.

\bibitem{Hurricane2023RMP}
OA~Hurricane, PK~Patel, R~Betti, DH~Froula, SP~Regan, SA~Slutz, MR~Gomez, and MA~Sweeney.
\newblock Physics principles of inertial confinement fusion and us program overview.
\newblock {\em Reviews of Modern Physics}, 95(2):025005, 2023.

\bibitem{NoraBetti2014}
R.~Nora, R.~Betti, K.~S. Anderson, A.~Shvydky, A.~Bose, K.~M. Woo, A.~R. Christopherson, J.~A. Marozas, T.~J.~B. Collins, P.~B. Radha, S.~X. Hu, R.~Epstein, F.~J. Marshall, R.~L. McCrory, T.~C. Sangster, and D.~D. Meyerhofer.
\newblock {Theory of hydro-equivalent ignition for inertial fusion and its applications to OMEGA and the National Ignition Facilitya)}.
\newblock {\em Physics of Plasmas}, 21(5):056316, 05 2014.

\bibitem{Landen2024}
O.~L. Landen, R.~C. Nora, J.~D. Lindl, A.~L. Kritcher, S.~W. Haan, M.~D. Rosen, A.~Pak, L.~Divol, K.~L. Baker, P.~A. Amendt, D.~D.-M. Ho, J.~L. Milovich, J.~E. Ralph, D.~S. Clark, K.~D. Humbird, M.~Hohenberger, C.~R. Weber, R.~Tommasini, D.~T. Casey, C.~V. Young, D.~J. Schlossberg, S.~A. Maclaren, E.~L. Dewald, P.~F. Schmit, T.~Chapman, D.~E. Hinkel, J.~D. Moody, V.~A. Smalyuk, O.~A. Hurricane, and R.~P.~J. Town.
\newblock {What next: Further implosion space exploration on the path to NIF extended yield capability}.
\newblock {\em Physics of Plasmas}, 31(6):062712, 06 2024.

\bibitem{Lindemuth2024}
Irvin~R. Lindemuth.
\newblock {A fusion-grade 20-MA liner-on-plasma implosion system without preheat and premagnetization}.
\newblock {\em Physics of Plasmas}, 31(7):072707, 07 2024.

\bibitem{Mead_IAEA_1990}
W~C Mead, B~Bezzerides, S~V Coggeshall, M~Cray, H~N Fisher, N~M Hoffman, and G~R Magelssen.
\newblock A current prediction of the {ICF} gain curve, its uncertainties, and implications for {ICF} strategy.
\newblock In {\em {Plasma Physics and Controlled Nuclear Fusion Research: Thirteenth Conference Proceedings}}, pages 159--164. IAEA, 1991.

\bibitem{Zimmerman1975}
G.~B. Zimmerman and W.~L. Kruer.
\newblock Numerical simulation of laser-initiated fusion.
\newblock {\em Comments Plasma Phys Controlled Fusion}, 2(2):51--60, 1975.

\bibitem{ClarkHero3D_PoP2015}
D.~S. Clark et~al.
\newblock {Radiation hydrodynamics modeling of the highest compression inertial confinement fusion ignition experiment from the National Ignition Campaign}.
\newblock {\em Phys. Plasmas}, 22(2), 2015.

\bibitem{Hurricane2014}
O.~A. Hurricane, D.~A. Callahan, D.~T. Casey, P.~M. Celliers, C.~Cerjan, E.~L. Dewald, T.~R. Dittrich, T.~Doppner, D.~E. Hinkel, L.~F.~Berzak Hopkins, J.~L. Kline, S.~Le~Pape, T.~Ma, A.~G. MacPhee, J.~L. Milovich, A.~Pak, H.~S. Park, P.~K. Patel, B.~A. Remington, J.~D. Salmonson, P.~T. Springer, and R.~Tommasini.
\newblock Fuel gain exceeding unity in an inertially confined fusion implosion.
\newblock {\em Nature}, 506(7488):343--348, 02 2014.

\bibitem{LePape2018}
S.~Le~Pape, L.~F. Berzak~Hopkins, L.~Divol, A.~Pak, E.~L. Dewald, S.~Bhandarkar, L.~R. Bennedetti, T.~Bunn, J.~Biener, J.~Crippen, D.~Casey, D.~Edgell, D.~N. Fittinghoff, M.~Gatu-Johnson, C.~Goyon, S.~Haan, R.~Hatarik, M.~Havre, D.~D-M. Ho, N.~Izumi, J.~Jaquez, S.~F. Khan, G.~A. Kyrala, T.~Ma, A.~J. Mackinnon, A.~G. MacPhee, B.~J. MacGowan, N.~B. Meezan, J.~Milovich, M.~Millot, P.~Michel, S.~R. Nagel, A.~Nikroo, P.~Patel, J.~Ralph, J.~S. Ross, N.~G. Rice, D.~Strozzi, M.~Stadermann, P.~Volegov, C.~Yeamans, C.~Weber, C.~Wild, D.~Callahan, and O.~A. Hurricane.
\newblock Fusion energy output greater than the kinetic energy of an imploding shell at the national ignition facility.
\newblock {\em Phys. Rev. Lett.}, 120:245003, Jun 2018.

\bibitem{Kritcher2018}
A.~L. Kritcher, D.~Clark, S.~Haan, S.~A. Yi, A.~B. Zylstra, D.~A. Callahan, D.~E. Hinkel, L.~F. Berzak~Hopkins, O.~A. Hurricane, O.~L. Landen, S.~A. MacLaren, N.~B. Meezan, P.~K. Patel, J.~Ralph, C.~A. Thomas, R.~Town, and M.~J. Edwards.
\newblock {Comparison of plastic, high density carbon, and beryllium as indirect drive NIF ablators}.
\newblock {\em Physics of Plasmas}, 25(5):056309, 04 2018.

\bibitem{Hurricane2016NaturePhysics}
O.~A. Hurricane, D.~A. Callahan, D.~T. Casey, E.~L. Dewald, T.~R. Dittrich, T.~Doppner, S.~Haan, D.~E. Hinkel, L.~F. Berzak~Hopkins, O.~Jones, A.~L. Kritcher, S.~Le~Pape, T.~Ma, A.~G. MacPhee, J.~L. Milovich, J.~Moody, A.~Pak, H.~S. Park, P.~K. Patel, J.~E. Ralph, H.~F. Robey, J.~S. Ross, J.~D. Salmonson, B.~K. Spears, P.~T. Springer, R.~Tommasini, F.~Albert, L.~R. Benedetti, R.~Bionta, E.~Bond, D.~K. Bradley, J.~Caggiano, P.~M. Celliers, C.~Cerjan, J.~A. Church, R.~Dylla-Spears, D.~Edgell, M.~J. Edwards, D.~Fittinghoff, M.~A. Barrios~Garcia, A.~Hamza, R.~Hatarik, H.~Herrmann, M.~Hohenberger, D.~Hoover, J.~L. Kline, G.~Kyrala, B.~Kozioziemski, G.~Grim, J.~E. Field, J.~Frenje, N.~Izumi, M.~Gatu~Johnson, S.~F. Khan, J.~Knauer, T.~Kohut, O.~Landen, F.~Merrill, P.~Michel, A.~Moore, S.~R. Nagel, A.~Nikroo, T.~Parham, R.~R. Rygg, D.~Sayre, M.~Schneider, D.~Shaughnessy, D.~Strozzi, R.~P.~J. Town, D.~Turnbull, P.~Volegov, A.~Wan, K.~Widmann, C.~Wilde, and C.~Yeamans.
\newblock Inertially confined fusion plasmas dominated by alpha-particle self-heating.
\newblock {\em Nat Phys}, advance online publication, 04 2016.

\bibitem{Patel2020HotSpot}
P.~K. Patel, P.~T. Springer, C.~R. Weber, L.~C. Jarrott, O.~A. Hurricane, B.~Bachmann, K.~L. Baker, L.~F. Berzak~Hopkins, D.~A. Callahan, D.~T. Casey, C.~J. Cerjan, D.~S. Clark, E.~L. Dewald, L.~Divol, T.~Döppner, J.~E. Field, D.~Fittinghoff, J.~Gaffney, V.~Geppert-Kleinrath, G.~P. Grim, E.~P. Hartouni, R.~Hatarik, D.~E. Hinkel, M.~Hohenberger, K.~Humbird, N.~Izumi, O.~S. Jones, S.~F. Khan, A.~L. Kritcher, M.~Kruse, O.~L. Landen, S.~Le~Pape, T.~Ma, S.~A. MacLaren, A.~G. MacPhee, L.~P. Masse, N.~B. Meezan, J.~L. Milovich, R.~Nora, A.~Pak, J.~L. Peterson, J.~Ralph, H.~F. Robey, J.~D. Salmonson, V.~A. Smalyuk, B.~K. Spears, C.~A. Thomas, P.~L. Volegov, A.~Zylstra, and M.~J. Edwards.
\newblock {Hotspot conditions achieved in inertial confinement fusion experiments on the National Ignition Facility}.
\newblock {\em Physics of Plasmas}, 27(5):050901, 05 2020.

\bibitem{RuizIFSA2023}
D.E. Ruiz et~al.
\newblock {Similarity scaling MagLIF loads to achieve high fusion yields (>200 MJ) in the laboratory, 12th International Conference on Inertial Fusion Sciences and Applications, Denver, CO}.
\newblock 2023.

\bibitem{Haines_2011}
M~G Haines.
\newblock A review of the dense z-pinch.
\newblock {\em Plasma Physics and Controlled Fusion}, 53(9):093001, jun 2011.

\bibitem{krishnan_2012}
Mahadevan Krishnan.
\newblock The dense plasma focus: A versatile dense pinch for diverse applications.
\newblock {\em IEEE Transactions on Plasma Science}, 40(12):3189--3221, 2012.

\bibitem{Klir_PPCF_2012}
D~Klir, P~Kubes, M~Paduch, T~Pisarczyk, T~Chodukowski, M~Scholz, Z~Kalinowska, B~Bienkowska, L~Karpinski, J~Kortanek, J~Kravarik, K~Rezac, I~Ivanova-Stanik, K~Tomaszewski, and E~Zielinska.
\newblock Search for thermonuclear neutrons in a mega-ampere plasma focus.
\newblock {\em Plasma Physics and Controlled Fusion}, 54(1):015001, dec 2011.

\bibitem{Schmidt_PoP_2014}
A.~Schmidt, A.~Link, D.~Welch, B.~T. Meehan, V.~Tang, C.~Halvorson, M.~May, and E.~C. Hagen.
\newblock {Fully kinetic simulations of megajoule-scale dense plasma focus}.
\newblock {\em Physics of Plasmas}, 21(10):102703, 10 2014.

\bibitem{Klir_PRL_2014}
D.~Klir, P.~Kubes, K.~Rezac, J.~Cikhardt, J.~Kravarik, O.~Sila, A.~V. Shishlov, B.~M. Kovalchuk, N.~A. Ratakhin, V.~A. Kokshenev, A.~Yu. Labetsky, R.~K. Cherdizov, F.~I. Fursov, N.~E. Kurmaev, G.~N. Dudkin, B.~A. Nechaev, V.~N. Padalko, H.~Orcikova, and K.~Turek.
\newblock Efficient neutron production from a novel configuration of deuterium gas-puff $z$-pinch.
\newblock {\em Phys. Rev. Lett.}, 112:095001, Mar 2014.

\bibitem{coverdale_2007}
C.~A. Coverdale, C.~Deeney, A.~L. Velikovich, J.~Davis, R.~W. Clark, Y.~K. Chong, J.~Chittenden, S.~Chantrenne, C.~L. Ruiz, G.~W. Cooper, A.~J. Nelson, J.~Franklin, P.~D. LePell, J.~P. Apruzese, J.~Levine, and J.~Banister.
\newblock {Deuterium gas-puff Z-pinch implosions on the Z acceleratora)}.
\newblock {\em Physics of Plasmas}, 14(5):056309, 04 2007.

\bibitem{velikovich_2007}
A.~L. Velikovich, R.~W. Clark, J.~Davis, Y.~K. Chong, C.~Deeney, C.~A. Coverdale, C.~L. Ruiz, G.~W. Cooper, A.~J. Nelson, J.~Franklin, and L.~I. Rudakov.
\newblock {Z-pinch plasma neutron sources}.
\newblock {\em Physics of Plasmas}, 14(2):022701, 02 2007.

\bibitem{Welch_PoP_2010}
D.~R. Welch, D.~V. Rose, C.~Thoma, R.~E. Clark, C.~B. Mostrom, W.~A. Stygar, and R.~J. Leeper.
\newblock {Kinetic simulation of thermonuclear-neutron production by a 107-A deuterium Z pinch}.
\newblock {\em Physics of Plasmas}, 17(7):072702, 07 2010.

\bibitem{Appelbe_PoP_2015}
B.~Appelbe and J.~Chittenden.
\newblock {Neutron spectra from beam-target reactions in dense Z-pinches}.
\newblock {\em Physics of Plasmas}, 22(10):102703, 10 2015.

\bibitem{Offerman_PRL_2016}
Dustin~T. Offermann, Dale~R. Welch, Dave~V. Rose, Carsten Thoma, Robert~E. Clark, Chris~B. Mostrom, Andrea E.~W. Schmidt, and Anthony~J. Link.
\newblock Transition from beam-target to thermonuclear fusion in high-current deuterium $z$-pinch simulations.
\newblock {\em Phys. Rev. Lett.}, 116:195001, May 2016.

\bibitem{Knapp_PoP_2013}
P.~F. Knapp, D.~B. Sinars, and K.~D. Hahn.
\newblock Diagnosing suprathermal ion populations in z-pinch plasmas using fusion neutron spectra.
\newblock {\em Physics of Plasmas (1994-present)}, 20(6):--, 2013.

\bibitem{Gomez_PRL_2014}
M.~R. Gomez, S.~A. Slutz, A.~B. Sefkow, D.~B. Sinars, K.~D. Hahn, S.~B. Hansen, E.~C. Harding, P.~F. Knapp, P.~F. Schmit, C.~A. Jennings, T.~J. Awe, M.~Geissel, D.~C. Rovang, G.~A. Chandler, G.~W. Cooper, M.~E. Cuneo, A.~J. Harvey-Thompson, M.~C. Herrmann, M.~H. Hess, O.~Johns, D.~C. Lamppa, M.~R. Martin, R.~D. McBride, K.~J. Peterson, J.~L. Porter, G.~K. Robertson, G.~A. Rochau, C.~L. Ruiz, M.~E. Savage, I.~C. Smith, W.~A. Stygar, and R.~A. Vesey.
\newblock Experimental demonstration of fusion-relevant conditions in magnetized liner inertial fusion.
\newblock {\em Phys. Rev. Lett.}, 113:155003, Oct 2014.

\bibitem{Hahn_2016}
K~D Hahn, G~A Chandler, C~L Ruiz, G~W Cooper, M~R Gomez, S~Slutz, A~B Sefkow, D~B Sinars, S~B Hansen, P~F Knapp, P~F Schmit, E~Harding, C~A Jennings, T~J Awe, M~Geissel, D~C Rovang, J~A Torres, J~A Bur, M~E Cuneo, V~Yu Glebov, A~J Harvey-Thompson, M~C Herrman, M~H Hess, O~Johns, B~Jones, D~C Lamppa, J~S Lash, M~R Martin, R~D McBride, K~J Peterson, J~L Porter, J~Reneker, G~K Robertson, G~A Rochau, M~E Savage, I~C Smith, J~D Styron, and R~A Vesey.
\newblock Fusion-neutron measurements for magnetized liner inertial fusion experiments on the z accelerator.
\newblock {\em Journal of Physics: Conference Series}, 717(1):012020, may 2016.

\bibitem{Yager-Elorriaga_2022}
D.A. Yager-Elorriaga, M.R. Gomez, D.E. Ruiz, S.A. Slutz, A.J. Harvey-Thompson, C.A. Jennings, P.F. Knapp, P.F. Schmit, M.R. Weis, T.J. Awe, G.A. Chandler, M.~Mangan, C.E. Myers, J.R. Fein, B.R. Galloway, M.~Geissel, M.E. Glinsky, S.B. Hansen, E.C. Harding, D.C. Lamppa, W.E. Lewis, P.K. Rambo, G.K. Robertson, M.E. Savage, G.A. Shipley, I.C. Smith, J.~Schwarz, D.J. Ampleford, K.~Beckwith, K.J. Peterson, J.L. Porter, G.A. Rochau, and D.B. Sinars.
\newblock An overview of magneto-inertial fusion on the z machine at sandia national laboratories.
\newblock {\em Nuclear Fusion}, 62(4):042015, feb 2022.

\bibitem{Gomez_PRL_2020}
M.~R. Gomez, S.~A. Slutz, C.~A. Jennings, D.~J. Ampleford, M.~R. Weis, C.~E. Myers, D.~A. Yager-Elorriaga, K.~D. Hahn, S.~B. Hansen, E.~C. Harding, A.~J. Harvey-Thompson, D.~C. Lamppa, M.~Mangan, P.~F. Knapp, T.~J. Awe, G.~A. Chandler, G.~W. Cooper, J.~R. Fein, M.~Geissel, M.~E. Glinsky, W.~E. Lewis, C.~L. Ruiz, D.~E. Ruiz, M.~E. Savage, P.~F. Schmit, I.~C. Smith, J.~D. Styron, J.~L. Porter, B.~Jones, T.~R. Mattsson, K.~J. Peterson, G.~A. Rochau, and D.~B. Sinars.
\newblock Performance scaling in magnetized liner inertial fusion experiments.
\newblock {\em Phys. Rev. Lett.}, 125:155002, Oct 2020.

\bibitem{Knapp_PoP_2019}
P.~F. Knapp, M.~R. Gomez, S.~B. Hansen, M.~E. Glinsky, C.~A. Jennings, S.~A. Slutz, E.~C. Harding, K.~D. Hahn, M.~R. Weis, M.~Evans, M.~R. Martin, A.~J. Harvey-Thompson, M.~Geissel, I.~C. Smith, D.~E. Ruiz, K.~J. Peterson, B.~M. Jones, J.~Schwarz, G.~A. Rochau, D.~B. Sinars, R.~D. McBride, and P.-A. Gourdain.
\newblock {Origins and effects of mix on magnetized liner inertial fusion target performance}.
\newblock {\em Physics of Plasmas}, 26(1):012704, 01 2019.

\bibitem{Harvey_PoP_2018}
A.~J. Harvey-Thompson, M.~R. Weis, E.~C. Harding, M.~Geissel, D.~J. Ampleford, G.~A. Chandler, J.~R. Fein, M.~E. Glinsky, M.~R. Gomez, K.~D. Hahn, S.~B. Hansen, C.~A. Jennings, P.~F. Knapp, R.~R. Paguio, L.~Perea, K.~J. Peterson, J.~L. Porter, P.~K. Rambo, G.~K. Robertson, G.~A. Rochau, D.~E. Ruiz, J.~Schwarz, J.~E. Shores, D.~B. Sinars, S.~A. Slutz, G.~E. Smith, I.~C. Smith, C.~S. Speas, and K.~Whittemore.
\newblock {Diagnosing and mitigating laser preheat induced mix in MagLIF}.
\newblock {\em Physics of Plasmas}, 25(11):112705, 11 2018.

\bibitem{lechien2023sirius}
KR~LeChien et~al.
\newblock {Sirius I: prototype of a prime-power source for future 1-10 GJ fusion-yield experiments}.
\newblock Technical report, Lawrence Livermore National Lab.(LLNL), Livermore, CA (United States), 2023.

\bibitem{Allen2023}
Raymond~J. Allen and David Hinshelwood.
\newblock Overview of the castle circuit simulator.
\newblock In {\em 2023 IEEE Pulsed Power Conference (PPC)}, pages 1--3, 2023.

\bibitem{welch2020hybrid}
Dale~R Welch, Nichelle~L Bennett, Thomas~C Genoni, Carsten Thoma, and David~V Rose.
\newblock Fast hybrid particle-in-cell technique for pulsed-power accelerators.
\newblock {\em Physics of Accelerators and Beams}, 23(110401), 2020.

\bibitem{Creedon1975}
John~M. Creedon.
\newblock {Relativistic Brillouin flow in the high $\nu/\gamma$ diode}.
\newblock {\em Journal of Applied Physics}, 46(7):2946--2955, 07 1975.

\bibitem{Creedon1977}
John~M. Creedon.
\newblock {Magnetic cutoff in high‐current diodes}.
\newblock {\em Journal of Applied Physics}, 48(3):1070--1077, 03 1977.

\bibitem{VanDevender1979}
J.~P. VanDevender.
\newblock {Long self‐magnetically insulated power transport experiments}.
\newblock {\em Journal of Applied Physics}, 50(6):3928--3934, 06 1979.

\bibitem{DiCapua1983}
Marco~S. Di~Capua.
\newblock Magnetic insulation.
\newblock {\em IEEE Transactions on Plasma Science}, 11(3):205--215, 1983.

\bibitem{Mendel1992}
C.~W. Mendel, S.~E. Rosenthal, and D.~B. Seidel.
\newblock Low-pressure relativistic electron flow.
\newblock {\em Phys. Rev. A}, 45:5854--5865, Apr 1992.

\bibitem{Mendel1996}
Jr. Mendel, C.~W. and S.~E. Rosenthal.
\newblock {Dynamic modeling of magnetically insulated transmission line systems}.
\newblock {\em Physics of Plasmas}, 3(11):4207--4219, 11 1996.

\bibitem{Ottinger2006}
P.~F. Ottinger and J.~W. Schumer.
\newblock {Rescaling of equilibrium magnetically insulated flow theory based on results from particle-in-cell simulations}.
\newblock {\em Physics of Plasmas}, 13(6):063109, 06 2006.

\bibitem{Stygar2005}
W.~A. Stygar, J.~A. Lott, T.~C. Wagoner, V.~Anaya, H.~C. Harjes, H.~C. Ives, Z.~R. Wallace, G.~R. Mowrer, R.~W. Shoup, J.~P. Corley, R.~A. Anderson, G.~E. Vogtlin, M.~E. Savage, J.~M. Elizondo, B.~S. Stoltzfus, D.~M. Andercyk, D.~L. Fehl, T.~F. Jaramillo, D.~L. Johnson, D.~H. McDaniel, D.~A. Muirhead, J.~M. Radman, J.~J. Ramirez, L.~E. Ramirez, R.~B. Spielman, K.~W. Struve, D.~E. Walsh, E.~D. Walsh, and M.~D. Walsh.
\newblock Improved design of a high-voltage vacuum-insulator interface.
\newblock {\em Phys. Rev. ST Accel. Beams}, 8:050401, 2005.

\bibitem{Stygar2004}
W.~A. Stygar, H.~C. Ives, T.~C. Wagoner, J.~A. Lott, V.~Anaya, H.~C. Harjes, J.~P. Corley, R.~W. Shoup, D.~L. Fehl, G.~R. Mowrer, Z.~R. Wallace, R.~A. Anderson, J.~D. Boyes, J.~W. Douglas, M.~L. Horry, T.~F. Jaramillo, D.~L. Johnson, F.~W. Long, T.~H. Martin, D.~H. McDaniel, O.~Milton, M.~A. Mostrom, D.~A. Muirhead, T.~D. Mulville, J.~J. Ramirez, L.~E. Ramirez, T.~M. Romero, J.~F. Seamen, J.~W. Smith, C.~S. Speas, R.~B. Spielman, K.~W. Struve, G.~E. Vogtlin, D.~E. Walsh, E.~D. Walsh, M.~D. Walsh, and O.~Yamamoto.
\newblock Flashover of a vacuum-insulator interface: A statistical model.
\newblock {\em Phys. Rev. ST Accel. Beams}, 7:070401, Jul 2004.

\bibitem{stygar2015conceptual}
WA~Stygar et~al.
\newblock Conceptual designs of two petawatt-class pulsed-power accelerators for high-energy-density-physics experiments.
\newblock {\em Physical Review Special Topics-Accelerators and Beams}, 18(11):110401, 2015.

\bibitem{Bennett2019transport}
Nichelle Bennett, Dale~R Welch, Christopher~A Jennings, Edmund Yu, Michael~H Hess, Brian~T Hutsel, George Laity, J~K Moore, David~V Rose, Kyle Peterson, and Michael~E Cuneo.
\newblock Current transport and loss mechanisms in the z accelerator.
\newblock {\em Physics of Accelerators and Beams}, 23(120401), 2019.

\bibitem{Gomez2017}
M.~R. Gomez, R.~M. Gilgenbach, M.~E. Cuneo, C.~A. Jennings, R.~D. McBride, E.~M. Waisman, B.~T. Hutsel, W.~A. Stygar, D.~V. Rose, and Y.~Maron.
\newblock Experimental study of current loss and plasma formation in the $z$ machine post-hole convolute.
\newblock {\em Phys. Rev. Accel. Beams}, 20:010401, Jan 2017.

\bibitem{welch2019hybrid}
Dale~R Welch, Nichelle~L Bennett, Thomas~C Genoni, David~V Rose, Carsten Thoma, Craig Miller, and William~A Stygar.
\newblock Electrode contaminant plasma effects in 107-a z pinch accelerators.
\newblock {\em Physics of Accelerators and Beams}, 22(070401), 2019.

\bibitem{welch2009hybrid}
Dale~R Welch, David~V Rose, Nichelle Bruner, Robert~E Clark, Bryan~V Oliver, Kelly~D Hahn, and Mark~D Johnston.
\newblock Hybrid simulation of electrode plasmas in high-power diodes.
\newblock {\em Physics of Plasmas}, 16(12), 2009.

\bibitem{Bennett_milestone:2023}
N.~Bennett, Derek Lamppa, Andrew Porwitzky, Christopher Jennings, Dale Welch, Evstati Evstatiev, Clayton Myers, Kathy Chandler, Jacob Banecek, Sonal Patel, Eric Watson, David Yager-Elorriaga, Mark Savage, Mark Johnston, Mark Hess, David Rose, and Michael Cuneo.
\newblock Mrt 7365: Power flow physics and key physics phenomena.
\newblock Technical Report {SAND}2023-nnnn, Sandia National Laboratories, 2023.

\bibitem{tummel2022flow}
Kurt Tummel, Dale~R Welch, David~V Rose, Anthony~J Link, and Keith~R LeChien.
\newblock Impact of power flow on z-pinch loads.
\newblock {\em Physics of Plasmas}, 29(113102), 2022.

\bibitem{Rose2015}
D.~V. Rose, E.~A. Madrid, D.~R. Welch, R.~E. Clark, C.~B. Mostrom, W.~A. Stygar, and M.~E. Cuneo.
\newblock Computational analysis of current-loss mechanisms in a post-hole convolute driven by magnetically insulated transmission lines.
\newblock {\em Phys. Rev. ST Accel. Beams}, 18:030402, Mar 2015.

\bibitem{LangmuirSCL}
Irving Langmuir.
\newblock The effect of space charge and residual gases on thermionic currents in high vacuum.
\newblock {\em Phys. Rev.}, 2:450--486, Dec 1913.

\bibitem{Rose2008}
D.~V. Rose, D.~R. Welch, T.~P. Hughes, R.~E. Clark, and W.~A. Stygar.
\newblock Plasma evolution and dynamics in high-power vacuum-transmission-line post-hole convolutes.
\newblock {\em Phys. Rev. ST Accel. Beams}, 11:060401, Jun 2008.

\bibitem{Blaugrund1977}
A.~E. Blaugrund, G.~Cooperstein, and Shyke~A. Goldstein.
\newblock {Relativistic electron beam pinch formation processes in low impedance diodes}.
\newblock {\em The Physics of Fluids}, 20(7):1185--1194, 07 1977.

\bibitem{Stygar2009}
W.~A. Stygar, P.~A. Corcoran, H.~C. Ives, R.~B. Spielman, J.~W. Douglas, B.~A. Whitney, M.~A. Mostrom, T.~C. Wagoner, C.~S. Speas, T.~L. Gilliland, G.~A. Allshouse, R.~E. Clark, G.~L. Donovan, T.~P. Hughes, D.~R. Humphreys, D.~M. Jaramillo, M.~F. Johnson, J.~W. Kellogg, R.~J. Leeper, F.~W. Long, T.~H. Martin, T.~D. Mulville, M.~D. Pelock, B.~P. Peyton, J.~W. Poukey, J.~J. Ramirez, P.~G. Reynolds, J.~F. Seamen, D.~B. Seidel, A.~P. Seth A.~W. Sharpe, R.~W. Shoup, J.~W. Smith, D.~M. Van~De Valde, and R.~W. Wavrik.
\newblock 55-{TW} magnetically insulated transmission-line system: {D}esign, simulations, and performance.
\newblock {\em Phys. Rev. ST Accel. Beams}, 12:120401, 2009.

\bibitem{Bennett2021transport}
Nichelle~L Bennett, Dale~R Welch, Christopher~A Jennings, Edmund Yu, Michael~H Hess, Brian~T Hutsel, George Laity, J~K Moore, David~V Rose, Kyle Peterson, and Michael~E Cuneo.
\newblock Magnetized particle transport in multi-ma accelerators.
\newblock {\em Physics of Accelerators and Beams}, 24(060401), 2021.

\bibitem{Bennett_milestone2023}
N.~Bennett, Derek Lamppa, Andrew Porwitzky, Christopher Jennings, Dale Welch, Evstati Evstatiev, Clayton Myers, Kathy Chandler, Jacob Banecek, Sonal Patel, Eric Watson, David Yager-Elorriaga, Mark Savage, Mark Johnston, Mark Hess, David Rose, and Michael Cuneo.
\newblock Mrt 7365: Power flow physics and key physics phenomena.
\newblock Technical Report {SAND}2023-nnnn, Sandia National Laboratories, 2023.

\bibitem{kilkenny2016national}
JD~Kilkenny, PM~Bell, DK~Bradley, DL~Bleuel, JA~Caggiano, EL~Dewald, WW~Hsing, DH~Kalantar, RL~Kauffman, DJ~Larson, et~al.
\newblock The national ignition facility diagnostic set at the completion of the national ignition campaign, september 2012.
\newblock {\em Fusion Science and Technology}, 69(1):420--451, 2016.

\bibitem{hartouni2023evidence}
EP~Hartouni, AS~Moore, AJ~Crilly, BD~Appelbe, PA~Amendt, KL~Baker, DT~Casey, DS~Clark, T~D{\"o}ppner, MJ~Eckart, et~al.
\newblock Evidence for suprathermal ion distribution in burning plasmas.
\newblock {\em Nature Physics}, 19(1):72--77, 2023.

\bibitem{volegov2015three}
Petr~Lvovich Volegov, Christopher~R Danly, Frank~Edward Merrill, Raspberry Simpson, and Carl~Huerstel Wilde.
\newblock On three-dimensional reconstruction of a neutron/x-ray source from very few two-dimensional projections.
\newblock {\em Journal of Applied Physics}, 118(20), 2015.

\bibitem{herrmann2010diagnosing}
Hans~W Herrmann, N~Hoffman, DC~Wilson, W~Stoeffl, L~Dauffy, YH~Kim, A~McEvoy, CS~Young, JM~Mack, CJ~Horsfield, et~al.
\newblock Diagnosing inertial confinement fusion gamma ray physics.
\newblock {\em Review of Scientific Instruments}, 81(10), 2010.

\bibitem{fry2023geant4}
Cathleen Fry, Samual Langendorf, Yongho Kim, Kevin Meaney, Johnny Goett, Christopher Prokop, Justin Jorgenson, Tana Morrow, Matthias Hochanadel, Ramon Leeper, et~al.
\newblock Geant4 simulation towards understanding signals from a new proton recoil telescope for z.
\newblock In {\em APS Division of Plasma Physics Meeting Abstracts}, volume 2023, pages BP11--047, 2023.

\bibitem{gatu2022high}
M~Gatu~Johnson, TM~Johnson, BJ~Lahmann, FH~S{\'e}guin, B~Sperry, N~Bhandarkar, RM~Bionta, E~Casco, DT~Casey, AJ~Mackinnon, et~al.
\newblock High-yield magnetic recoil neutron spectrometer on the national ignition facility for operation up to 60 mj.
\newblock {\em Review of Scientific Instruments}, 93(8), 2022.

\bibitem{bleuel2012neutron}
Darren~L Bleuel, CB~Yeamans, LA~Bernstein, RM~Bionta, JA~Caggiano, DT~Casey, GW~Cooper, OB~Drury, JA~Frenje, CA~Hagmann, et~al.
\newblock Neutron activation diagnostics at the national ignition facility.
\newblock {\em Review of Scientific Instruments}, 83(10), 2012.

\bibitem{khan2012measuring}
SF~Khan, PM~Bell, DK~Bradley, SR~Burns, JR~Celeste, LS~Dauffy, MJ~Eckart, MA~Gerhard, C~Hagmann, DI~Headley, et~al.
\newblock Measuring x-ray burn history with the streaked polar instrumentation for diagnosing energetic radiation (spider) at the national ignition facility (nif).
\newblock In {\em Target Diagnostics Physics and Engineering for Inertial Confinement Fusion}, volume 8505, pages 33--40. SPIE, 2012.

\bibitem{chandler1999filtered}
GA~Chandler, C~Deeney, M~Cuneo, DL~Fehl, JS~McGurn, RB~Spielman, JA~Torres, JL~McKenney, J~Mills, and KW~Struve.
\newblock Filtered x-ray diode diagnostics fielded on the z accelerator for source power measurements.
\newblock {\em Review of Scientific Instruments}, 70(1):561--565, 1999.

\bibitem{datte2020design}
P~Datte, J~Baker, D~Bliss, N~Butler, P~Celliers, S~Cohen, M~Crosley, J~Edwards, D~Erskine, D~Fratanduono, et~al.
\newblock The design of a line velocity interferometer for any reflector for inertial confinement experiments on the z-machine.
\newblock {\em Review of Scientific Instruments}, 91(4), 2020.

\bibitem{Knapp_2023}
P.~F. Knapp and W.~E. Lewis.
\newblock {Advanced data analysis in inertial confinement fusion and high energy density physics}.
\newblock {\em Review of Scientific Instruments}, 94(6):061103, 06 2023.

\bibitem{Knapp_JPP_2023}
P.F. Knapp, W.E. Lewis, V.R. Joseph, C.A. Jennings, and M.E. Glinsky.
\newblock Optimizing the configuration of plasma radiation detectors in the presence of uncertain instrument response and inadequate physics.
\newblock {\em Journal of Plasma Physics}, 89(1):895890101, 2023.

\bibitem{FPA2010}
K.~Matzen, M.~Herrmann, M.~Cuneo, and et~al.
\newblock Pulsed power inertial fusion energy.
\newblock \url{https://fire.pppl.gov/IFE_NAS2_SNL_Cuneo_Pulse_Power.pdf}, 2011.

\bibitem{CSI}
CSI.
\newblock Blue catalog - high voltage capacitors 500 to 250,000 volts.
\newblock General Catalog, 2024.

\bibitem{slutz2003low}
SA~Slutz, CL~Olson, and Per Peterson.
\newblock Low mass recyclable transmission lines for z-pinch driven inertial fusion.
\newblock {\em Physics of Plasmas}, 10(2):429--437, 2003.

\bibitem{masters2016debris}
ND~Masters, A~Fisher, D~Kalantar, J~St{\"o}lken, C~Smith, R~Vignes, S~Burns, T~Doeppner, A~Kritcher, and HS~Park.
\newblock Debris and shrapnel assessments for national ignition facility targets and diagnostics.
\newblock In {\em Journal of Physics: Conference Series}, volume 717, page 012108. IOP Publishing, 2016.

\bibitem{masters2014evaluation}
ND~Masters, A~Fisher, D~Kalantar, R~Prasad, JS~St{\"o}lken, and C~Wlodarczyk.
\newblock Evaluation of observed blast loading effects on nif x-ray diagnostic collimators.
\newblock {\em Review of Scientific Instruments}, 85(11), 2014.

\bibitem{yager2021optical}
David~A Yager-Elorriaga, Michael~M Montoya, David Bliss, Christopher~R Ball, Phillip Atencio, Brian~C Carpenter, Kyle Fuerschbach, Karin Fulford, Derek~C Lamppa, Michael~C Lowinske, et~al.
\newblock Optical imaging on z ldrd: Design and development of self-emission and debris imagers.
\newblock Technical report, Sandia National Lab.(SNL-NM), Albuquerque, NM (United States), 2021.

\bibitem{de2003recyclable}
JS~De~Groot, Craig~Lee Olson, Kyle~Robert Cochrane, Stephen~A Slutz, Roger~Alan Vesey, and Per~F Peterson.
\newblock Recyclable transmission line concept for z-pinch driven inertial fusion energy.
\newblock Technical report, Sandia National Laboratories (SNL), Albuquerque, NM, and Livermore, CA~…, 2003.

\bibitem{sharpe2007recyclable}
Robin~Arthur Sharpe, Alexander~S Kingsep, David~Lewis Smith, Craig~Lee Olson, Paul~F Ottinger, Joseph~Wade Schumer, Dale~Robert Welch, Alexander Kim, Gerald~L Kulcinski, Daniel~C Kammer, et~al.
\newblock Recyclable transmission line (rtl) and linear transformer driver (ltd) development for z-pinch inertial fusion energy (z-ife) and high yield.
\newblock Technical report, Sandia National Laboratories (SNL), Albuquerque, NM, and Livermore, CA~…, 2007.

\bibitem{blink1985high}
James~A Blink, WJ~Hogam, Jack Hovingh, ER~Meier, and John~H Pitts.
\newblock High-yield lithium-injection fusion-energy (hylife) reactor.
\newblock Technical report, Lawrence Livermore National Lab.(LLNL), Livermore, CA (United States), 1985.

\bibitem{moir1991hylife}
Ralph~W Moir.
\newblock Hylife-ii inertial confinement fusion reactor design.
\newblock {\em Fusion Technology}, 19(3P2A):617--624, 1991.

\bibitem{ogando2024preliminary}
Francisco Ogando, Michael~T Tobin, Wayne~R Meier, Gonzalo Farga-Ni{\~n}oles, Jaime Marian, Susana Reyes, Javier Sanz, and Conner Galloway.
\newblock Preliminary nuclear analysis of hylife-iii: A thick-liquid-wall chamber for inertial fusion energy.
\newblock {\em Fusion Engineering and Design}, 202:114333, 2024.

\bibitem{olson2004z}
Craig~L Olson et~al.
\newblock Z-pinch inertial fusion energy.
\newblock {\em Landholt-Boernstein Handbook on Energy Technologies (Editor in chief; W. Martienssen)}, 8:3, 2004.

\bibitem{anderson2006z}
Mark Anderson, Gerald Kulcinski, Haihua Zhao, Benjamin~B Cipiti, Craig~Lee Olson, Dannelle~P Sierra, Wayne Meier, Paul~E McConnell, M~Ghiaasiaan, Brian Kern, et~al.
\newblock Z-inertial fusion energy: power plant final report fy 2006.
\newblock Technical report, Sandia National Laboratories (SNL), Albuquerque, NM, and Livermore, CA~…, 2006.

\bibitem{meier2004analyses}
W~Meier, R~Abbott, J~Latkowski, R~Moir, S~Reyes, and R~Schmitt.
\newblock Analyses in support of z-ife: Llnl progress report for fy-04.
\newblock Technical report, Lawrence Livermore National Lab.(LLNL), Livermore, CA (United States), 2004.

\bibitem{olson2005development}
C~Olson, G~Rochau, S~Slutz, C~Morrow, R~Olson, M~Cuneo, D~Hanson, G~Bennett, T~Sanford, J~Bailey, et~al.
\newblock Development path for z-pinch ife.
\newblock {\em Fusion Science and Technology}, 47(3):633--640, 2005.

\bibitem{abbott2005dynamic}
Ryan~P Abbott.
\newblock The dynamic response of thick-liquid shielding in z-ife reactors.
\newblock In {\em 21st IEEE/NPS Symposium on Fusion Engineering SOFE 05}, pages 1--3. IEEE, 2005.

\bibitem{NationalAcademySci_Fusion_2021}
National~Academy of~Engineering, Engineering National Academies~of Sciences, and Medicine.
\newblock {\em Bringing Fusion to the U.S. Grid}.
\newblock The National Academies Press, Washington, DC, 2021.

\bibitem{NNSA24Enterprise}
NNSA.
\newblock {\em NNSA 2024 Enterprise Blueprint}.
\newblock October 2024.

\bibitem{Betti2010IgnitionMetrics}
R.~Betti, P.~Y. Chang, B.~K. Spears, K.~S. Anderson, J.~Edwards, M.~Fatenejad, J.~D. Lindl, R.~L. McCrory, R.~Nora, and D.~Shvarts.
\newblock {Thermonuclear ignition in inertial confinement fusion and comparison with magnetic confinementa)}.
\newblock {\em Physics of Plasmas}, 17(5):058102, 04 2010.

\end{thebibliography}

\end{document}